%% file: paper1.tex
\documentclass[tighten,twocolappendix,twocolumn]{aastex631}

\usepackage{amsmath,footmisc}
\usepackage[T1]{fontenc}

\newcommand{\ab}[1]{\left|#1\right|}
\newcommand{\av}[1]{\left\langle#1\right\rangle}
\newcommand{\br}[1]{\left[#1\right]}
\newcommand{\cu}[1]{\left\{#1\right\}}
\newcommand{\pa}[1]{\left(#1\right)}
\newcommand{\ed}{\mathop{}\!\mathrm{d}}
\newcommand{\pd}{\mathop{}\!\partial}
\DeclareMathOperator\sign{sign}

\graphicspath{{figures/}}

\defcitealias{PaperI}{EHTC~I}
\defcitealias{PaperV}{EHTC~V}
\defcitealias{PaperVI}{EHTC~VI}
\defcitealias{PaperVII}{EHTC~VII}
\defcitealias{PaperVIII}{EHTC~VIII}
\defcitealias{Paper2}{Paper II}
\defcitealias{Paper3}{Paper III}

\begin{document}

\title{Black Hole Polarimetry I: A Signature of Electromagnetic Energy Extraction}

\author[0000-0003-2966-6220]{Andrew Chael}
\correspondingauthor{Andrew Chael}
\email{achael@princeton.edu}
\affiliation{Princeton Gravity Initiative, Princeton University, Princeton NJ 08544, USA}

\author[0000-0002-1559-6965]{Alexandru Lupsasca}
\affiliation{Department of Physics \& Astronomy, Vanderbilt University, Nashville TN 37212, USA}

\author[0000-0001-6952-2147]{George~N.~Wong}
\affiliation{Princeton Gravity Initiative, Princeton University, Princeton NJ 08544, USA}
\affiliation{School of Natural Sciences, Institute for Advanced Study, Princeton NJ 08540, USA}

\author[0000-0001-9185-5044]{Eliot Quataert}
\affiliation{Princeton Gravity Initiative, Princeton University, Princeton NJ 08544, USA}
\affiliation{Department of Astrophysical Sciences, Princeton University, Princeton NJ 08544, USA}

\begin{abstract}
In 1977, Blandford and Znajek showed that the electromagnetic field surrounding a rotating black hole can harvest its spin energy and use it to power a collimated astrophysical jet, such as the one launched from the center of the elliptical galaxy M87.
Today, interferometric observations with the Event Horizon Telescope (EHT) are delivering high-resolution, event-horizon-scale, polarimetric images of the supermassive black hole M87* at the jet launching point.
These polarimetric images offer an unprecedented window into the electromagnetic field structure around a black hole.
In this paper, we show that a simple polarimetric observable---the phase $\angle\beta_2$ of the second azimuthal Fourier mode of the linear polarization in a near-horizon image---depends on the sign of the electromagnetic energy flux and therefore provides a direct probe of black hole energy extraction.
In Boyer-Lindquist coordinates, the Poynting flux for axisymmetric electromagnetic fields is proportional to the product $B^\phi B^r$.
The phase $\angle\beta_2$ likewise depends on the ratio $B^\phi/B^r$, 
thereby enabling an observer to experimentally determine the direction of electromagnetic energy flow in the near-horizon environment.
Data from the 2017 EHT observations of M87* are consistent with electromagnetic energy outflow.
Currently envisioned multi-frequency observations of M87* will achieve higher dynamic range and angular resolution, and hence deliver measurements of $\angle\beta_2$ closer to the event horizon as well as better constraints on Faraday rotation.
Such observations will enable a definitive test for energy extraction from the black hole M87*.
\end{abstract}

\nocite{PaperI}
\nocite{PaperV}
\nocite{PaperVI}
\nocite{PaperVII}
\nocite{PaperVIII}

\section{Introduction}

The Event Horizon Telescope (EHT) has produced resolved images of the hot, magnetized, synchrotron-emitting plasma around the supermassive black holes in the Galactic Center \citep[Sgr A*;][]{PaperI_SgrA} and the elliptical galaxy M87 \citep[M87*;][]{PaperI} on scales comparable to their respective projected black hole event horizons.
The 230\,GHz images of both sources feature similar ring-like morphologies.
Despite a difference of three orders of magnitude in the black hole mass, both rings have diameters $d\approx10\,GM/Dc^2$,\footnote{For M87*, which will be the main focus of this paper, EHT observations infer a mass $M=6.5\pm 0.7\times10^9 M_\odot$ at a distance $D=16.8\pm0.8\,$Mpc \citep{PaperVI}.
These values imply a gravitational radius for M87* of $r_{\rm g}=GM/c^2=9.6\times10^{14}\,{\rm cm}=64\,$AU and an angular gravitational radius $\theta_{\rm g}=r_{\rm g}/D = 3.8\,\mu$as.
For a more recent measurement of the mass of M87*, see \citet{Liepold2023}.}
consistent with predictions from analytic and numerical models of low-luminosity accretion flows in the extreme near-horizon regime, where the effects of strong gravitational lensing and redshift become important \citep[][hereafter EHTC~V]{PaperV}. 

Near-horizon polarimetric images carry more information than total-intensity images and more stringently constrain models of the accretion flow and jet-launching region.
In particular, EHT images of M87* reveal a helical pattern in the electric vector position angles (EVPAs) of the linearly polarized intensity around the emission ring \citep[][hereafter EHTC~VII]{PaperVII}. 
While EHT images of M87* in total intensity are consistent with a wide variety of astrophysical models \citepalias{PaperV}, the polarimetric images strongly prefer models of magnetically arrested (MAD) accretion flows \citep[][hereafter EHTC~VIII]{PaperVIII}.
In MAD flows, the horizon-scale magnetic field is strong, ordered, and dynamically important \citep{Bisnovatyi1974,Narayan2003}.

Strong horizon-scale magnetic fields near black holes naturally produce powerful electromagnetic outflows.
\citet{Penrose1969} showed that energy can be extracted from a black hole's spin via particle interactions within the ergosphere.
This version of the Penrose process may have astrophysical implications for high-energy emission \citep[e.g.,][]{Williams1995,Schnittman2015}, but the primary astrophysical channel for tapping a black hole's spin energy is widely believed to be the Blandford-Znajek \citep[BZ;][]{BlandfordZnajek1977} mechanism, whereby electromagnetic energy is extracted from a black hole's spin via magnetic fields that thread its event horizon.

The BZ mechanism is the dominant source of energy outflow in general-relativistic magnetohydrodynamic (GRMHD) simulations of MAD accretion flows around spinning black holes (e.g., \citealt{Tchekhovskoy2011}; \citealt{McKinney2012}; \citetalias{PaperV}).
The BZ mechanism is thus a natural candidate for the launching power of extragalactic jets \citep{Begelman1984}, including the famous jet from M87, which has been observed in over a century of multi-frequency observations to deliver energy from horizon to galactic scales \citep[e.g.,][]{EHTMWL,Lu2023}.
GRMHD simulations of MAD flows featuring BZ powered jets provide good matches to both the observed 230\,GHz image morphology of M87* near the event horizon (\citetalias{PaperV}; \citetalias{PaperVIII}), as well as the jet power, profile, and core-shift farther downstream \citep{Chael2019,Mizuno2021,Cruz2022}.

Polarimetric EHT observations were critical for constraining the space of astrophysical models for M87*'s accretion in \citetalias{PaperVIII}, and for providing evidence for strong, ordered, horizon-scale magnetic fields.
However, it has not been clear if or how EHT observations can directly probe the electromagnetic energy flow close to the black hole or, more ambitiously, test whether M87's jet is truly powered by the BZ mechanism.

In this paper, we will show that, provided that the sense of rotation of magnetic field lines is known, and provided that Faraday rotation of linear polarization in the emission region or by an external screen is not too severe, the local direction of electromagnetic energy flow is directly mapped onto near-horizon polarimetric images observed by the EHT.
In particular, the observable $\angle\beta_2$, a quantity that characterizes the helicity of the spiral of polarization vectors around the black hole image \citep{Palumbo2020}, can be used to infer the direction of the Poynting flux.
As we will show, the EHT's polarimetric image of M87* \citepalias{PaperVII} is consistent with outflowing electromagnetic energy on scales of $\sim5\,GM/c^2$.  
While this energy outflow on horizon scales near the jet-launching point suggests that the black hole spin of M87* truly powers the extragalactic jet, current EHT observations cannot yet conclusively rule out that the rotation power of the accretion disk \citep{BlandfordPayne1982} also plays a significant role.

In the following sections, we develop the connection between $\angle\beta_2$ and the direction of electromagnetic energy flux using simple arguments, and we test this connection in both simple analytic models and complex numerical simulations.
In \autoref{sec:Formalism}, we review the main features of degenerate, stationary, axisymmetric magnetospheres in the Kerr spacetime.
We show that electromagnetic energy outflow requires a certain sign of the ratio of magnetic field components $B^\phi/B^r$ in Boyer-Lindquist coordinates.
We argue that, for synchrotron-emitting plasmas, this ratio is probed by the observed polarization helicity encoded in the observable $\angle\beta_2$.
In \autoref{sec:RingModels}, we investigate semi-analytic models of the emission close to M87* and show that relativistic effects of aberration, parallel transport, and lensing do not substantially change the relationship between $\angle\beta_2$ and the sign of energy outflow.
Furthermore, using the \citet{BlandfordZnajek1977} monopole solution to model the magnetic fields close to M87*, we show that there is a clear relationship between black hole spin and $\angle\beta_2$ in observed images, which results from larger spins ``winding up'' the magnetic fields more rapidly and thereby increasing the ratio $|B^\phi/B^r|$. 

In \autoref{sec:GRMHD}, we move from simple analytic models to an analysis of a full library of GRMHD-simulated images.
Remarkably, these reveal that the connection between the helicity of the EHT polarization spiral and the direction of electromagnetic energy flow continues to hold even when considering a time-dependent, turbulent, Faraday-rotating plasma.
While we will show that published EHT observations are consistent with an outflowing electromagnetic energy flux in M87*, they still do not unambiguously connect the energy flux with the extraction of black hole spin energy.
To do this and test the BZ mechanism, we argue in \autoref{sec:Discussion}, will require pushing EHT measurements of the linear polarization pattern closer to the apparent boundary of the black hole event horizon, or ``inner shadow'' \citep{Chael2021}.
With ongoing and planned upgrades to the EHT expected to increase its resolution and dynamic range, these tests may become feasible within the next decade.
We summarize our findings in \autoref{sec:Conclusion}.

In the Appendices, we provide a detailed review of properties of degenerate electromagnetic fields around black holes.
While these results are not new (and draw heavily on previous reviews such as \citealt{Gralla2014}), they present a unified treatment of the field structure and energetics around black holes that we use throughout this paper and will reference in later papers in this series on black hole polarimetric images.
In particular, in Paper II \citep{Paper2}, we will use these results to quantify the expected near-horizon polarization pattern for field lines that thread the event horizon, providing a direct future observational test of black hole energy extraction.
In Paper III \citep{Paper3}, we will develop simplified analytic models for the dependence of the near-horizon polarization on black hole spin, providing a physical interpretation of existing GRMHD simulation results and a path toward quantitatively constraining black hole spin using polarimetry.

\newpage

\section{Black Hole Energy Extraction and the Sign of \texorpdfstring{$\angle\beta_2$}{argBeta2}}
\label{sec:Formalism}

\subsection{Kerr metric in Boyer-Lindquist coordinates}

We use units where $G=c=1$.
We work in the Kerr spacetime of a black hole with mass $M$ and angular momentum $J=aM=a_*M^2$ in Boyer-Lindquist (BL) coordinates $(t,r,\theta,\phi)$.
The metric (\autoref{eq:Kerr}) is expressed in terms of three poloidal functions
\begin{gather}
    \Delta=r^2-2Mr+a^2,\quad
    \Sigma=r^2+a^2\cos^2{\theta},\nonumber\\
    \Pi=\pa{r^2+a^2}^2-a^2\Delta\sin^2{\theta}.
\end{gather}
The metric determinant is 
\begin{align}
    \sqrt{-g}=\Sigma\sin{\theta}.
\end{align}
The event horizons are located at radii where $\Delta=0$; in particular, the outer Kerr event horizon lies at a radius
\begin{align}
    r_+=M+\sqrt{M^2-a^2}.
\end{align}
Normal observers in BL coordinates have a four-velocity $\eta_\mu=(-\alpha,0,0,0)$, where $\alpha=1/\sqrt{-g^{tt}}=\sqrt{\Delta\Sigma/\Pi}$ is the lapse function.
For $a\neq0$, BL normal observers have a nonzero angular velocity 
\begin{align}
    \omega=\frac{\eta^\phi}{\eta^t}=\frac{2aMr}{\Pi}.
\end{align}
At the event horizon, the angular velocity of the normal observer equals that of the horizon: 
\begin{align}
    \Omega_{\rm H}=\frac{a}{2Mr_+}
    =\omega|_{r=r_+}.
\end{align}

\subsection{Stationary, axisymmetric, and degenerate electromagnetic fields}

We consider electromagnetic fields $F_{\mu\nu}$ in Kerr 
that are both degenerate and magnetically dominated:
\begin{subequations}
\label{eq:Conditions}
\begin{align}
    \star F_{\mu\nu}F^{\mu\nu}&=0&&\text{(degenerate)},\\
    F^{\mu\nu}F_{\mu\nu}&>0&&\text{(magnetically dominated)}.
\end{align}    
\end{subequations}
As shown in \autoref{app:Degeneracy} and \ref{app:DegenerateElectromagnetism}, for such fields there exists an infinite family of timelike frames $u^\mu$ in which the electric field vanishes: $e^\mu=F^{\mu\nu}u_\nu=0$.\footnote{This family of timelike frames can be parameterized by the Lorentz factor $\gamma\ge\gamma_\perp$ relative to a normal observer. 
The frame with minimal Lorentz factor $\gamma_\perp=\pa{1-E^2/B^2}^{-1/2}$ has a four-velocity that is perpendicular to the normal observer's magnetic field.
See \autoref{app:Degeneracy}.}
In particular, electromagnetic fields obeying the equations of ideal GRMHD are constrained to be degenerate and magnetically dominated.
In GRMHD, the fluid four-velocity picks out a unique frame $u^\mu$ from among the family of frames in which $e^\mu$ vanishes.
Fields obeying the equations of general-relativistic force-free electrodynamics (GRFFE) are also required to be degenerate, though without additional constraints they can evolve into states in which magnetic domination does not hold ($F^2<0$).

The time-space components of the Maxwell tensor $\star F$ define the ``lab-frame'' magnetic field: $B^i=\star F^{i0}$.\footnote{Note that we use Lorentz-Heaviside units for the magnetic field, which are related to Gaussian units by $B_{\rm LH}=B_{\rm G}/\sqrt{4\pi}$.}
In \autoref{app:KerrFields} and \ref{app:KerrDegenerateFields}, we review \citep[see, e.g.,][]{Phinney1983,Gralla2014} how degenerate electromagnetic fields $F$ that are also stationary ($\pd_t F\to0$) and axisymmetric ($\pd_\phi F\to0)$ are completely characterized by three quantities: the magnetic flux function (or poloidal potential) $\psi(r,\theta)=A_\phi(r,\theta)$, the poloidal current $I(\psi)$, and the field-line angular velocity $\Omega_F(\psi)$.
In 
particular, the lab-frame magnetic field $B^i$ in BL coordinates is determined by the coordinate-independent flux $\psi$ and current $I$ as
\begin{align}
\label{eq:ComponentsB}
    B^r=\frac{\pd_\theta\psi}{\sqrt{-g}},\quad
    B^\theta=-\frac{\pd_r\psi}{\sqrt{-g}},\quad
    B^\phi=\frac{I}{2\pi\Delta\sin^2{\theta}}.
\end{align}

\begin{figure*}[ht!]
\centering
\includegraphics[width=\linewidth]{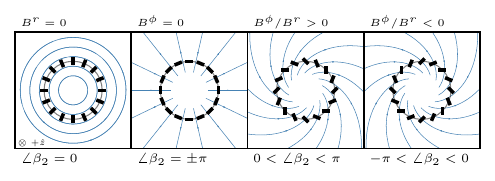}
\caption{
Schematic diagram illustrating the connection between different simple equatorial axisymmetric magnetic field configurations around a black hole and the polarimetric observable $\angle\beta_2$.
No relativistic or propagation effects (e.g., parallel transport, aberration, Faraday rotation) are included.
The $+\hat{z}$ axis is into the page, and the local polarization direction is given by $\hat{z}\times\vec{B}$.
The range of values obtained by the EHT for M87* ($-163\deg<\angle\beta_2<-129\deg$; \citetalias{PaperVIII}) is consistent with a ratio $B^\phi/B^r<0$ in the emission region.
Assuming that the black hole spin is aligned with $\hat{z}$, that the magnetic field is axisymmetric and degenerate, and that the magnetic field lines co-rotate with the black hole, this field structure produces an outward Poynting flux by \autoref{eq:EnergyFlux}. 
}
\label{fig:Cartoon}
\end{figure*}

\subsection{The Blandford-Znajek mechanism}

For stationary, axisymmetric, degenerate fields in BL coordinates (\autoref{eq:ComponentsB}), the radial electromagnetic energy flux density is (see \autoref{app:Efluxes})
\begin{align}
    \label{eq:EnergyFlux}
    \mathcal{J}_\mathcal{E}^r\equiv-{T^r}_{t\;\mathrm{EM}}
    =-\Omega_F B^r B^\phi\Delta\sin^2{\theta}.
\end{align}
In BL coordinates, $B^\phi$ diverges on the horizon as $1/\Delta$, but the product $\Delta B^\phi$, and hence the energy flux density $\mathcal{J}_\mathcal{E}^r$, remain finite. 
In particular, \autoref{app:KerrDegenerateFields} shows how by considering the field in Kerr-Schild coordinates, which are regular on the horizon, one can derive the Znajek condition \citep{Znajek1977} that fixes the current $I$ on the horizon as a function of $\Omega_F$ (see also \citealt{MacDonald1982}, \citealt{Thorne1986}).
We can also express the Znajek condition as a condition on the ratio $\Delta B^\phi/B^r$ in BL coordinates:
\begin{align}
    \left.\frac{\Delta B^\phi}{B^r}\right|_{r=r_+}=2Mr_+\pa{\Omega_F-\Omega_{\rm H}}.
\end{align}
Hence, the radial Poynting flux through the horizon is 
\begin{align}
    \label{eq:RadialHorizonFlux}
    \mathcal{J}_\mathcal{E}^r\big|_{r=r_+}=2Mr_+\Omega_F\pa{\Omega_{\rm H}-\Omega_{F}}\pa{B^r}^2\sin^2{\theta}.
\end{align}
From \autoref{eq:RadialHorizonFlux}, it is clear that the energy extraction from a black hole is maximized when the field-line angular velocity is $\Omega_F=\frac{1}{2}\Omega_{\rm H}$ \citep{BlandfordZnajek1977}.
This result holds for all values of the black hole spin $a$.
The total outward electromagnetic energy flux through a sphere of radius $r$ is 
\begin{align}
    \label{eq:Edot}
    \dot{\mathcal{E}}=2\pi\int_0^\pi\mathcal{J}_\mathcal{E}^r\sqrt{-g}\ed\theta.
\end{align}
Assuming that $\Omega_F=\Omega_{\rm H}/2$ and expanding in powers of $\Omega_F$, plugging \autoref{eq:RadialHorizonFlux} into \autoref{eq:Edot} gives
\begin{align}
    \label{eq:FluxScaling}
    \dot{\mathcal{E}}=k\,\Omega_{\rm H}^2\Phi^2_{\rm H}+\mathcal{O}\!\pa{\Omega^4_{\rm H}},
\end{align}
recovering the familiar result that the total energy extracted from a spinning black hole by electromagnetic fields is proportional to the square of $\Omega_{\rm H}$ and the square of the horizon magnetic flux $\Phi_{\rm H}=\int\left[B^r\sqrt{-g}\right]_{r=r_+}\mathrm{d}\theta\mathrm{d}\phi$.
For the BZ monopole solution, the proportionality constant is $k=1/24\pi$.\footnote{\citet{Tchekhovskoy2010} use $\Phi$ to represent the flux through one hemisphere of the horizon, in which case $k=1/6\pi$.}
With a modified constant $k$ to reflect differences in the field geometry, the scaling in \autoref{eq:FluxScaling} is observed in force-free \citep[e.g.,][]{Tchekhovskoy2010} and GRMHD \citep[e.g.,][]{Tchekhovskoy2012,Narayan2022} simulations of black hole jets, even up to large values of the black hole spin.

By inspection of \autoref{eq:EnergyFlux}, it is clear that if the field lines co-rotate with the black hole ($\Omega_F>0$), then outward energy flow $\mathcal{J}_{\mathcal{E}}^r>0$ requires a ratio of the BL lab-frame fields $B^\phi/B^r<0$.
This preferred orientation of the azimuthal field should manifest in images of synchrotron-emitting electrons close to the event horizon, such as those obtained by the EHT of M87* at 230\,GHz \citepalias{PaperVII,PaperVIII}.\footnote{While $B^\phi$ and $B^r$ are both coordinate-dependent quantities, both the total energy flux 
and the observed polarization pattern from the near-horizon accretion flow are coordinate-invariant.
Thus, while we motivate and derive the connection between the observed linear polarization and the direction of energy flux using BL coordinates, our main conclusions are frame-independent.}

\subsection{Polarization and energy flux in near-horizon images}

The direction of the polarization vector produced locally by synchrotron emission is perpendicular to its wavevector and to the magnetic field.
In the absence of the complicating effects of Faraday rotation and parallel transport through curved spacetime, the pattern of linear polarization in a resolved image can be used to directly infer the magnetic field geometry around a black hole.
We consider these important complications in more realistic models in \autoref{sec:RingModels} and \autoref{sec:GRMHD}.

\autoref{fig:Cartoon} shows a cartoon picture of the synchrotron polarization pattern for several simple axisymmetric magnetic field configurations viewed face-on.
Here, we have assumed that the emitting matter is in the equatorial plane and that there is no vertical magnetic field, $B^\theta=0$.   
In \autoref{fig:Cartoon}, we orient the $+\hat{z}$ axis into the page, corresponding to an observer inclination $\theta_{\rm o}\approx\pi$. 
We further assume that the spin of the black hole is oriented along $+\hat{z}$ into the plane of the sky, matching the spin orientation inferred for M87* by the EHT \citepalias{PaperV}. 

As in \citetalias{PaperVIII}, we use the $\beta_2$ statistic \citep{Palumbo2020} to quantify the structure of near-horizon linear polarization in EHT images.
$\beta_2$ is the second azimuthal Fourier mode of the linear polarization image:
\begin{align}
    \label{eq:Beta2}
    \beta_2=\frac{\int_{\rho_{\rm min}}^{\rho_{\rm max}}\int_0^{2\pi}\mathcal{P}(\rho,\varphi)e^{-2i\varphi}\rho\ed\rho\ed\varphi}{\int_{\rho_{\rm min}}^{\rho_{\rm max}}\int_0^{2\pi}I(\rho,\varphi)\rho\ed\rho\ed\varphi}.
\end{align}
In \autoref{eq:Beta2}, $\mathcal{I}$ is the total intensity brightness, $\mathcal{P}=\mathcal{Q}+i\mathcal{U}$ is the complex linearly polarized brightness, and ($\rho$,$\varphi$) are polar coordinates in the image plane.
For narrow, axisymmetric rings of polarized emission viewed face-on, $\angle\beta_2$ is equal to twice the value of the EVPA at $\varphi=0$.\footnote{We measure both $\varphi$ and EVPA East of North (i.e., counterclockwise from the positive $y$-axis).
We fix the range of EVPA to be $(-\pi/2,\pi/2]$ and the range of $\angle\beta_2$ to be $(-\pi,\pi]$.} 

Neglecting all relativistic effects and assuming a face-on observer and a narrow axisymmetric emission ring of radius $r$, the observed value of $\angle\beta_2$ can be simply related to the BL field components in the emission region by
\begin{align}
    \angle\beta_2&\approx2\arctan\pa{\frac{B^r}{rB^\phi}}\quad
    (\text{observer at }\theta_{\rm o}=\pi).
\end{align}
In the cartoon picture of \autoref{fig:Cartoon}, where the $+\hat{z}$ axis points into the page, observed $\angle\beta_2$ values in the range $(-\pi,0)$ correspond to fields with $B^\phi/B^r<0$.  

If we assume that the black hole spin is oriented along $+\hat{z}$ and that magnetic field lines co-rotate with the black hole $(\Omega_F>0)$, then negative values of $\angle\beta_2$ correspond to electromagnetic energy outflow for axisymmetric, stationary, degenerate fields.
By contrast, $\angle\beta_2$ values in the range $(0,\pi)$ correspond to fields with $B^\phi/B^r>0$ and would therefore represent a field configuration with energy flowing into the black hole. 

In this simple picture, we can thus constrain the sign of the near-horizon electromagnetic energy flux $\mathcal{J}_\mathcal{E}^r$ by measuring $\angle\beta_2$, assuming that we know the signs of the inclination $\cos\theta_{\rm o}$ and field-line angular velocity $\Omega_F$.
In the next subsection, we turn to more general orientations of the black hole spin, field-line angular velocity, and observer inclination.

\subsection{General relationship between \texorpdfstring{$\angle\beta_2$}{argBeta2} and \texorpdfstring{$\mathcal{J}_\mathcal{E}$}{JE}}
\label{sec:Conventions}

In a black hole magnetosphere, the black hole spin, magnetic field lines, and emitting matter all have angular momenta that can in principle take independent orientations.
Throughout this work, we assume all of these are either aligned or anti-aligned with the $+\hat{z}$ coordinate axis ($\theta=0$) in BL coordinates. 
We thus use signed scalars to represent the black hole spin $a$, field-line angular velocity $\Omega_F$, and fluid angular velocity $v^\phi=u^\phi/u^t$.
Each scalar is either positive or negative depending on whether it is aligned (positive) or anti-aligned (negative) with the $+\hat{z}$-axis.
The observer at infinity can view the system at any inclination angle in the range $0\leq\theta_{\rm o}\leq\pi$.

In principle, we could follow a convention in which the black hole spin is always positive by aligning the $+\hat{z}$ axis of our coordinate system with the black hole spin.
In GRMHD simulations, however, the community's convention for simulating \emph{retrograde} accretion flows is to use an initial equilibrium torus far from the black hole with a positive angular velocity $v^\phi>0$, and to set the black hole spin along the $-\hat{z}$ direction, $a<0$.
For this reason, in this work, we also treat retrograde accretion flows in both analytic models and simulations by setting $a<0$.

From \autoref{eq:EnergyFlux}, for arbitrary signs of black hole spin and field-line angular velocity, it is always true for stationary axisymmetric magnetic fields that 
\begin{align}
    \label{eq:GeneralJ}
    \sign\pa{\mathcal{J}_\mathcal{E}^r}=-\sign\pa{\Omega_F B^r B^\phi}.
\end{align}
Furthermore, a rule of thumb for the sign of $\angle\beta_2$ that accounts for arbitrary observer inclination is
\begin{align}
    \label{eq:GeneralBeta}
    \sign\pa{\angle\beta_2}\simeq-\sign\pa{B^r B^\phi\cos{\theta_{\rm o}}}.
\end{align}
Combining \autoref{eq:GeneralJ} and \autoref{eq:GeneralBeta} yields an approximate relation valid for all inclinations $\theta_{\rm o}$ and values of $\Omega_F$:
\begin{align}
    \label{eq:GeneralFlow}
    \sign\pa{\mathcal{J}_\mathcal{E}^r}\simeq\sign\pa{\beta_2}\times\sign\pa{\Omega_F\cos{\theta_{\rm o}}}.
\end{align}

\begin{figure*}[t!]
\centering
\includegraphics[width=\textwidth]{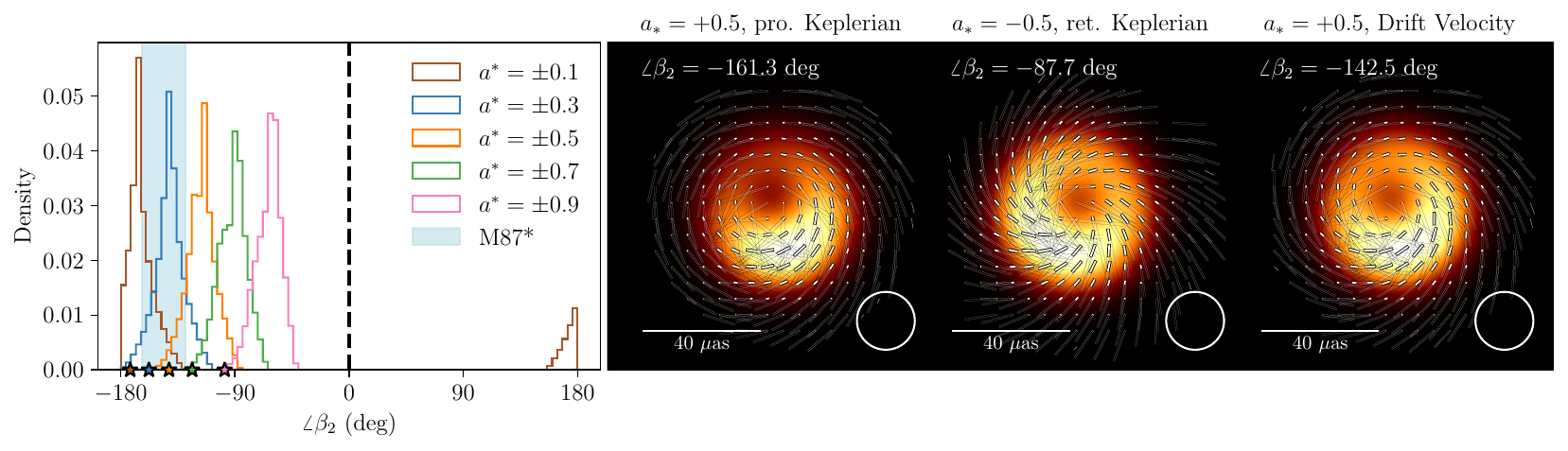}
\caption{
(Left) Histograms of $\angle\beta_2$ covering different velocity prescriptions in axisymmetric, equatorial models of emission around M87*.
In these models, the emission radius is fixed to $r=2r_+$ and the lab-frame magnetic field $B^i$ given by the BZ monopole solution.
For each value of black hole spin, the histograms cover uniform priors in the velocity parameters $\xi$, $\beta_r$, $\beta_\phi$ in the range $[0,1]$ for both prograde ($a_*>0$) and retrograde ($a_*<0$) orbits.
The observer inclination $\theta_{\rm o}$ is $163\deg$ relative to the spin axis for $a_*>0$ and $17\deg$ for $a_*<0$.
The stars on the $x$ axis indicate the values of $\angle\beta_2$ for the same fields $B^i$ assuming the self-consistent BZ drift-frame velocity.
The range of $\angle\beta_2$ values consistent with EHT observations of M87* in 2017 \citepalias{PaperVII} is indicated with the blue band.
(Right) Representative images of the equatorial BZ monopole model for $|a_*|=0.5$ with a prograde Keplerian velocity model, a retrograde Keplerian velocity model, and the outflowing drift-frame velocity from the complete BZ monopole solution.
}
\label{fig:AnalyticModels}
\end{figure*}

\subsection{Application to M87*}
\label{sec:application}

In order to match the M87* total-intensity image asymmetry, \citetalias{PaperV} found that the fluid angular velocity in the emission region must be directed \emph{into} the plane of the sky.
Furthermore, \citetalias{PaperV} and \citet{Wong2021} found that in nearly all GRMHD simulations, the material producing the 230\,GHz emission observed by the EHT corotates with both the black hole spin and the magnetic field lines (see also \autoref{fig:SimulationFields}). 

That is, for \emph{prograde} accretion flows ($a>0$), both $v^\phi>0$ and $\Omega_F>0$ in the emitting region.
In order for the image to match the observed brightness asymmetry, the observer inclination must then be $\theta_{\rm o}\approx\pi$.
Electromagnetic energy outflow in this case requires $B^\phi/B^r<0$ (\autoref{eq:EnergyFlux}), which by \autoref{fig:Cartoon} is observable as $-\pi<\angle\beta_2<0$.

For \emph{retrograde} accretion models ($a<0$), instead both $v^\phi<0$ and $\Omega_F<0$ in the emitting region.
To match the brightness asymmetry in this case requires $\theta_{\rm o}\approx0$.
For there to be energy outflow in this case, the azimuthal-to-radial field ratio in BL coordinates must be positive, $B^\phi/B^r>0$.
However, because $\theta_{\rm o}\approx 0$ in this case, outflowing electromagnetic energy still corresponds to an observable $\angle\beta_2$ in the range $-\pi<\angle\beta_2<0$.

In summary, we do not know either the sign of $\cos{\theta_{\rm o}}$ or $\Omega_F$ in M87* independently.
From \citetalias{PaperV}, however, we do know that $\sign\left(\Omega_F\cos{\theta_{\rm o}}\right)=-1$.
Thus, we can use the value of $\angle\beta_2$ measured by the EHT to constrain the direction of the near-horizon electromagnetic energy flow in the idealized picture of \autoref{fig:Cartoon}. 
The observed range of $\angle\beta_2$ for M87* from the EHT is $-163\deg<\angle\beta_2<-129\deg$ \citepalias{PaperVII}. 
We thus find from \autoref{eq:GeneralFlow} that the EHT results are consistent with an outward Poynting flux in M87*'s black hole magnetosphere. 

This correspondence between the observed $\angle\beta_2$ with the direction of Poynting flux in M87*'s magnetosphere is complicated by several effects. 
First, relativistic parallel transport (from the curved spacetime close to the horizon) and aberration effects (from the rapidly rotating emitting plasma) may result in observed polarization directions that are not perpendicular to the projected magnetic fields in the emission region.
We explore these effects in analytic models in \autoref{sec:RingModels}.
Second, realistic electromagnetic fields around supermassive black holes are not stationary or axisymmetric, and propagation effects through the turbulent plasma (especially Faraday rotation) may significantly alter the emitted polarization direction.
We explore these effects in full GRMHD simulation images of M87* in \autoref{sec:GRMHD}.

\section{\texorpdfstring{$\angle\beta_2$}{argBeta2} and Energy Flux in Semi-Analytic Models}
\label{sec:RingModels}

In this section, we examine whether the straightforward connection between the observed $\angle\beta_2$ in polarized images of synchrotron radiation, the ratio $B^\phi/B^r$, and the direction of electromagnetic energy flux we established in \autoref{sec:Formalism}
persists in semi-analytic models of the near-horizon emission from M87*. 

We restrict ourselves to axisymmetric, equatorial emission models in this section. We describe the degenerate electromagnetic field in these simple models using the same variables common in GRMHD simulations \citep[e.g.,][]{Gammie2003}.
That is, we use the fluid-frame four-velocity $u^\mu$ and the ``lab-frame'' magnetic field $B^i=\star F^{i0}$. 

As a toy model of a degenerate electromagnetic field configuration in the Kerr spacetime with
outward Poynting flux, we adopt the perturbative Blandford-Znajek monopole solution \citep{BlandfordZnajek1977} for the potential $\psi$ and current $I$ (see \autoref{app:Monopole}). We then  determine $B^r$ and $B^\phi$ by \autoref{eq:ComponentsB}; $B^\theta$ vanishes in the equatorial plane.
While the BZ monopole solution does not capture the detailed structure of a black hole magnetosphere as seen in GRMHD, GRFFE, or particle-in-cell simulations, it has been shown to match certain time-averaged properties of these simulations well \citep[e.g.,][]{McKinney2004,Parfrey2019}.

Real black hole magnetospheres may have a significant vertical magnetic field component in the equatorial plane \citep{Blandford2022}; the paraboloidal BZ solution may provide a better match to the near-horizon magnetic field structure in these systems \citep[e.g.,][]{McKinney2007,Penna2013}.
We will explore polarized images of the paraboloidal solution in a future work.
For now, we take the BZ monopole as a toy model that captures both an outgoing energy flux and a field ratio $B^\phi/B^r$ that depends on black hole spin, similar to what is observed in GRMHD simulations \citep[e.g.,][]{Palumbo2020,Emami2022}.

For the fluid velocity $u^\mu$, we explore two different models.
First, we use the unique field-perpendicular or \emph{drift-frame} velocity $u_\perp^\mu$ determined by the full BZ solution for both the lab-frame magnetic field $B^i=\star F^{i0}$ and electric field $E^i=F^{0i}$ (see \autoref{eq:DriftVelocity}, \autoref{app:DegenerateElectromagnetism}). 
The drift-frame velocity in BL coordinates produces an outflow in the equatorial plane; while we can include arbitrary boosts along the magnetic field and not change the observed polarization pattern, this family of four-velocities is not fully representative of black hole accretion flows.

Consequently, we also explore a family of parametrized equatorial inflow models for $u^\mu$
from \cite{CardenasAvendano2022}.
This velocity model is described by three parameters $(\xi,\beta_r,\beta_\phi)$ taking values between 0 and 1.
The parameter $\xi$ represents the ratio of the fluid angular momentum to the Keplerian value, while $\beta_r$ and $\beta_\phi$ represent the relative magnitude of the Keplerian and infall velocities, respectively.
In addition to the parameters $(\xi,\beta_r,\beta_\phi$), we can also choose the (sub)Keplerian velocity to be either prograde or retrograde with respect to the black hole spin. 
We summarize this four-velocity model in \autoref{app:AART}.

After fixing the magnetic field and velocity profile, we model the 230\,GHz emitting region as a narrow ring of constant rest-frame emissivity $J(r_{\rm eq})$ in the equatorial plane centered at $r_{\rm ring}=2r_{+}$.
We generate images from this model using the \texttt{kgeo} code \citep{kgeo}. We present more details of the model construction in \autoref{app:ModelDescription}. 

\begin{figure*}[t!]
\centering
\includegraphics[width=0.75\textwidth]{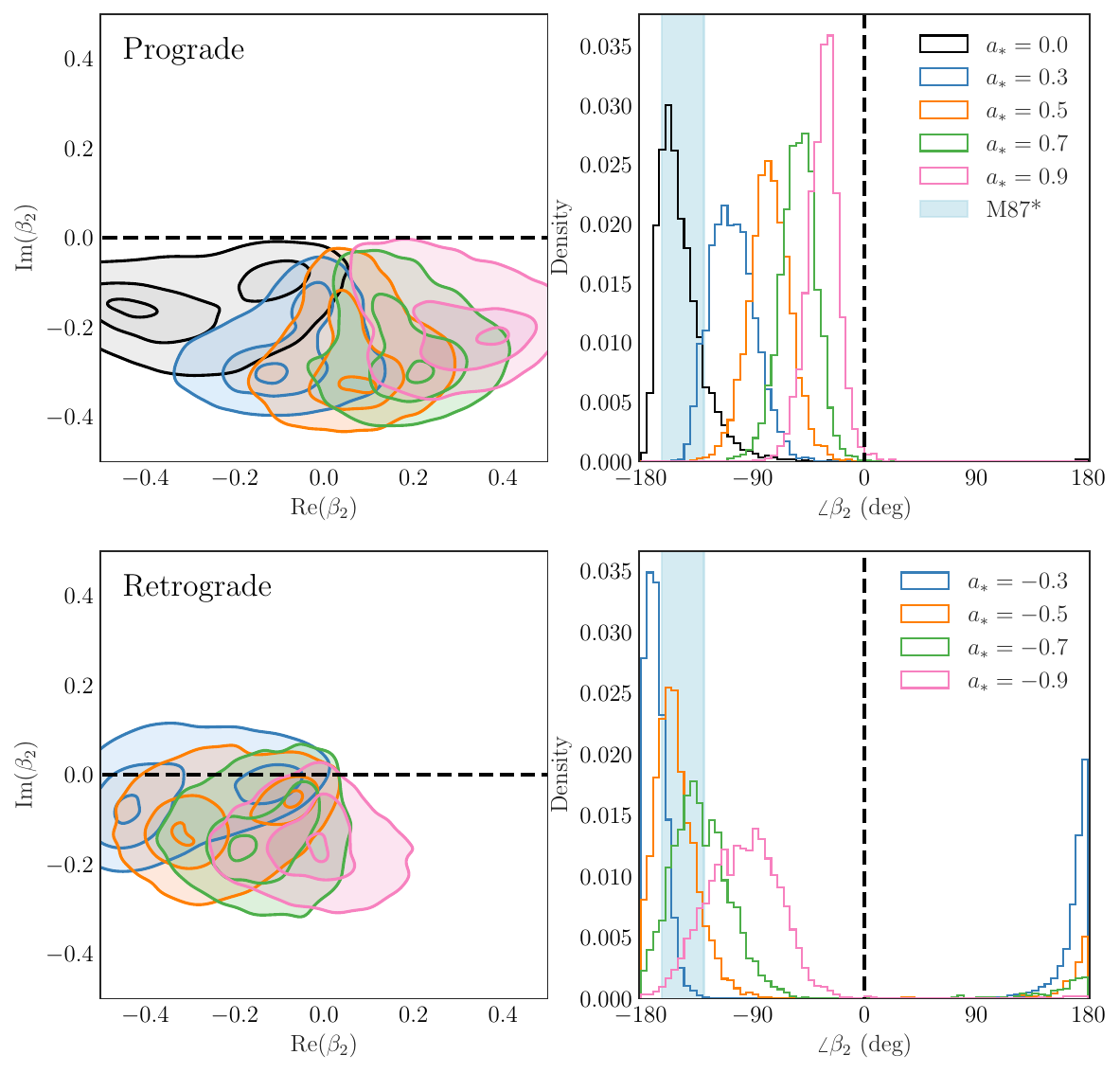}
\caption{
(Left column) Histograms of snapshot values of $\beta_2$ in the complex plane for images from nine MAD GRMHD simulations raytraced with M87* parameters.
Prograde models ($a_*>0$) are shown in the top row, and retrograde models ($a_*<0$) in the bottom row.
The histograms include all snapshot images at different simulation time slices between $50,000\,t_{\rm g}$ and $100,000\,t_{\rm g}$. 
Simulation images were blurred to the EHT resolution of $20\,\mu$as before computing $\beta_2$.
The histograms cover six different models for the low-magnetization ion-to-electron temperature ratio $R_{\rm high}$.
(Right column) Histograms of $\angle\beta_2$ from the same simulation images.
With a few exceptions for predominantly low-spin, retrograde simulation snapshots, nearly all MAD GRMHD snapshots have $\angle\beta_2<0$, corresponding to energy outflow in the simple picture of \autoref{fig:Cartoon}.
The range of $\angle\beta_2$ values consistent with EHT observations of M87* in 2017 is shown in the blue band \citepalias{PaperVII}.
}
\label{fig:SimulationModels}
\end{figure*}

We assume uniform priors on the velocity model parameters $(\xi,\beta_r,\beta_\phi)$ and generate model images covering 250,000 combinations of these parameters for each black hole spin $a_*\in[\pm0.1,\pm0.3,\pm0.5,\pm0.7,\pm0.9]$, where negative spins correspond to retrograde flows.
We fix the observer inclination $\theta_{\rm o}=163\deg$ for prograde flows ($\theta_{\rm o}=17\deg$ for retrograde flows), as measured for M87* \citep{Mertens2016}.
We also generate a single image from each model using the outflowing drift-frame velocity $u_\perp^\mu$.
We then blur the images to the EHT resolution of $20\,\mu$as using a circular Gaussian kernel and compute $\angle\beta_2$ using (\autoref{eq:Beta2})

We plot histograms of the results for $\angle\beta_2$ over the velocity model parameter space in \autoref{fig:SimulationModels}.
The star symbols show the results for the drift-frame BZ velocity $u^\mu_\perp$ for each black hole spin.
With the exception of a few retrograde, low-spin ($a_*=-0.1$) models, nearly all models in this parameter space have $-\pi<\angle\beta_2<0$, corresponding to energy outflow in the simple picture of \autoref{fig:Cartoon}.
Unlike our arguments in \autoref{sec:Formalism}, these models include effects from parallel transport, aberration and nonzero inclination.
They demonstrate that the sign of $\angle\beta_2$ is a robust probe of energy extraction from black holes, relatively immune to uncertainty in the kinematics of the emitting plasma. 

We also show blurred EHT-resolution images of three extreme cases corresponding to fully prograde and retrograde Keplerian orbiting material at radius $r=2r_+$ for a spin $|a_*|=0.5$ and the outflowing BZ drift velocity solution.
While the images from a uniform emission ring for these two velocity models are quite different (in particular, the overall Stokes $\mathcal{I}$ image asymmetry in the retrograde model is substantially reduced from the prograde model), both have $\angle\beta_2$ values less than zero.

There is a clear spin-dependence to the value of $\angle\beta_2$ in \autoref{fig:AnalyticModels}.
In particular, higher-spin black holes produce values of $\angle\beta_2$ closer to zero.
In the BZ monopole solution, higher-spin black holes ``wind up'' magnetic field lines closer to the black hole.
As a result, they produce more radial polarization patterns when observed nearly face-on.
This same spin-dependence is present in GRMHD simulation images \citep{Palumbo2020} (see also \autoref{fig:SimulationModels}).
We will discuss the comparison of analytic models for this spin dependence with GRMHD simulations and EHT observations in more detail in \citetalias{Paper3}.

\section{\texorpdfstring{$\angle\beta_2$}{argBeta2} and Energy Flux in GRMHD Simulations}
\label{sec:GRMHD}

In this section, we investigate snapshots from magnetically arrested (MAD) GRMHD simulations of M87*.
Unlike the analytic models in \autoref{sec:RingModels}, GRMHD simulations feature complex, non-axisymmetric, time-dependent structure in their magnetic fields and fluid velocities.
Furthermore, the radiative transfer from GRMHD simulations fully accounts for a non-uniform synchrotron emitting region, self-absorption, and Faraday rotation and conversion of the emitted polarization along the photon trajectory. 
We find that the picture presented in \autoref{fig:Cartoon} and \autoref{fig:AnalyticModels} remains valid even when subject to these astrophysically important complications.
In particular, when observing MAD simulations of M87* at 230\,GHz with the black hole spin oriented into the plane of the sky, nearly all simulation snapshots have $\angle\beta_2<0$, consistent with outward electromagnetic energy flux, as in the simple picture of \autoref{fig:Cartoon}. 

We use nine MAD GRMHD simulations from \citet{Narayan2022} performed using the code \texttt{KORAL} \citep{Sadowski2013,Sadowski2014}. These simulations consider both prograde and retrograde accretion disks at five values of the dimensionless black hole spin $a_*\in[0,\pm0.3,\pm0.5,\pm0.7,\pm0.9]$.
The simulations where $|a_*|>0$ have substantial outward Poynting flux driven by BZ energy extraction \citep[see Figure 4 of][]{Narayan2022}.
The $a_*=0$ simulation also has an outward Poynting flux, which is launched by the rotational energy of the accretion flow rather than the black hole spin \citep{BlandfordPayne1982}.
The total energy in the outflow from the $a_*=0$ simulation is much smaller in than in the simulations with $|a_*|>0$ \citep{Narayan2022}.

We generate 230\,GHz images from these simulations using the GR radiative transfer code \texttt{ipole} \citep{Noble2007,Moscibrodzka2018} following the parameters discussed in \autoref{app:GRMHDLibrary}.
In \autoref{fig:SimulationModels}, we show the distributions of $\beta_2$ in the complex plane and $\angle\beta_2$ for all snapshot images in the library, colored by the black hole spin.
The distributions for a given spin cover a wide range of simulation snapshots and electron-to-ion temperature ratio $R_{\rm high}$ \citep{Moscibrodzka2016}.
Nearly all snapshots have $-\pi<\angle\beta_2<0$, matching the intuition from \autoref{sec:Formalism} that negative $\angle\beta_2$ corresponds to outward electromagnetic energy flow in M87*, assuming that the black hole spin vector is into the plane of the sky.

As in the semi-analytic models of M87* using the BZ monopole (\autoref{fig:AnalyticModels}), there is also a clear spin-dependence among the distributions of $\angle\beta_2$ in the GRMHD images (\autoref{fig:SimulationModels}).
Higher-spin simulations produce $\angle\beta_2$ values closer to zero (see also \citealt{Palumbo2020}, \citetalias{PaperV}, \citealt{Jia2022} for examples of this same trend in other simulation image libraries).
This trend of $\angle\beta_2$ with spin in GRMHD simulations is a result of more rapidly spinning black holes ``winding up'' the magnetic field lines more rapidly, producing more toroidal fields in the emission region \citep{Emami2022}.
The black hole spin in GRMHD simulations of MAD accretion flows thus significantly alters the magnetic field structure of the flow in the 230\,GHz emission region, with the same qualitative dependence on spin as in the BZ monopole model.
In \citetalias{Paper3}, we will investigate in more detail the connection between $\angle\beta_2$ and spin in both simple models and GRMHD simulations.  

It was not immediately obvious that the relationship between the sign of $\angle\beta_2$ and the direction of electromagnetic energy flux should be as robust in full GRMHD simulations as it is in the cartoon picture of \autoref{sec:Formalism} or the simple models of \autoref{sec:RingModels}.
For one, the instantaneous dynamics in GRMHD simulations are neither time-stationary nor axisymmetric, as assumed in \autoref{sec:Formalism} and \autoref{sec:RingModels}. 
Perhaps more surprisingly, the simulations also frequently have significant Faraday rotation in the emission region (\citealt{Ricarte2020}, \citetalias{PaperVIII}). 
A significant degree of internal Faraday rotation is in fact necessary to sufficiently de-polarize 230\,GHz simulation images to match EHT observations \citepalias{PaperVIII}, 
and one might expect Faraday rotation to shift $\angle\beta_2$ substantially from its ``intrinsic'' value without Faraday rotation.
Nonetheless, we find that Faraday rotation in the MAD GRMHD simulation images of M87* used here depolarize the 230\,GHz images but do not produce an overall EVPA rotation severe enough to change the sign of $\angle\beta_2$.

\begin{figure*}[t!]
\centering
\includegraphics[width=\textwidth]{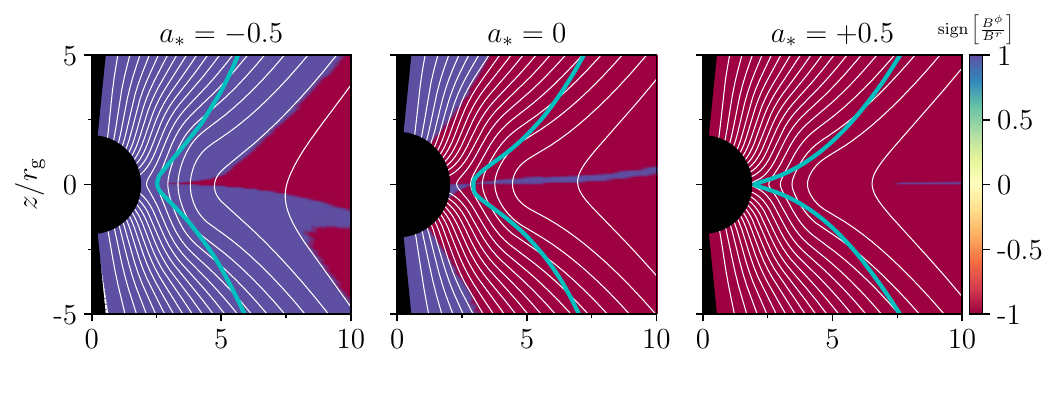} \\
\includegraphics[width=\textwidth]{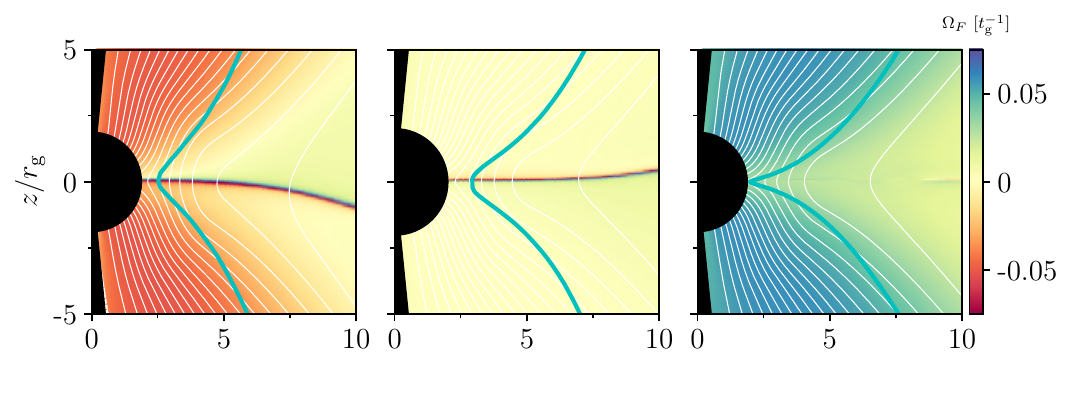} \\
\includegraphics[width=\textwidth]{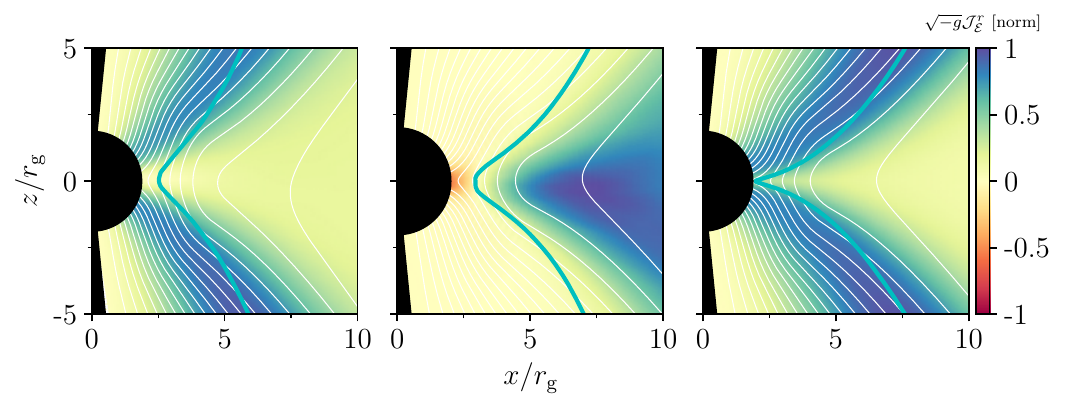}
\caption{
(Top) The sign of $B^\phi/B^r$ for three MAD GRMHD simulations averaged in time and azimuth.
From left to right, the GRMHD simulations correspond to a retrograde accretion disk around a spinning black hole ($a_*=-0.5$), a Schwarzschild disk ($a_*=0)$, and a prograde disk $(a_*=+0.5)$.
(Middle) The angular frequency of the magnetic field lines $\Omega_{F}$ in the same simulations in natural units of $t_{\rm g}^{-1}$.
(Bottom) The averaged radial Poynting flux $\sqrt{-g}\mathcal{J}^r_{\mathcal{E}}$ in BL coordinates in the same simulations.
In the bottom panel, all three panels are independently normalized and are not plotted in the same units; the energy flux in the $a_*=0$ simulation (driven by the accretion disk) is an order of magnitude smaller than in the $a_*=\pm0.5$ simulations (driven by black hole spin).
In all plots, the cyan contour indicates the surface where the magnetization $\sigma=1$.
White contours show surfaces of constant vector potential $\psi\equiv A_\phi$.
}
\label{fig:SimulationFields}
\end{figure*}

\begin{figure*}[t!]
\centering
\includegraphics[width=\linewidth]{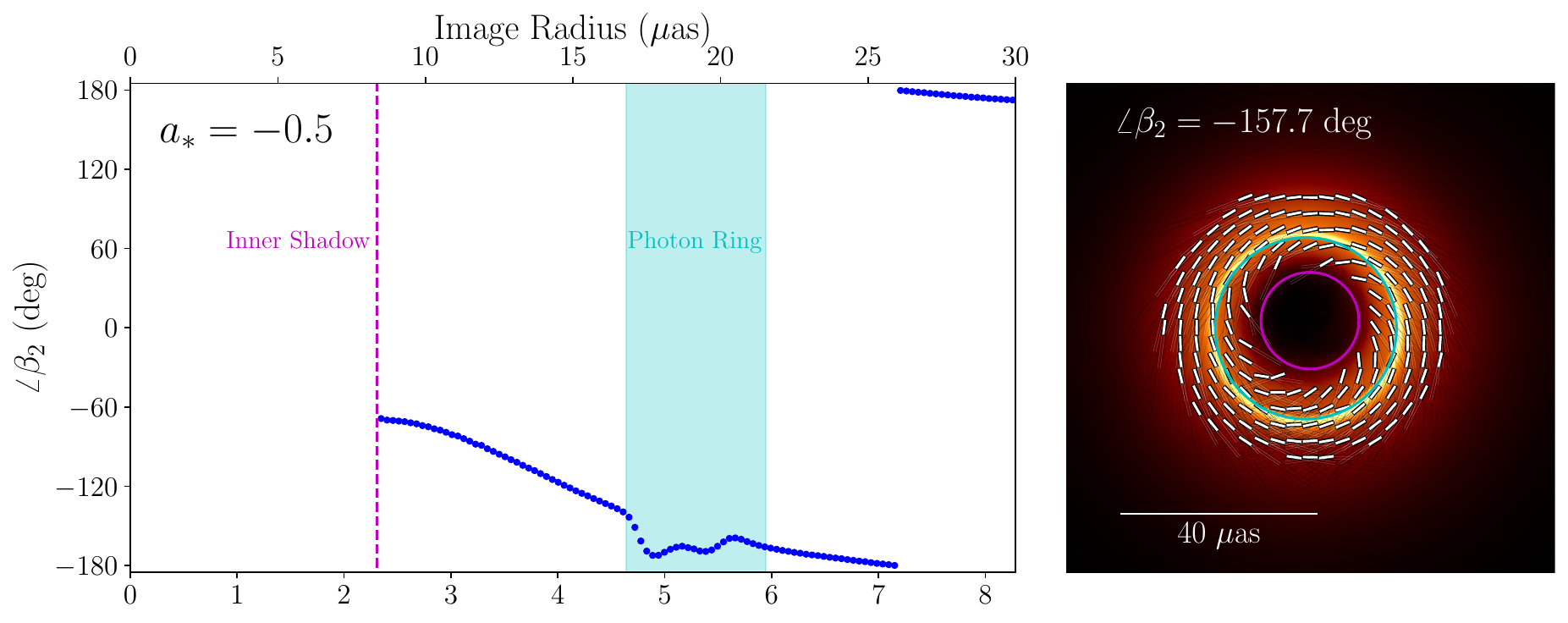} \\
\includegraphics[width=\linewidth]{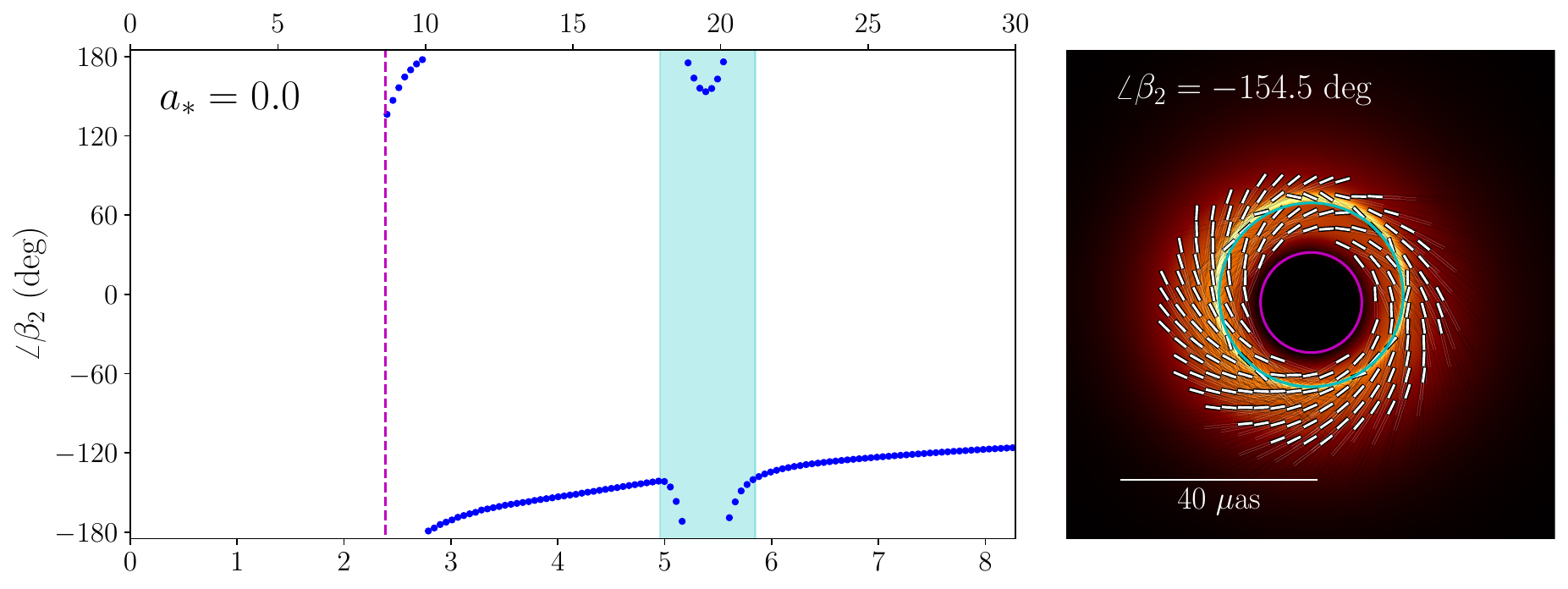} \\
\includegraphics[width=\linewidth]{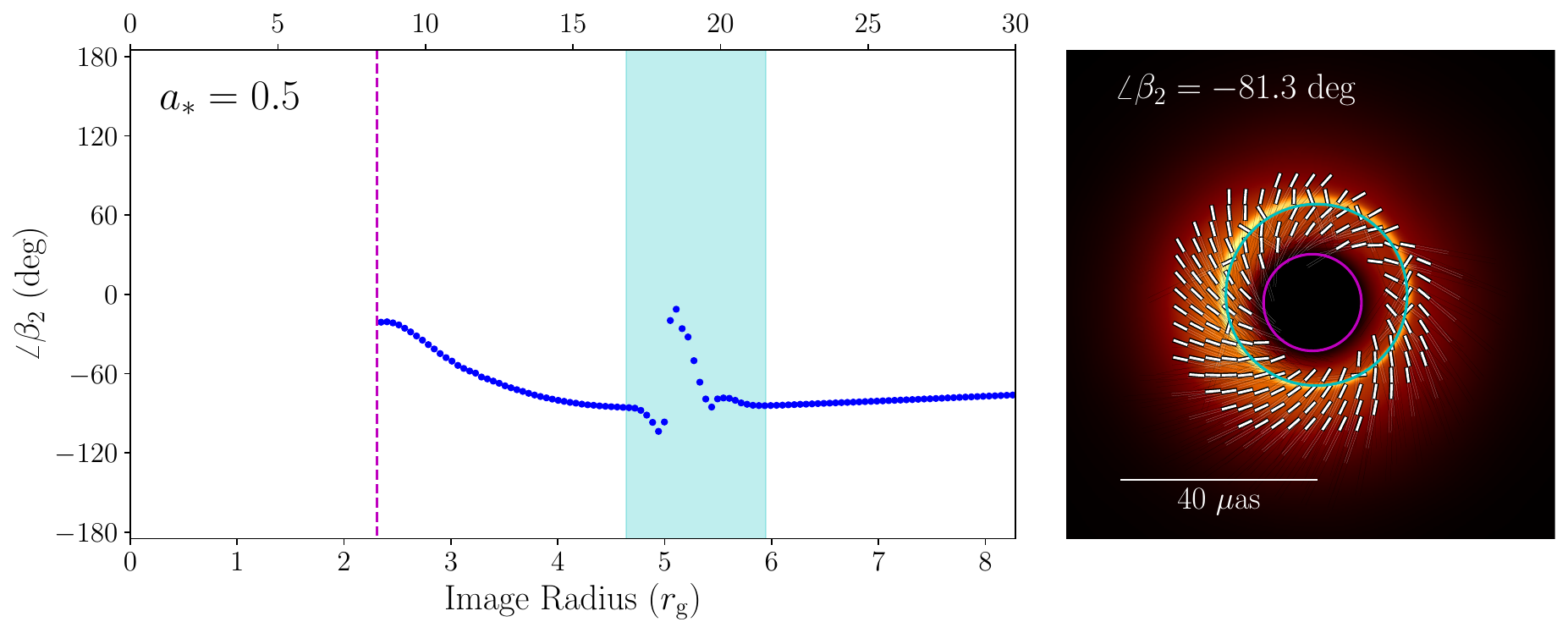}
\caption{
(Left column) $\angle\beta_2$ as a function of image radius in time-averaged images of M87* from MAD GRMHD simulations.
The top row shows the time-averaged image for $a_*=-0.5$, the middle row for $a_*=0$, and the bottom row for $a_*=0.5$.
All images use $R_{\rm high}=1$ to set the electron temperature in radiative transfer.
The vertical magenta line shows the outermost radius of the ``inner shadow,'' or the lensed image of the equatorial event horizon; the cyan band shows a range of radii corresponding to the $n=1$ photon ring.
(Right column) The time-averaged images used in computing $\angle\beta_2$ in the left panels.
Images are displayed in a gamma color scale and are not blurred to EHT resolution.
The black hole shadow/critical curve is indicated by the cyan contour, and the inner shadow with the magenta contour.
The image-average value of $\angle\beta_2$ is displayed on each panel.
A version of the top row of this plot was first produced in \cite{Ricarte2022}, Figure 8.
}
\label{fig:RadialBeta2}
\end{figure*}

\newpage
\section{Discussion}
\label{sec:Discussion}
 
We have established that, assuming we know the sign of the magnetic field line angular velocity $\Omega_F$, the helicity of linear polarization in near-horizon synchrotron radiation around a black hole---quantified by the polarimetric Fourier mode $\angle\beta_2$---is a probe of the direction of electromagnetic energy flow in the system.
We tested this relationship in both simple axisymmetric models and in images from full 3D, turbulent GRMHD simulations.
The latter include non-axisymmetric and non-equatorial emission regions, relativistic parallel transport, aberration, and Faraday rotation.
Remarkably, none of the complicating effects present in GRMHD simulations changed the qualitative insight of \autoref{fig:Cartoon}.
For M87*, negative values of $\angle\beta_2$ such as those observed by the EHT correspond to an outward electromagnetic energy flux on spatial scales of $\sim5GM/c^2$.

Here, we probe the relationship between $\angle\beta_2$, the magnetic field structure, the electromagnetic energy flow, and spin more deeply in the magnetically arrested GRMHD simulations from \autoref{sec:GRMHD}. 
In \autoref{fig:SimulationFields}, we show time- and azimuth-averaged quantities (see \autoref{app:Averaging}) from the simulations for three values of the black hole spin: $a_*=-0.5$ (left column), $a_*=0$ (middle column), and $a_*=+0.5$ (right column).
In all panels, the cyan contour shows the surface where the magnetization $\sigma=b^2/\rho=1$ and the white contours show surfaces of constant potential $\psi\equiv A_\phi$, corresponding to magnetic field lines in the axisymmetrized data.
Direct synchrotron emission observed by the EHT can arise in the equatorial plane or at higher latitudes, but it typically originates at characteristic radii in the range $r_{\rm emis}\approx3-5\,r_{\rm g}$ \citepalias{PaperV}.

The top row of \autoref{fig:SimulationFields} shows the sign of the ratio $B^\phi/B^r$ in BL coordinates in the averaged data from the three simulations.
The middle row shows the angular frequency of field lines $\Omega_F$ (in units of $t_{\rm g}^{-1}$), and the bottom row shows the outward Poynting flux $\sqrt{-g}\mathcal{J}_\mathcal{E}^r$.
In the bottom row, each panel is normalized independently relative to its maximum value, as the overall magnitude of the electromagnetic energy flux in the $a_*=0$ simulation is far smaller than in the two simulations with nonzero spin.

In the prograde simulation average, the field ratio $B^\phi/B^r<0$ everywhere.
In the retrograde simulation, $B^\phi/B^r> 0$ inside and somewhat exterior to the $\sigma=1$ contour where the field corotates with the black hole.
At larger radii, $B^\phi/B^r<0$ and the field rotates with the disk in the opposite sense to the black hole.
Both panels reinforce the physical insight of \autoref{eq:EnergyFlux}; outflowing Poynting flux requires the field to be wound up such that $\sign(\Omega_F B^r B^\phi)<0$ (\autoref{eq:GeneralJ}).
Both spinning simulations have field-line angular velocity $\Omega_F$ that are much larger on field lines connected to the black hole horizon than on field lines that are unconnected to the black hole.
The value of $|\Omega_F|$ on horizon-penetrating field lines is close to the value predicted by \citet{BlandfordZnajek1977}: $|\Omega_F|\approx\frac{1}{2}|\Omega_{\rm H}|=0.067\,t_{\rm g}^{-1}$ for $|a_*|=0.5$.
The field lines that are connected to the black hole are also where the outward electromagnetic energy flux $\sqrt{-g}\mathcal{J}^r_{\mathcal{E}}$ is strongest.
It is this outward electromagnetic energy flux, extracted from the black hole spin energy, that powers the large-scale jets in these simulations.

The behavior of $B^\phi/B^r$ is more complicated in the $a_*=0$ model.
In this simulation, electromagnetic energy flows outward at large radii on field lines disconnected from the black hole via the \citet{BlandfordPayne1982} mechanism; that is, outflow in this simulation is driven by accretion power rather than black hole spin energy.
At smaller radii and on field lines connected to the black hole horizon, electromagnetic energy flows \emph{into} the black hole in the $a_*=0$ simulation.
In order to have electromagnetic energy outflow (inflow), the product $\Omega_F B^r B^\phi$ must be negative (positive).
Close inspection reveals that these relationships are always satisfied in the averaged data from the $a_*=0$ simulation.
On field lines disconnected from the black hole, $B^\phi/B^r<0$ and $\Omega_F>0$, producing a weak energy outflow.
On field lines close to and connected to the black hole horizon, the ratio $B^\phi/B^r$ and the angular velocity $\Omega_F$ do not have fixed signs, but the total product $\Omega_FB^rB^\phi$ is always positive.
In individual snapshots from the $a_*=0$ simulation, the field ratio and angular velocity are less ordered and much more turbulent than in the $a_*=\pm0.5$ simulations, which largely resemble their time-averaged structure.

In \autoref{fig:RadialBeta2}, we investigate the behavior of $\angle\beta_2$ as a function of radius in these three simulations.
Here, we fix $R_{\rm high}=1$ and consider the time-averaged 230\,GHz images from the simulations, computing $\angle\beta_2$ from the time-average image at native resolution in annuli of width $0.2\,\mu$as.
When considering $\angle\beta_2$ as a function of radius in the $a_*=-0.5$ simulation, there is an obvious sign change at an image radius of $7r_{\rm g}\approx25\,\mu$as, corresponding to where the fields transition from counter-rotating to corotating with the black hole in \autoref{fig:SimulationFields}.
The image-averaged value of $\beta_2$ in this case is dominated by emission close to the black hole where the magnetic field lines corotate; as a result, the majority of retrograde snapshots still have image-averaged $\angle\beta_2<0$ in \autoref{fig:SimulationModels}.
Changes in the sign of $\angle\beta_2$ with radius can be used to diagnose changes in the sign of $B^\phi/B^r$ and $\Omega_F$ as the flow begins to corotate near the horizon \citep{Ricarte2022}.

All three simulations show interesting behavior in $\angle\beta_2$ in the range of image radii corresponding to the $n=1$ photon ring, where there are contributions from photons that have executed a half-orbit or more around the black hole \citep{JohnsonLupsasca2020}.
Shifts in $\angle\beta_2$ in the $n=1$ photon ring from parallel transport and radiative transfer effects \citep{Himwich2020,Jimenez2021,Palumbo2022} may be detectable in EHT targets on long 345\,GHz baselines \citep{Palumbo2023}.

Furthermore, in \autoref{fig:RadialBeta2} there is a clear radial trend in $\angle\beta_2$ in all three simulations as the image radius $\rho$ approaches the value corresponding to the directly lensed image of the event horizon (the ``inner shadow''; \citealt{Chael2021}).
Close to the horizon, the ratio $B^\phi/B^r$ changes rapidly with radius.
The precise behavior of $\angle\beta_2$ with radius close to the inner shadow depends strongly on black hole spin.
Interestingly, in both simulations with nonzero black hole spin, $\angle\beta_2$ trends to zero or positive values as the image radius approaches the inner shadow.
In the $a_*=0$ simulation, $\angle\beta_2$ becomes more negative, likely due to the necessarily inward-flowing energy flux very close to the horizon.
In \citetalias{Paper2} and \citetalias{Paper3}, we will explore the radial dependence of $\angle\beta_2$ and its value at the inner shadow in analytic models and GRMHD images in detail.

\autoref{fig:SimulationFields} indicates that the electromagnetic energy outflow that powers the jet in MAD simulations originates via the BZ mechanism and is concentrated on field lines that thread the black hole event horizon.
In this paper, we have argued that the observed $\angle\beta_2$ from EHT images of M87* also indicates electromagnetic energy outflow.
However, current observations are not conclusive as to whether or not the observed 230\,GHz synchrotron emission originates on field lines that thread the event horizon (thereby extracting energy via the BZ mechanism) or if they only thread the accretion disk at larger radii, extracting energy from the accretion disk's rotation (as in the $a_*=0$ simulation in \autoref{fig:SimulationFields}). 

Future EHT observations with more sites and at 345\,GHz frequency will have higher resolution and significantly more dynamic range than the pioneering 2017 observations of M87*~\citep{Doeleman2019,Raymond2021}.
By imaging the faint linear polarization signal closer to the event horizon, these observations may be able to conclusively determine if the observed emission originates on horizon-threading field lines that extract black hole spin energy.
With more work on calibrating simulation and analytic models, and accounting fully for the effects of Faraday rotation, these observations have the potential to conclusively determine whether or not the extragalactic jet from M87 is powered by the spin of the M87* supermassive black hole. 

\section{Conclusion}
\label{sec:Conclusion}

In this paper, we have investigated the link between the structure of resolved polarimetric images of synchrotron radiation near a black hole event horizon and the direction of electromagnetic energy flux in the black hole magnetosphere.
We have shown that: 
\begin{itemize}
    \item The sign of a polarimetric observable from EHT images of supermassive black holes---$\angle\beta_2$, which quantifies the helicity of the observed linear polarization---can be used to infer the sign of the ratio $B^\phi/B^r$ in simple axisymmetric models.
    \item Provided the orientation of the angular velocity of magnetic field lines around a black hole is known, the sign of the radial electromagnetic energy flux $\mathcal{J}_\mathcal{E}^r$ is determined by the sign of the ratio $B^\phi/B^r$. 
    \item Full GRMHD simulations including Faraday effects and non-axisymmetric fields (\autoref{fig:SimulationModels}) show the same overall trends in $\angle\beta_2$ as we expect from idealized analytic arguments (\autoref{fig:Cartoon}) and see in simple semi-analytic models (\autoref{fig:AnalyticModels}).
    These results provide strong support for connecting $\angle\beta_2$ and the direction of energy flux in real polarized images of black holes.
    \item If the emission comes from a magnetic field line threading the horizon, then the sign of $\angle\beta_2$ and its dependence on image radius measures whether energy is flowing into or being extracted from the black hole itself.
    \item Applied to published EHT observations of  M87* \citepalias{PaperVII,PaperVIII}, these conclusions are consistent with electromagnetic energy flowing outward in M87*'s magnetosphere.
    \item As first noted in \citet{Palumbo2020}, GRMHD values of $\angle\beta_2$ show a strong dependence on black hole spin, with larger spins featuring more azimuthal magnetic fields in the synchrotron-emitting region.
    This trend (although not the exact values of $\angle\beta_2$) is well-reproduced in images from the Blandford-Znajek monopole model (\autoref{fig:AnalyticModels}).
    Larger spins ``wrap up'' the magnetic field to be more azimuthal in the 230\,GHz emission region.
    This physical intuition and correspondence between models suggests a path forward for robustly measuring black hole spin with $\beta_2$.
    \item It is unclear from current EHT images whether or not the observed emission lies on field lines that thread the black hole event horizon and thus extract energy directly from the black hole spin.
    Future observations with an expanded EHT array will have more resolution and dynamic range and should be able to observe polarized emission from close to the projected event horizon, or ``inner shadow'' \citep{Chael2021}.
    These observations will be critical for distinguishing whether the outward energy flow observed by the current EHT is a result of the Blandford-Znajek process \citep{BlandfordZnajek1977} or whether it is powered by the accretion flow somewhat farther out from the event horizon \citep{BlandfordPayne1982}. 
\end{itemize}

\software{\texttt{kgeo} \citep{kgeo}, \texttt{KORAL} \citep{Sadowski2013,Sadowski2014}, \texttt{ipole} \citep{Noble2007,Moscibrodzka2018}, \texttt{eht-imaging} \citep{Chael2016,eht-imaging}, Numpy \citep{NumPy2020},  Scipy \citep{SciPy2020}, Matplotlib \citep{Matplotlib}}

\begin{acknowledgements}
The authors would like to thank Charles Gammie for serving as the EHT collaboration internal referee of this paper; his comments significantly improved the manuscript.
We thank Ramesh Narayan for providing the \texttt{KORAL} simulation data presented in \autoref{sec:GRMHD} and \autoref{sec:Discussion}, and for his useful comments and feedback.
We further thank Daniel Palumbo, Michael Johnson, Peter Galison, Dominic Chang, Razi Emami, and Angelo Ricarte for helpful discussions.   

AC was supported by the Princeton Gravity Initiative.
GNW was supported by the Taplin Fellowship.
AL was supported in part by NSF grant 2307888.
EQ was supported in part by a Simons Investigator award from the Simons Foundation. 
\end{acknowledgements}

\clearpage
\def\subsectionautorefname{section}
\appendix
\include{paper1_appendices}

\bibliography{paper1}
\bibliographystyle{aasjournal}

\end{document}

%% file: paper1_appendices.tex
\section*{Outline}
\label{app:Outline}

These appendices review key properties of degenerate electromagnetic fields in 4D spacetime, and more particularly in the background of a Kerr black hole.
Each appendix begins with a detailed summary of its results.

First, in \autoref{app:Degeneracy}, we define degenerate 2-forms and describe equivalent ways of characterizing them.
Our discussion is mathematical, but the results that we derive are then shown in \autoref{app:DegenerateElectromagnetism} to be relevant for electromagnetic fields in ideal magnetohydrodynamics (GRMHD) and force-free electrodynamics (FFE).

In \autoref{app:KerrFields}, we focus on stationary, axisymmetric electromagnetic fields in the Kerr spacetime.
We give explicit forms for such fields in both Boyer-Lindquist and Kerr-Schild coordinates.
We also derive their energy and angular momentum fluxes in both coordinate systems.

Then, these two threads join in \autoref{app:KerrDegenerateFields}, where we specialize to degenerate, stationary, axisymmetric fields in Kerr.
We review how such fields may be completely described by three functions in the poloidal $(r,\theta)$ plane: the magnetic flux function $\psi(r,\theta)$, the field-line current $I(\psi)$, and the field-line angular velocity $\Omega(\psi)$.
We derive the \cite{Znajek1977} condition for regularity of these fields on the horizon, and we describe its implications for the horizon fluxes of energy and angular momentum.
Our treatment covers similar ground as \citet[e.g.,][]{Carter1979, Gralla2014}.

Next, \autoref{app:KerrSolutions} reviews the known solutions to FFE in the Kerr background, with a particular emphasis on the \cite{BlandfordZnajek1977} monopole solution that we use in \autoref{sec:RingModels}.
In \autoref{app:AART}, we review the fluid velocity model
that we used to produce the simulated images shown in \autoref{sec:RingModels}; \autoref{app:ModelDescription} explains our procedure for generating these images in detail.
Finally, \autoref{app:GRMHDLibrary} describes the GRMHD simulations that we used in \autoref{sec:GRMHD}, and our strategy for time- and azimuth-averaging the GRMHD simulation data is described in \autoref{app:Averaging}.

\section{Degenerate forms}
\label{app:Degeneracy}

In this appendix, we first review general 2-forms in 4 dimensions before specializing to degenerate 2-forms, for which we present several characterizations and then prove their equivalence (sections \ref{app:GeneralForms} and \ref{app:DegenerateForms}).

We show that if a degenerate 2-form is magnetically dominated---the case of astrophysical relevance---then it must admit a timelike vector in its kernel (\autoref{app:MagneticallyDominatedForms}), and that if it is closed---as in electromagnetism---then it can be described in terms of two scalar Euler potentials (\autoref{app:ClosedDegenerateForms}).
We also introduce an ``electromagnetic'' decomposition for general 2-forms, which simplifies for degenerate fields (\autoref{app:ElectromagneticDecomposition}), and an associated local frame (\autoref{app:LocalFrames}).

At this stage in our discussion, these statements are purely mathematical, but their physical significance will become clearer in \autoref{app:DegenerateElectromagnetism}, where we will examine degenerate 2-forms that also obey Maxwell's equations.

\subsection{General 2-forms in 4 dimensions}
\label{app:GeneralForms}

Consider a 4-dimensional spacetime $\mathcal{M}$ with metric $g$, and let $v\cdot w=g_{\mu\nu}v^\mu w^\mu$ denote its inner product.
The Levi-Civita tensor is defined as $\epsilon_{\mu\nu\alpha\beta}=\sqrt{-g}[\mu\nu\alpha\beta]$, where $g=\det{g_{\mu\nu}}$ is the metric determinant and $[\mu\nu\alpha\beta]$ denotes the completely antisymmetric symbol, which is defined to be $\pm1$ according to whether $\mu\nu\alpha\beta$ is an even or odd permutation of the coordinates $\cu{0,1,2,3}$, and zero otherwise.\footnote{Likewise, $\epsilon^{\mu\nu\alpha\beta}=-\frac{1}{\sqrt{-g}}[\mu\nu\alpha\beta]$.}
The volume form is the contraction\footnote{The wedge product $A\wedge B$ of a $p$-form $A$ with a $q$-form $B$ is the $(p+q)$-form given by the antisymmetrized tensor product
\begin{align}
    (A \wedge B)_{\mu_1\cdots\mu_{p+q}}=\frac{(p+q)!}{p!q!}A_{[\mu_1\cdots\mu_p}B_{\mu_{p+1}\cdots\mu_{p+q}]}.
\end{align}}
\begin{subequations}
\label{eq:VolumeForm}
\begin{align}
    \epsilon &\equiv \epsilon_{\mu\nu\alpha\beta}\ed x^\mu\ed x^\nu\ed x^\alpha\ed x^\beta \\
    &=\frac{1}{4!}\epsilon_{\mu\nu\alpha\beta}\ed x^\mu\wedge\ed x^\nu\wedge\ed x^\alpha\wedge\ed x^\beta\\
    &=\sqrt{-g}\ed x^0\wedge\ed x^1\wedge\ed x^2\wedge\ed x^3.
\end{align}
\end{subequations}
A 2-form $F=F_{\mu\nu}\ed x^\mu\ed x^\nu=\frac{1}{2}F_{\mu\nu}\ed x^\mu\wedge\ed x^\nu$ has a dual
\begin{align}
    \label{eq:Dual}
    (\star F)_{\mu\nu}\equiv\frac{1}{2}\epsilon_{\mu\nu\alpha\beta}F^{\alpha\beta},
\end{align}
which, by definition, is the 2-form such that\footnote{The Hodge dual $\star\zeta$ of a $p$-form $\zeta$ is the $(4-p)$-form such that $\eta\wedge\star\zeta=\av{\eta,\zeta}\epsilon$ for any $p$-form $\eta$, where $\av{\eta,\zeta}=\eta_{\mu_1\cdots\mu_p}\zeta^{\mu_1\cdots\mu_p}$ is the inner product on $p$-forms.
Since $\epsilon$ is the only completely antisymmetric 4-tensor, $(F\wedge\star F)_{\mu\nu\alpha\beta}=6F_{[\mu\nu}(\star F)_{\alpha\beta]}\propto\epsilon_{\mu\nu\alpha\beta}$, and indeed the proportionality factor is $6F_{[01}(\star F)_{23]}=\frac{1}{2}F^2$.\label{fn:HodgeDual}}$^,$\footnote{This manifestly implies that $(\star F)^2\equiv(\star F)_{\mu\nu}(\star F)^{\mu\nu}=-F^2$.}
\begin{align}
    F\wedge\star F=\frac{F^2}{2}\epsilon,\quad
    F^2\equiv F_{\mu\nu}F^{\mu\nu}.
\end{align}
Meanwhile, $F\wedge F=6F_{[\mu\nu}F_{\alpha\beta]}\ed x^\mu\ed x^\nu\ed x^\alpha\ed x^\beta$, where brackets denote antisymmetrized indices, and
\begin{align}
    \label{eq:Antisymmetrization}
    F_{[\mu\nu}F_{\alpha\beta]}=\frac{1}{3}(F_{\mu\nu}F_{\alpha\beta}-F_{\mu\alpha}F_{\nu\beta}+F_{\mu\beta}F_{\nu\alpha}).
\end{align}
The same manipulations as in \eqref{eq:VolumeForm} reveal that
\begin{subequations}
\label{eq:DegenerateVolume}
\begin{align}
    F\wedge F&=\frac{1}{4}F_{[\mu\nu}F_{\alpha\beta]}\ed x^\mu\wedge\ed x^\nu\wedge\ed x^\alpha\wedge\ed x^\beta\\
    &=6F_{[01}F_{23]}\ed x^0\wedge\ed x^1\wedge\ed x^2\wedge\ed x^3\\
    &=-\frac{1}{2}F_{\mu\nu}(\star F)^{\mu\nu}\epsilon.
\end{align}
\end{subequations}
Finally, a direct computation shows that\footnote{Similarly, $\det{F^{\mu\nu}}=\det{(\star F)^{\mu\nu}}=\frac{1}{4\sqrt{-g}}F_{\mu\nu}(\star F)^{\mu\nu}$, since $T^{\mu\nu}=g^{\mu\alpha}g^{\nu\beta}T_{\alpha\beta}$, $\det{AB}=\det{A}\det{B}$, and $\det{g^{-1}}=\frac{1}{\det{g}}$.}
\begin{align}
    \label{eq:Determinants}
    \det{F_{\mu\nu}}=\det{(\star F)_{\mu\nu}}=\br{\frac{\sqrt{-g}}{4}F_{\mu\nu}(\star F)^{\mu\nu}}^2\!.
\end{align}

\subsection{Degenerate 2-forms}
\label{app:DegenerateForms}

A 2-form $F$ is \textit{degenerate} if $F\wedge F=0$, or explicitly, if
\begin{align}
    F_{[\mu\nu}F_{\alpha\beta]}=0.
\end{align}
By \eqref{eq:DegenerateVolume}--\eqref{eq:Determinants}, the degeneracy of $F$ is also equivalent to
\begin{align}
    \det{F_{\mu\nu}}=\det{(\star F)_{\mu\nu}}
    =F_{\mu\nu}(\star F)^{\mu\nu}
    =0,
\end{align}
which implies that $F$ is degenerate if and only if $\star F$ is.

A 2-form $F$ has a \textit{nontrivial kernel} if there exists some vector $v=v^\mu\pd_\mu$ such that $v^\mu F_{\mu\nu}=0$.

A 2-form $F$ is \textit{simple} if it is a wedge product $F=v\wedge w$ of two 1-forms $v$ and $w$, or explicitly,
\begin{align}
    \label{eq:Simple}
    F_{\mu\nu}=2v_{[\mu}w_{\nu]}
    =v_\mu w_\nu-v_\nu w_\mu.
\end{align}
This decomposition is far from unique, since $F$ is left invariant under shifts $v\to v+f(w)$ or $w\to w+g(v)$.
As a result of this freedom, we can assume without loss of generality that $v$ and $w$ are orthogonal.\footnote{Suppose that $F=v'\wedge w'$ for some $v'$ and $w'$ with nonzero overlap $v'\cdot w'\neq0$.
Then $v=v'-cw'$ with $c=\pa{v'\cdot w'}/\pa{w'\cdot w'}$ is manifestly orthogonal to $w=w'$ and $v\wedge w=F$ by construction.}

We now show that these three properties are in fact all pointwise equivalent, that is, that they all imply each other provided that we work locally at a fixed point in spacetime, treating $F_{\mu\nu}$ as a matrix rather than a tensor field.\footnote{The equivalence almost holds at the level of fields, except that degeneracy only implies simplicity \textit{locally}: there is no guarantee that the vectors $x$ and $y$ in \eqref{eq:Contraction}, which are defined in each tangent space separately, can be glued into smooth vector fields across spacetime; \S(3.5.35) of \cite{Penrose1984} presents a counter-example that illustrates this phenomenon.}
Our treatment mirrors the discussion of degeneracy in \cite{Gralla2014}.

First, suppose that $F$ is simple.
Then $F=v\wedge w$ for some 1-forms $v$ and $w$.
As such, $F\wedge F=v\wedge w\wedge v\wedge w=0$ (by antisymmetry of the wedge product) and hence, $F$ is degenerate.
Conversely, suppose that $F$ is degenerate, so that $F_{[\mu\nu}F_{\alpha\beta]}=0$.
There must nonetheless exist at least two nonzero vectors $x$ and $y$ such that $f^2\equiv F_{\mu\nu}x^\mu y^\nu\neq0$ (or else, $F=0$ identically).
From \eqref{eq:Antisymmetrization}, it follows that
\begin{align}
    \label{eq:Contraction}
    \pa{F_{\mu\nu}F_{\alpha\beta}-F_{\mu\alpha}F_{\nu\beta}+F_{\mu\beta}F_{\nu\alpha}}x^\alpha y^\beta=0.
\end{align}
Letting $fv_\nu\equiv x^\mu F_{\mu\nu}$ and $fw_\nu\equiv y^\mu F_{\mu\nu}$, this shows that
\begin{align}
    F_{\mu\nu}=\frac{\pa{F_{\mu\alpha}F_{\nu\beta}-F_{\mu\beta}F_{\nu\alpha}}x^\alpha y^\beta}{F_{\alpha\beta}x^\alpha y^\beta}
    =2v_{[\mu}w_{\nu]}.
\end{align}
Hence, \eqref{eq:Simple} holds and we conclude that $F$ is simple.
This proves that degeneracy $\Leftrightarrow$ (local) simplicity.

Next, suppose that $F\neq0$ is simple, so that $F=v\wedge w$ for some (orthogonal) 1-forms $v$ and $w$.
Then at every spacetime point $p$, the vectors $v$ and $w$ define a 2-plane in the tangent space $T_p\mathcal{M}\cong\mathbb{R}^4$.
Every point on this codimension-2 surface is intersected by a perpendicular 2-plane spanned by vectors $x$ and $y$, such that the set of vectors $(v,w,x,y)$ forms an orthogonal basis of $\mathbb{R}^4$.\footnote{Recall that $F$ is simple $\Leftrightarrow$ $\star F$ is simple.
By a suitable rescaling of $x$ and $y$, one can arrange to have $F=v\wedge w$ and $\star F=x\wedge y$.}
Thus, $x\cdot v=x\cdot w=y\cdot v=y\cdot w=0$, and as such, it follows from \eqref{eq:Simple} that $x^\mu F_{\mu\nu}=y^\mu F_{\mu\nu}=0$.
Hence, $F$ admits a nontrivial (2-dimensional) kernel spanned by $x$ and $y$.
Conversely, suppose that $F$ admits a nontrivial kernel.
Then by assumption, $\det{F_{\mu\nu}}=0$, and so it follows from \eqref{eq:Determinants} that $F_{\mu\nu}(\star F)^{\mu\nu}=0$.
By \eqref{eq:DegenerateVolume}, this implies that $F$ is degenerate and therefore simple.
Thus, a nontrivial kernel $\Leftrightarrow$ (local) simplicity.
This concludes our proof.

\subsection{Magnetically dominated, degenerate 2-forms}
\label{app:MagneticallyDominatedForms}

A 2-form $F$ is \textit{magnetically dominated} if $F^2>0$, \textit{null} if $F^2=0$, and \textit{electrically dominated} if $F^2<0$.\footnote{This nomenclature is justified by the invariant sign of \eqref{eq:ScalarContraction}.}

We now argue that if $F$ is a magnetically dominated, degenerate 2-form, then there exists a timelike vector $u$ such that $u^\mu F_{\mu\nu}=0$.
By the preceding discussion, since $F$ is degenerate, it is also (locally) simple, and so at every point in spacetime it can be written as $F=v\wedge w$ for some orthogonal 1-forms $v$ and $w$.
Then by \eqref{eq:Simple},
\begin{align}
    \label{eq:Norm}
    F^2=2v^2w^2.
\end{align}
Since $F^2>0$, the signs of $v^2$ and $w^2$ must be identical.
Moreover, this sign must be positive, so that $v$ and $w$ are both spacelike (since they must be orthogonal and there is only one independent timelike direction in spacetime).
As a result, the nontrivial kernel of $F$, which is spanned by $x$ and $y$, must admit a timelike vector $u$ (some linear combination $ax+by$) such that $u^\mu F_{\mu\nu}=0$.
By a suitable rescaling, we may always assume $u$ to have unit norm.

Conversely, if $F$ is a degenerate 2-form with a timelike vector $u$ in its kernel, then it must be magnetically dominated.
Indeed, given that there is a unique timelike direction $t$ in spacetime, if $t$ lies in the kernel of $F=v\wedge w$ (which is spanned by $x$ and $y$), then $t$ cannot lie within the perpendicular 2-plane spanned by the orthogonal vectors $v$ and $w$, which must therefore be spacelike.
By \eqref{eq:Norm}, it then follows that $F^2>0$.
This shows that for a degenerate $F$, magnetic domination ($F^2>0$) $\Leftrightarrow$ the existence of a (unit-norm) timelike vector $u$ in its kernel ($u^\mu F_{\mu\nu}=0$).\footnote{In principle, this analysis only applies at individual spacetime points, but in our applications, the vectors $u$ form a smooth field.}
Evidently, a degenerate $F$ always has a spacelike vector in its kernel.

\subsection{Euler potentials for closed, degenerate 2-forms}
\label{app:ClosedDegenerateForms}

Given any two scalars $\lambda_1$ and $\lambda_2$, the simple 2-form $F=\ed\lambda_1\wedge\ed\lambda_2$ is manifestly degenerate and also closed: $\ed F=0$ because $\ed^2=0$.\footnote{If $A$ is a $p$-form, then $\ed\pa{A\wedge B}=\ed A\wedge B+(-1)^pA\wedge\ed B$.}
Conversely, if a degenerate 2-form $F$ is also closed, then there (locally) exist scalars $\lambda_1$ and $\lambda_2$---known as \textit{Euler potentials}\footnote{In plasma physics, they are also known as \textit{Clebsch coordinates}.}---such that
\begin{align}
    \label{eq:EulerPotentials}
    F=\ed\lambda_1\wedge\ed\lambda_2.
\end{align}
\cite{Gralla2014} present multiple proofs of this fact in their \S3.2. 
As discussed therein, the pair of Euler potentials $(\lambda_1,\lambda_2)$ is not unique, as there are infinitely many other pairs $(\lambda_1',\lambda_2')$ such that $F=\ed\lambda_1'\wedge\ed\lambda_2'$, but the intersections of hypersurfaces or constant $\lambda_1$ and $\lambda_2$ are well-defined (i.e., independent of the choice of pair).

If $F$ solves the Maxwell equations \eqref{eq:Maxwell}, then these 2-dimensional surfaces are called \textit{field sheets}, and if $F$ is magnetically dominated, then these sheets are timelike.

\subsection{Local ``electromagnetic'' decomposition}
\label{app:ElectromagneticDecomposition}

Given a 2-form $F$, any vector $u$ defines projections\footnote{By \eqref{eq:Dual}, $b^\mu=-\frac{1}{2}\epsilon^{\mu\nu\alpha\beta}u_\nu F_{\alpha\beta}$, as in (8) of \cite{Noble2006}.}
\begin{align}
	\label{eq:LocalFields}
	e^\nu=-u_\mu F^{\mu\nu},\quad
	b^\nu=u_\mu(\star F)^{\mu\nu}.
\end{align}
By the antisymmetry of $F_{\mu\nu}$ and $(\star F)_{\mu\nu}$, $u\cdot e=u\cdot b=0$.
The inverse relations provide a general decomposition of $F$ and its dual $\star F$ in terms of three vectors $u$, $e$ and $b$,
\begin{subequations}
\label{eq:Decomposition}
\begin{align}
    -u^2F_{\mu\nu}&=u_\mu e_\nu-u_\nu e_\mu-\epsilon_{\mu\nu\alpha\beta}b^\alpha u^\beta,\\
    -u^2(\star F)_{\mu\nu}&=-u_\mu b_\nu+u_\nu b_\mu-\epsilon_{\mu\nu\alpha\beta}e^\alpha u^\beta,
\end{align}
\end{subequations}
in agreement with \cite{Baumgarte2003} and (only when $u^2=-1$) with (4)--(5) of \cite{McKinney2006}.

Physically, if $F$ is an electromagnetic field strength and $u$ is a unit-norm timelike vector, then the electric and magnetic fields in the local frame of an observer with 4-velocity $u$ are given by $e$ and $b$, respectively.
As such, we will call $e$ and $b$ the electric and magnetic fields, even when $F$ does not obey the Maxwell equations \eqref{eq:Maxwell}.

In terms of $e$ and $b$, the scalar invariants $F^2=F_{\mu\nu}F^{\mu\nu}$ and $(\star F)^2=(\star F)_{\mu\nu}(\star F)^{\mu\nu}$ take the form\footnote{Use $\epsilon_{\mu\nu\alpha\beta}\epsilon^{\mu\nu\kappa\lambda}=-2\pa{\delta_\alpha^\kappa\delta_\beta^\lambda-\delta_\alpha^\lambda\delta_\beta^\kappa}$ and $u\cdot e=u\cdot b=0$.}
\begin{subequations}
\begin{align}
    \label{eq:ScalarContraction}
    -\frac{u^2}{2}F^2=\frac{u^2}{2}(\star F)^2
    &=b^2-e^2,\\
    -\frac{u^2}{4}F_{\mu\nu}(\star F)^{\mu\nu}&=e\cdot b,
\end{align}
\end{subequations}
which is manifestly independent of the choice of $u$ in the decomposition \eqref{eq:Decomposition}.
By \eqref{eq:DegenerateVolume}--\eqref{eq:Determinants}, the degeneracy of $F$ is equivalent to the frame-invariant property
\begin{align}
    \label{eq:DegeneracyInvariant}
    e\cdot b=0.
\end{align}
Likewise, magnetic domination $F^2>0$ is equivalent to
\begin{align}
    \label{eq:DominationInvariant}
    b^2>e^2.
\end{align}
If $F$ is both degenerate and magnetically dominated, then (as we have shown) there must exist a unit-norm timelike vector $u$ in its kernel, so that in the frame of an observer with 4-velocity $u$, the electric field $e^\mu=u_\nu F^{\mu\nu}$ vanishes.
In other words, a 2-form $F$ satisfies both \eqref{eq:DegeneracyInvariant} and \eqref{eq:DominationInvariant} if and only if there exists a vector $u$ such that $e=0$, in which case \eqref{eq:Decomposition} simplifies to\footnote{
The antisymmetric matrix $F_{\mu\nu}$ has 6 independent degrees of freedom (dofs), now encoded in $u$ (3 dofs after normalization) and $b$ (3 dofs transverse to $u$), as in (11)--(12) of \cite{McKinney2006}.}
\begin{subequations}
\label{eq:GRMHD}
\begin{align}
    F_{\mu\nu}&=-\epsilon_{\mu\nu\alpha\beta}b^\alpha u^\beta,\\
    (\star F)_{\mu\nu}&=b_\mu u_\nu-b_\nu u_\mu,
\end{align}
\end{subequations}
making it manifest that $\star F=b\wedge u$ is a simple 2-form.
We reiterate that the choice of $u$ is not unique.
We will explicitly construct all such 4-velocities in \eqref{eq:TimelikeVector} below.

\subsection{Local frames for degenerate 2-forms}
\label{app:LocalFrames}

The antisymmetry of $F_{\mu\nu}$ and $(\star F)_{\mu\nu}$ guarantees that the projections \eqref{eq:LocalFields} satisfy $u\cdot e=u\cdot b=0$.
Moreover, if $F$ is degenerate, then \eqref{eq:DegeneracyInvariant} also holds, and so the vectors $u$, $e$ and $b$ are mutually orthogonal.
In that case, the following vector fields define an orthogonal frame (provided that $u$ is not in the kernel of $F$, so $e\neq0$):
\begin{align}
    \pa{u^\mu,e^\mu,b^\mu,z^\mu},\quad
    z^\mu\equiv-\epsilon^{\mu\nu\alpha\beta}u_\nu e_\alpha b_\beta.
\end{align}
Using \eqref{eq:Decomposition}, we can compute the projections
\begin{subequations}
\label{eq:FrameContractions}
\begin{gather}
    u^\mu F_{\mu\nu}=-e_\nu,\quad
    e^\mu F_{\mu\nu}=\frac{z_\nu+e^2u_\nu}{u^2},\\
    b^\mu F_{\mu\nu}=0,\quad
    z^\mu F_{\mu\nu}=b^2e_\nu.
\end{gather}
\end{subequations}

If $u$ is timelike ($u^2<0$), then $e$, $b$ and $z$ are spacelike, and we can form an orthonormal frame with frame fields
\begin{align}
    \label{eq:TimelikeFrame}
    \pa{T^\mu,X^\mu,Y^\mu,Z^\mu}=\pa{\frac{u^\mu}{\sqrt{-u^2}},\frac{e^\mu}{\sqrt{e^2}},\frac{b^\mu}{\sqrt{b^2}},\frac{z^\mu}{\sqrt{z^2}}}\!,
\end{align}
where $z^2=-u^2e^2b^2+u^2\pa{e\cdot b}^2=-u^2e^2b^2$.\footnote{Use $\epsilon^{\mu\nu\alpha\beta}\epsilon_{\mu\rho\kappa\lambda}=-3!\delta_{\rho\kappa\lambda}^{[\nu\alpha\beta]}$ with $\delta_{\rho\kappa\lambda}^{\nu\alpha\beta}\equiv\delta_\rho^\nu\delta_\kappa^\alpha\delta_\lambda^\beta$.
Explicitly, $\epsilon^{\mu\nu\alpha\beta}\epsilon_{\mu\rho\kappa\lambda}=-\delta_{\rho\kappa\lambda}^{\nu\alpha\beta}+\delta_{\rho\kappa\lambda}^{\nu\beta\alpha}-\delta_{\rho\kappa\lambda}^{\beta\nu\alpha}+\delta_{\rho\kappa\lambda}^{\beta\alpha\nu}-\delta_{\rho\kappa\lambda}^{\alpha\beta\nu}+\delta_{\rho\kappa\lambda}^{\alpha\nu\beta}$.\label{fn:Contraction}}

\section{Degenerate electromagnetism}
\label{app:DegenerateElectromagnetism}

In this appendix, we now consider degenerate 2-forms that also describe physical electromagnetic fields.
Such 2-forms $F$ must obey the Maxwell equations, which in terms of the current $J$ sourcing the field take the form\footnote{The last equation can also be written as the Bianchi identity $\nabla_{[\rho}F_{\mu\nu]}=0$, which expresses the closure $\ed F=0$ of the exact 2-form $F=\ed A$, where the 1-form $A$ is the gauge potential.}
\begin{align}
    \label{eq:Maxwell}
    \nabla_\nu F^{\mu\nu}=J^\mu,\quad
    \nabla_\nu(\star F)^{\mu\nu}=0.
\end{align}

We give expressions for the electric and magnetic fields in the frames of the ``normal observer,'' of the ``lab,'' and of a fluid.
We also construct the unique one-parameter family of 4-velocities for which the electric field vanishes, and we show explicitly that velocity components parallel to the electric field do not enter the stress-energy tensor.

\subsection{Electromagnetic stress-energy tensor}

A 2-form $F$ that obeys the Maxwell equations \eqref{eq:Maxwell} describes an electromagnetic field with stress-energy
\begin{align}
    \label{eq:StressTensor}
    T_{\mu\nu}^{\rm EM}\equiv F_{\mu\rho}{F_\nu}^\rho-\frac{1}{4}g_{\mu\nu}F^2.
\end{align}
The change in the energy-momentum is equal to the force exerted, so the Lorentz force density is\footnote{This is the relativistic generalization of $\vec{f}=\rho\vec{E}+\vec{J}\times\vec{B}$.}
\begin{align}
    \label{eq:LorentzForce}
    f_\nu\equiv\nabla^\mu T_{\mu\nu}^{\rm EM}
    =J^\mu F_{\mu\nu},
\end{align}
where the last step follows from \eqref{eq:Maxwell} and \eqref{eq:StressTensor}.
The general projection \eqref{eq:LocalFields}--\eqref{eq:Decomposition} decomposes $T_{\mu\nu}^{\rm EM}$ as\footref{fn:Contraction}$^,$\footnote{This agrees with (7) of \cite{McKinney2006} when $u^4=1$.}
\begin{align}
    u^4T_{\mu\nu}^{\rm EM}&=\frac{b^2+e^2}{2}\pa{u_\mu u_\nu+P_{\mu\nu}}+u^2\pa{b_\mu b_\nu+e_\mu e_\nu}\notag\\
    &\phantom{=}+u_\mu z_\nu+u_\nu z_\mu, 
\end{align}
where $u_\mu z_\nu+u_\nu z_\mu=-\pa{u_\mu\epsilon_{\nu\rho\kappa\lambda}+u_\nu\epsilon_{\mu\rho\kappa\lambda}}u^\rho e^\kappa b^\lambda$, and
\begin{align}
    \label{eq:Projection}
    P_{\mu\nu}\equiv u_\mu u_\nu-u^2g_{\mu\nu},\quad
    P_{\mu\nu}u^\mu u^\nu=0,
\end{align}
is a projection operator onto hypersurfaces normal to $u$.

\subsection{Application to GRMHD}

A prime example of degenerate electromagnetism is general-relativistic magnetohydrodynamics (GRMHD).
In GRMHD, $u$ is the 4-velocity of the plasma sourcing the electromagnetic field strength $F$, and $e$ and $b$ are the electric and magnetic fields in the fluid rest frame.
In the ideal GRMHD approximation, the plasma is assumed to be a perfect conductor.
As a consequence, the electric field in its rest frame is completely screened (assuming that there is sufficient free charge): this is the ``ideal MHD condition'' $e^\mu=u_\nu F^{\mu\nu}=0$ \citep{Gammie2003}.
By the preceding discussion, $e=0$ implies that $F$ is both degenerate ($e\cdot b=0\cdot b=0$) and magnetically dominated ($b^2>0=e^2$), so that it can be decomposed as in \eqref{eq:GRMHD}.

The fluid 4-velocity $u$ is not the only timelike vector field in the kernel of $F$.
In fact, as we will explicitly show in \eqref{eq:TimelikeVector}, there are infinitely many such vectors $u'$ (each with its own associated magnetic field $b'$), allowing for infinitely many decompositions of $\star F=b'\wedge u'$.

When $F$ is degenerate and magnetically dominated (as in GRMHD), its kernel contains a unit-norm, timelike $u$ (the fluid 4-velocity). Then in the rest frame of $u$, $e=0$ and the stress-energy tensor \eqref{eq:StressTensor} takes the simple form 
\begin{align}
    \label{eq:DegenerateStressTensor}
    T_{\mu\nu}^{\rm EM}=b^2u_\mu u_\nu+\frac{b^2}{2}g_{\mu\nu}-b_\mu b_\nu.
\end{align}

\subsection{Normal observer and lab frame}

Given a time coordinate $t$ and spatial coordinates $x^i$, the metric $ds^2=g_{\mu\nu}\ed x^\mu\ed x^\nu$ decomposes as
\begin{align}
    ds^2=-\alpha^2\ed t^2+g_{ij}\pa{\ed x^i+\beta^i\ed t}\!\pa{\ed x^j+\beta^j\ed t},
\end{align}
where the \textit{lapse} $\alpha$ and \textit{shift vector} $\beta^i$ are defined by
\begin{align}
    \label{eq:LapseShift}
    \alpha=\frac{1}{\sqrt{-g^{tt}}},\quad
    \beta^i=-\frac{g^{ti}}{g^{tt}}=\alpha^2g^{ti}.
\end{align}
Here, Latin indices run only over the spatial coordinates, and we reserve Greek indices for spacetime coordinates.

The \textit{normal observer} is defined to have the 4-velocity%
\begin{subequations}
\label{eq:NormalObserver}
\begin{align}
    \eta&=\eta_\mu\ed x^\mu
    =-\alpha\ed t\\
    &=\eta^\mu\pd_\mu
    =\frac{1}{\alpha}\pa{\pd_t-\beta^i\pd_i},
\end{align}
\end{subequations}
which is manifestly timelike with unit norm, $\eta\cdot\eta=-1$.

Numerical GRMHD codes often work with the normal observer's electric and magnetic fields \citep{Noble2006}%
\begin{subequations}
\label{eq:NormalFields}
\begin{align}
    \mathcal{E}^i&=-\eta_\mu F^{\mu i}
    =\alpha F^{0i},\\
	\mathcal{B}^i&=\eta_\mu(\star F)^{\mu i}
    =-\alpha(\star F)^{0i},
\end{align}
\end{subequations}
where we have now replaced the Greek spacetime index $\nu$ in \eqref{eq:LocalFields} by a Latin spatial index $i$ since by construction, $\mathcal{E}$ and $\mathcal{B}$ have vanishing time components.
Occasionally, instead of working in the frame of the normal observer \eqref{eq:NormalObserver}, GRMHD codes use as primitive variables
\begin{align}
	\label{eq:LabFrameFields}
	E^i=\frac{\mathcal{E}^i}{\alpha}
    =F^{0i},\quad
	B^i=\frac{\mathcal{B}^i}{\alpha}
    =-(\star F)^{0i},
\end{align}
which may be regarded as the electromagnetic fields in the ``lab frame'' defined by the (non-normalized) vector
\begin{align}
    \label{eq:LabFrame}
    \zeta=\zeta_\mu\ed x^\mu=-\ed t.
\end{align}
In flat spacetime, the lab frame becomes normal ($\zeta=\eta$), and $E^i=\mathcal{E}^i$ and $B^i=\mathcal{B}^i$ reduce to the usual electric and magnetic fields.
In asymptotically flat spacetimes, the lab frame describes an observer ``at rest at infinity'' since $\zeta=\pd_t+\mathcal{O}(1/r)$ and $\zeta^2=-1+\mathcal{O}(1/r)$.

\subsection{Relation between fields in different frames}

We now assume that $F$ is degenerate and magnetically dominated, and let $u$ denote a unit-norm, timelike vector in its kernel.
We can then use \eqref{eq:GRMHD} to relate the electric and magnetic fields $e$ and $b$ in the frame of $u$, given by \eqref{eq:LocalFields}, to the lab-frame fields $E$ and $B$ given by \eqref{eq:LabFrameFields}:
\begin{subequations}
\begin{align}
    B^i&=(\star F)^{i0}=b^iu^t-b^tu^i,\\
    E^i&=F^{0i}=-\epsilon^{0i\alpha\beta}b_\alpha u_\beta.
\end{align}
\end{subequations}
Since $u\cdot B=\zeta_\mu(\star F)^{\mu\nu}u_\nu=-\zeta\cdot b=b^t$, the inverse is
\begin{align}
    b^t=g_{i\mu}B^iu^\mu,\quad
    b^i=\frac{B^i+b^tu^i}{u^t},\quad
    e^\mu=0.
\end{align}
In terms of the projection tensor \eqref{eq:Projection}, which in this case is simply $P_{\mu\nu}=g_{\mu\nu}+u_\mu u_\nu$, this can be recast as
\begin{align}
    \label{eq:MagneticProjection}
    b^\mu=-\frac{1}{\zeta\cdot u}{P^\mu}_\nu B^\nu
    =\frac{1}{\gamma}{P^\mu}_\nu\mathcal{B}^\nu,
\end{align}
where $\zeta\cdot u=-u^t$, while $\gamma$ denotes the Lorentz factor of the flow relative to the normal observer frame,
\begin{align}
    \label{eq:LorentzFactor}
    \gamma\equiv-\eta\cdot u
    =\alpha u^t.
\end{align}

\subsection{Explicit timelike frames for magnetically dominated, degenerate fields}

We showed that if $F$ is degenerate and magnetically dominated, then there must exist a unit-norm, timelike vector $u$ in its kernel.
In fact, this vector is not unique.

We now explicitly construct all possible such vectors.
The answer takes the general form\footnote{We note that we can freely replace the normal fields \eqref{eq:NormalFields} by the lab-frame fields \eqref{eq:LabFrameFields} in this expression, leaving it invariant.\label{fn:NormalLabFrame}}
\begin{gather}
    u_{(\gamma)}^\mu=\gamma\pa{\eta^\mu-\frac{\epsilon^{\mu\nu\alpha\beta}\eta_\nu\mathcal{E}_\alpha\mathcal{B}_\beta}{\mathcal{B}^2}\pm\sqrt{1-\frac{\mathcal{E}^2}{\mathcal{B}^2}-\frac{1}{\gamma^2}}\frac{\mathcal{B}^\mu}{\sqrt{\mathcal{B}^2}}}\!,\notag\\
    \label{eq:TimelikeVector}
    \gamma\ge\sqrt{\frac{\mathcal{B}^2}{\mathcal{B}^2-\mathcal{E}^2}}>1,
\end{gather}
where the subscript $(\gamma)$ indicates that the Lorentz factor \eqref{eq:LorentzFactor} of this frame relative to the normal observer is $\gamma$.

To prove this, we work in the orthonormal frame \eqref{eq:TimelikeFrame} associated with the normal observer, with frame fields
\begin{subequations}
\label{eq:NormalFrame}
\begin{gather}
    T^\mu=\eta^\mu,\quad
    X^\mu=\frac{\mathcal{E}^\mu}{\sqrt{\mathcal{E}^2}},\quad
    Y^\mu=\frac{\mathcal{B}^\mu}{\sqrt{\mathcal{B}^2}},\\
    Z^\mu=-\frac{\epsilon^{\mu\nu\alpha\beta}\eta_\nu\mathcal{E}_\alpha\mathcal{B}_\beta}{\sqrt{\mathcal{E}^2\mathcal{B}^2}}.
\end{gather}
\end{subequations}
The general form of a vector $u$ expanded in this basis is
\begin{align}
    \label{eq:FlowDecomposition}
    u=\gamma\eta+\pa{u\cdot X}X+\pa{u\cdot Y}Y+\pa{u\cdot Z}Z.
\end{align}
Since $\mathcal{E}^\mu=F^{\mu\nu}\eta_\nu$, a vector $u$ in the kernel of $F$ must by definition satisfy $u\cdot X=u\cdot\mathcal{E}/\sqrt{\mathcal{E}^2}=u_\mu F^{\mu\nu}\eta_\nu/\sqrt{\mathcal{E}^2}=0$. Thus, $u$ takes must take the form
\begin{align}
    u=\gamma\eta+\pa{u\cdot Y}Y+\pa{u\cdot Z}Z.
\end{align}
This condition is necessary but not sufficient, since\footnote{By \eqref{eq:FrameContractions} and \eqref{eq:NormalFrame}, $Y_\mu F^{\mu\nu}=0$ and $Z_\mu F^{\mu\nu}=\sqrt{\mathcal{B}^2}X^\nu$.}
\begin{align}
    u_\mu F^{\mu\nu}=\br{-\gamma+\pa{u\cdot Z}\sqrt{\frac{\mathcal{B}^2}{\mathcal{E}^2}}}\mathcal{E}^\nu
\end{align}
is generically nonzero.
Thus, to ensure that $u$ really lies in the kernel of $F$, we must also demand that
\begin{align}
    u\cdot Z=\gamma\sqrt{\frac{\mathcal{E}^2}{\mathcal{B}^2}}.
\end{align}
This then leaves us with the general linear combination
\begin{align}
    \label{eq:IntermediateStep}
    u=\gamma\eta+\pa{u\cdot Y}Y+\gamma\sqrt{\frac{\mathcal{E}^2}{\mathcal{B}^2}}Z.
\end{align}
Lastly, we impose the normalization condition $u\cdot u=-1$.
Since the frame \eqref{eq:NormalFrame} is orthonormal, this amounts to
\begin{align}
    -1=-\gamma^2+\pa{u\cdot Y}^2+\gamma^2\frac{\mathcal{E}^2}{\mathcal{B}^2}.
\end{align}
Solving this equation for $\gamma$ results in
\begin{align}
    \label{eq:LorentzBoost}
    \gamma^2=\frac{1+\pa{u\cdot Y}^2}{1-\mathcal{E}^2/\mathcal{B}^2}.
\end{align}
Since $\pa{u\cdot Y}^2\ge0$ and $\mathcal{B}^2>\mathcal{E}^2$ (magnetic domination), this implies that there is a minimum Lorentz factor $\gamma_0$:
\begin{align}
    \gamma\ge\gamma_0,\quad
    \gamma_0\equiv\frac{1}{\sqrt{1-\mathcal{E}^2/\mathcal{B}^2}}>1.
\end{align}
At last, rewriting \eqref{eq:LorentzBoost} as $\gamma^2=\gamma_0^2\br{1+\pa{u\cdot Y}^2}$, \eqref{eq:IntermediateStep} gives us the most general unit-norm timelike vector $u$ in the kernel of $F$, parameterized by its Lorentz factor $\gamma$ relative to the normal observer frame \eqref{eq:NormalFrame} and a sign:
\begin{align}
    \label{eq:FlowFamily}
    u_{(\gamma)}=\gamma\eta\pm\sqrt{\frac{\gamma^2}{\gamma_0^2}-1}Y+\gamma\sqrt{1-\frac{1}{\gamma_0^2}}Z.
\end{align}
This concludes the derivation of \eqref{eq:TimelikeVector}.
When $\gamma=\gamma_0$, this expression agrees with (17) of \cite{McKinney2006}.

\subsection{Field-perpendicular and field-parallel velocities}

The 4-velocity decomposition \eqref{eq:FlowDecomposition} can be recast as
\begin{align}
    \label{eq:VelocityDecomposition}
    u=\gamma\eta+\gamma\tilde{v},
\end{align}
where $\gamma\tilde{v}=\pa{u\cdot X}X+\pa{u\cdot Y}Y+\pa{u\cdot Z}Z$ is a purely spatial 3-velocity.
In fact, $\tilde{v}$ lies fully within the spatial slices normal to the timelike vector $\eta$, and it is the spacelike vector obtained by projecting $u$ onto these surfaces:
\begin{align}
    \label{eq:SpatialFlow}
    \tilde{v}^\mu=\frac{u^\mu}{\gamma}-\eta^\mu
    =\frac{1}{\gamma}{P^\mu}_\nu u^\nu,\quad
    P_{\mu\nu}=g_{\mu\nu}+\eta_\mu\eta_\nu.
\end{align}
Indeed, $\tilde{v}^t=0$ by definition (as $\gamma=\alpha u^t$ and $\eta^t=1/\alpha$) and since $\eta\cdot\eta=u\cdot u=-1$ while $\eta\cdot u=-\gamma$, we have
\begin{align}
    \label{eq:SpatialBoost}
    \tilde{v}^2=\pa{\frac{u}{\gamma}-\eta}^2
    =1-\frac{1}{\gamma^2}
    >0.
\end{align}

Typically, numerical GRMHD codes \citep[e.g.,][]{McKinney2004} do not evolve the 4-velocity $u$ of the flow directly, but rather its spatial projection \eqref{eq:SpatialFlow}, which by \eqref{eq:SpatialBoost} obeys the nice relation
\begin{align}
    \label{eq:}
    \gamma=\frac{1}{\sqrt{1-\tilde{v}^2}}.
\end{align}

If $F$ is degenerate and magnetically dominated (as in GRMHD), then its associated timelike flow must lie in the family \eqref{eq:FlowFamily}, and hence $\tilde{v}$ admits a decomposition
\begin{align}
    \label{eq:SpatialVelocity}
    \tilde{v}=\tilde{v}_\perp+\tilde{v}_\parallel,
\end{align}
where the (magnetic) field-perpendicular velocity is\footnote{With $\tilde{\epsilon}^{ijk}$ the antisymmetric symbol, $\tilde{v}_\perp^i=\frac{\alpha}{\sqrt{-g}\mathcal{B}^2}\tilde{\epsilon}^{ijk}\mathcal{E}_j\mathcal{B}_k$.\label{fn:PerpendicularVelocity}}
\begin{align}
    \label{eq:FieldPerpendicularVelocity}
    \tilde{v}_\perp^\mu\equiv\sqrt{1-\frac{1}{\gamma_0^2}}Z^\mu
    =-\frac{\epsilon^{\mu\nu\alpha\beta}\eta_\nu\mathcal{E}_\alpha\mathcal{B}_\beta}{\mathcal{B}^2},
\end{align}
while the (magnetic) field-parallel velocity is
\begin{align}
    \tilde{v}_\parallel^\mu\equiv\pm\sqrt{\frac{1}{\gamma_0^2}-\frac{1}{\gamma^2}}Y^\mu
    =\pm\sqrt{\frac{1}{\gamma_0^2}-\frac{1}{\gamma^2}}\frac{\mathcal{B}^\mu}{\sqrt{\mathcal{B}^2}}.
\end{align}
Since $\tilde{v}_\parallel$ vanishes when $\gamma=\gamma_0$, \eqref{eq:SpatialBoost} implies that
\begin{align}
    \tilde{v}_\perp^2=1-\frac{1}{\gamma_0^2}
    =\frac{\mathcal{E}^2}{\mathcal{B}^2},
\end{align}
so that the purely field-perpendicular flow (with no field-parallel component) has the minimal Lorentz factor
\begin{align}
    \label{eq:MinimalLorentzBoost}
    \gamma_\perp\equiv\frac{1}{\sqrt{1-\tilde{v}_\perp^2}}
    =\gamma_0.
\end{align}
Conversely, when the Lorentz factor diverges ($\gamma\to\infty$), the field-parallel velocity is maximized:
\begin{align}
    \tilde{v}_\parallel^\mu\to\pm\tilde{v}_\parallel^{\rm max}\frac{\mathcal{B}^\mu}{\sqrt{\mathcal{B}^2}},\quad
    \tilde{v}_\parallel^{\rm max}\equiv\frac{1}{\gamma_0}
    =\sqrt{1-\frac{\mathcal{E}^2}{\mathcal{B}^2}}.
\end{align}
We can thus parameterize the spatial velocity \eqref{eq:SpatialVelocity} as
\begin{align}
    \tilde{v}^i=\tilde{v}_\perp^i+\xi\tilde{v}_\parallel^{\rm max}\frac{\mathcal{B}^i}{\sqrt{\mathcal{B}^2}},\quad
    \xi\in(-1,1).
\end{align}
Lastly, we note---as in (18) of \cite{McKinney2006}---that the coordinate 3-velocity is
\begin{align}
    \label{eq:AvoidConfusion}
    \frac{u_{(\gamma)}^i}{u_{(\gamma)}^t}=\frac{\eta^i+\tilde{v}^i}{\eta^t}
    =-\beta^i+\frac{\alpha^2}{\sqrt{-g}\mathcal{B}^2}\tilde{\epsilon}^{ijk}\mathcal{E}_j\mathcal{B}_k.
\end{align}
For instance, when $i=\phi$ is an azimuthal coordinate, the flow \eqref{eq:TimelikeVector} has angular velocity $\Omega_{(\gamma)}\equiv u_{(\gamma)}^\phi/u_{(\gamma)}^t$.

\subsection{Representations of the stress-tensor}

When $F$ is degenerate and magnetically dominated, its energy-momentum tensor \eqref{eq:DegenerateStressTensor} can be written as
\begin{align}
    \label{eq:DriftStressTensor}
    T_{\rm EM}^{\mu\nu}=b_{(\gamma)}^2u_{(\gamma)}^\mu u_{(\gamma)}^\nu+\frac{b_{(\gamma)}^2}{2}g^{\mu\nu}-b_{(\gamma)}^\mu b_{(\gamma)}^\nu,
\end{align}
where $u_{(\gamma)}$ is any one of the unit-norm timelike vectors \eqref{eq:FlowFamily} in its kernel, with associated magnetic field \eqref{eq:MagneticProjection}:
\begin{align}
    \label{eq:DriftFrameMagneticField}
    b_{(\gamma)}^\mu=\frac{1}{\gamma}\br{\mathcal{B}^\mu+\pa{u_{(\gamma)}\cdot\mathcal{B}}u_{(\gamma)}^\mu}\!.
\end{align}
In particular, the norm of this magnetic field is
\begin{align}
    b_{(\gamma)}^2=\frac{1}{\gamma^2}\br{\mathcal{B}^2+\pa{u_{(\gamma)}\cdot\mathcal{B}}^2}\!.
\end{align}
Among all the flows \eqref{eq:FlowFamily}, one of them is preferred given our choice of coordinate system,
\begin{align}
    u_\perp\equiv u_{(\gamma_0)}
    =\gamma_\perp\pa{\eta+\tilde{v}_\perp},
\end{align}
because it has minimal Lorentz factor $\gamma_0=\gamma_\perp$ relative to our normal observer, and it has no component aligned with the magnetic field (so $\tilde{v}_\parallel=0$).
Since $u_\perp\cdot\mathcal{B}=0$,
\begin{align}
    \label{eq:Perpendicular}
    b_\perp^\mu\equiv b_{(\gamma)}^\mu
    =\frac{\mathcal{B}^\mu}{\gamma_\perp},\quad
    b_\perp^2=\frac{\mathcal{B}^2}{\gamma_\perp^2}
    =\mathcal{B}^2-\mathcal{E}^2,
\end{align}
and the resulting representation of stress-energy tensor,
\begin{align}
    \label{eq:PerpendicularDriftStressTensor}
    T_{\rm EM}^{\mu\nu}=b_\perp^2u_\perp^\mu u_\perp^\nu+\frac{b_\perp^2}{2}g^{\mu\nu}-b_\perp^\mu b_\perp^\nu,
\end{align}
is manifestly independent of the flow velocity parallel to the magnetic field, depending only on its perpendicular velocity (whereas $u_{(\gamma)}=\gamma(u_\perp/\gamma_\perp+\tilde{v}_\parallel)$ in general.)

\subsection{Application to FFE}
\label{app:FFE}

Force-free electrodynamics, or FFE, is a limit of ideal MHD in which the plasma becomes dilute and the local energy is dominated by the electromagnetic energy,
\begin{align}
    \label{eq:DilutePlasma}
    T_{\mu\nu}\approx T_{\mu\nu}^{\rm EM}.
\end{align}
In that case, the Bianchi identity for the Riemann tensor $\nabla^\mu T_{\mu\nu}=0$ ensures the conservation of electromagnetic energy: $\nabla^\mu T_{\mu\nu}^{\rm EM}\approx0$.\footnote{The force-free condition can hold even when \eqref{eq:DilutePlasma} does not.}
This approximation is known as the \textit{force-free} condition, since by \eqref{eq:LorentzForce}, it implies that the Lorentz force density vanishes.
That is, $f_\nu=0$, or
\begin{align}
    \label{eq:ForceFree}
    J^\mu F_{\mu\nu}=0.
\end{align}
This implies that the current is always perpendicular to the electric field $e^\mu=F^{\mu\nu}u_\nu$, for any 4-velocity $u$:\footnote{Thus, a force-free field obeys both $\rho\vec{E}+\vec{J}\times\vec{B}=0$ and $\vec{J}\cdot\vec{E}=0$.}
\begin{align}
    \label{eq:ForceFreeCurrent}
    J\cdot e=J_\mu F^{\mu\nu}u_\nu
    =0.
\end{align}
In particular, in the frame \eqref{eq:NormalFrame}, $J\cdot X=J\cdot\mathcal{E}=0$, so
\begin{align}
    \label{eq:CurrentExpansion}
    J=-\pa{J\cdot\eta}\eta+\pa{J\cdot Y}Y+\pa{J\cdot Z}Z.
\end{align}
To be more explicit, first note that we can write
\begin{align}
    J\cdot Z=-\frac{\epsilon^{\mu\nu\alpha\beta}J_\mu\eta_\nu\mathcal{E}_\alpha\mathcal{B}_\beta}{\sqrt{\mathcal{E}^2\mathcal{B}^2}}
    =-\frac{\epsilon^{\mu\nu\alpha\beta}J_\mu\mathcal{E}_\nu\mathcal{B}_\alpha\eta_\beta}{\sqrt{\mathcal{E}^2\mathcal{B}^2}}.
\end{align}
By \eqref{eq:Decomposition}, $F^{\mu\nu}=\eta^\mu\mathcal{E}^\nu-\eta^\nu\mathcal{E}^\mu-\epsilon^{\mu\nu\alpha\beta}\mathcal{B}_\alpha\eta_\beta$ and so
\begin{align}
    J\cdot Z=\frac{\pa{F^{\mu\nu}-\eta^\mu\mathcal{E}^\nu+\eta^\nu\mathcal{E}^\mu}J_\mu\mathcal{E}_\nu}{\sqrt{\mathcal{E}^2\mathcal{B}^2}}
    =-\pa{J\cdot\eta}\sqrt{\frac{\mathcal{E}^2}{\mathcal{B}^2}},\notag
\end{align}
where the first term vanished by \eqref{eq:ForceFree} and the last by \eqref{eq:ForceFreeCurrent}.
Hence, we can decompose the current \eqref{eq:CurrentExpansion} into
field-parallel and field-perpendicular currents
\begin{align}
    J=J_\perp+J_\parallel,
\end{align}
which are explicitly given by
\begin{subequations}
\begin{align}
    J_\parallel^\mu&=\pa{J\cdot Y}Y^\mu
    =\frac{J\cdot\mathcal{B}}{\mathcal{B}^2}\mathcal{B}^\mu,\\
    J_\perp^\mu&=-\pa{J\cdot\eta}\pa{\eta^\mu+\sqrt{\frac{\mathcal{E}^2}{\mathcal{B}^2}}Z^\mu}\\
    &=\alpha J^t\pa{\eta^\mu-\frac{\epsilon^{\mu\nu\alpha\beta}\eta_\nu\mathcal{E}_\alpha\mathcal{B}_\beta}{\mathcal{B}^2}},
\end{align}
\end{subequations}
and have norms
\begin{align}
    J_\parallel^2=\frac{\pa{J\cdot\mathcal{B}}^2}{\mathcal{B}^2},\quad
    J_\perp^2=-\pa{J\cdot\eta}^2\!\pa{1-\frac{\mathcal{E}^2}{\mathcal{B}^2}}\!.
\end{align}
Thus, $J_\parallel$ is always spacelike, while $J_\perp$ is timelike if and only if $F$ is magnetically dominated.
We also note that
\begin{align}
    J_\perp^\mu=\frac{\alpha J^t}{\gamma}u_\perp^\mu
    =\frac{J^t}{u^t}u_\perp^\mu,
\end{align}
where we used \eqref{eq:LorentzFactor}, so we can interpret $J_\perp$ as a charge density $J^t/u^t$ flowing allowing $u_\perp$.

In GRMHD, the field strength $F$ is always guaranteed to be both degenerate and magnetically dominated by the ideal MHD condition, which requires $F$ to have a timelike vector in its kernel: the flow 4-velocity $u$.

In FFE, there is no longer such a flow $u$.
Nevertheless, the force-free condition \eqref{eq:ForceFree} still requires $F$ to have a nontrivial kernel (containing $J$), which implies that $F$ is degenerate.
However, since $J$ need not be timelike,\footnote{In fact, $J$ is often spacelike, as in the split monopole solution: physically, the plasma consists of two oppositely charged species (electrons and positrons, say) with timelike velocities $u_+$ and $u_-$ producing a spacelike net current $J\propto u_+-u_-$.}
$F$ is not automatically magnetically dominated.
Instead, magnetic domination is a separate assumption, which in fact \textit{must} be imposed in order to ensure well-posedness of the initial value problem, that is, to ensure that the evolution equations are hyperbolic (rather than elliptic, as in the electrically dominated case where the charges making up the plasma are accelerated away).
For further discussion, see \cite{Komissarov2002} or \cite{Gralla2014} and other references therein.

\section{Stationary and axisymmetric electromagnetic fields on Kerr}
\label{app:KerrFields}

\subsection{Kerr spacetime in Boyer-Lindquist coordinates}

The Kerr metric in Boyer-Lindquist coordinates is
\begin{subequations}
\label{eq:Kerr}
\begin{gather}
	ds^2=-\frac{\Delta}{\Sigma}\pa{\ed t-a\sin^2{\theta}\ed\phi}^2+\frac{\Sigma}{\Delta}\ed r^2+\Sigma\ed\theta^2\notag\\
	+\frac{\sin^2{\theta}}{\Sigma}\br{\pa{r^2+a^2}\ed\phi-a\ed t}^2,\\
	\Delta=r^2-2Mr+a^2,\quad
	\Sigma=r^2+a^2\cos^2{\theta}.
\end{gather}	
\end{subequations}
The outer/inner event horizons $r_\pm$ are the zeros of $\Delta$,
\begin{align}
	\label{eq:HorizonRadius}
	r_\pm=M\pm\sqrt{M^2-a^2},
\end{align}
and the angular velocity of the (outer) event horizon is
\begin{align}
    \label{eq:HorizonVelocity}
    \Omega_{\rm H}=\frac{a}{2Mr_+}.
\end{align}
The metric determinant is
\begin{align}
    \label{eq:MetricDeterminant}
    \sqrt{-g}=\Sigma\sin{\theta}.
\end{align}
The lapse $\alpha$ and shift vector $\beta^i$ defined in \eqref{eq:LapseShift} are
\begin{align}
    \label{eq:KerrLapse}
    \alpha^2=\frac{\Delta\Sigma}{\Pi},\quad
    \beta^\phi=-\frac{2aMr}{\Pi},
\end{align}
and $\beta^r=\beta^\theta=0$, where we introduced
\begin{align}
    \Pi=\frac{\Sigma}{\sin^2{\theta}}g_{\phi\phi}
    =\pa{r^2+a^2}^2-a^2\Delta\sin^2{\theta}.
\end{align}
The Kerr geometry is symmetric under time translations and rotations about the spin axis.
These two isometries are respectively generated by the Killing vector fields
\begin{align}
    \label{eq:Killing}
    K=\pd_t,\quad
    R=\pd_\phi,
\end{align}
which leave the metric invariant: letting $\mathcal{L}_\xi$ denote a Lie derivative along $\xi$, $K$ and $R$ obey the Killing equation\footnote{Explicitly, the Killing equation $\mathcal{L}_\xi g=0$ is $\nabla_\mu\xi_\nu+\nabla_\nu\xi_\mu=0.$\label{fn:Killing}}
\begin{align}
    \mathcal{L}_Kg=\mathcal{L}_Rg
    =0.
\end{align}
An object with 4-momentum $p$ has energy $-p_t=-p\cdot K$ and spin angular momentum $p_\phi=p\cdot R$.

The normal observer with respect to Boyer-Lindquist coordinates, whose unit-norm timelike 4-velocity $\eta$ is given by \eqref{eq:NormalObserver} together with \eqref{eq:KerrLapse}, has zero angular momentum ($\eta_\phi=0$) and is therefore often referred to as ``the ZAMO'' \cite{Bardeen1972}.
However, due to frame-dragging effects ($g^{t\phi}\neq0$), the ZAMO nonetheless has a nonzero angular velocity (which equals $\Omega_{\rm H}$ at $r_+$)
\begin{align}
    \label{eq:ZAMO}
    \omega=\frac{\eta^\phi}{\eta^t}
    =-\beta^\phi
    =\frac{2aMr}{\Pi},
\end{align}
in terms of which $\eta=\alpha^{-1}\pa{\pd_t+\omega\pd_\phi}$.
The ``lab frame'' is associated with the (non-normalized) vector $\zeta=-\ed t$ defined in \eqref{eq:LabFrame}.
The ZAMO and the lab frame are both ``at rest at infinity'' ($r\to\infty$), where $\eta=\pd_t+\mathcal{O}(1/r)=\zeta$.

\subsection{Kerr spacetime in Kerr-Schild coordinates}

The Boyer-Lindquist coordinates $x^\mu=(t,r,\theta,\phi)$ and the Kerr-Schild coordinates $\bar{x}^\mu=(\bar{t},r,\theta,\bar{\phi})$ share the same poloidal coordinates $(r,\theta)$, whereas their toroidal coordinates are related by shifts
\begin{align}
    \label{eq:CoordinateTransformation}
	\ed\bar{t}=\ed t+\frac{2Mr}{\Delta}\ed r,\quad
	\ed\bar{\phi}=\ed\phi+\frac{a}{\Delta}\ed r,
\end{align}
which cancel the divergence of the Boyer-Lindquist line element \eqref{eq:Kerr} as $\Delta\to0$.
This is the main advantage of the Kerr-Schild coordinates: they remove the spurious coordinate singularities across the event horizons \eqref{eq:HorizonRadius}.

Using \eqref{eq:CoordinateTransformation}, one can convert the line element \eqref{eq:Kerr} to Kerr-Schild coordinates, or find the Jacobian for the coordinate transformation $x^\mu\to\bar{x}^\mu$.
For instance, in Kerr-Schild coordinates, the Killing vectors \eqref{eq:Killing} are
\begin{align}
    K=\pd_{\bar{t}},\quad
    R=\pd_{\bar{\phi}},
\end{align}
while the Boyer-Lindquist normal observer becomes
\begin{align}
    \eta=-\alpha\pa{\ed\bar{t}-\frac{2Mr}{\Delta}\ed r}\!,
\end{align}
which is no longer normal with respect to Kerr-Schild coordinates; instead, the Kerr-Schild normal observer is
\begin{align}
    \bar{\eta}=-\bar{\alpha}\ed\bar{t}
    =\frac{1}{\bar{\alpha}}\pa{\pd_{\bar{t}}-\bar{\beta}^r\pd_r}
    \neq\eta,
\end{align}
where the Kerr-Schild lapse $\bar{\alpha}$ and shift vector $\bar{\beta}^i$ are
\begin{align}
    \bar{\alpha}^2=\frac{\Sigma}{\Sigma+2Mr},\quad
    \bar{\beta}^r=\frac{2Mr}{\Sigma+2Mr},
\end{align}
and $\bar{\beta}^\theta=\bar{\beta}^{\bar{\phi}}=0$.
This normal observer $\bar{\eta}$ has zero angular momentum ($\bar{\eta}\cdot R=\bar{\eta}_{\bar{\phi}}=0$) and is thus also a ZAMO.
Moreover, this ZAMO also has zero angular velocity, since $\bar{\eta}^{\bar{\phi}}/\bar{\eta}^{\bar{t}}=-\bar{\beta}^{\bar{\phi}}=0$.
However, this ZAMO also has negative radial velocity $\bar{\eta}^r/\bar{\eta}^{\bar{t}}=-\bar{\beta}^r<0$, and is therefore infalling.
Lastly, the Kerr-Schild ``lab frame'' is defined by the (non-normalized) vector $\bar{\zeta}=-\ed\bar{t}\neq\zeta$.

\subsection{Stationary and axisymmetric fields on Kerr}

A Kerr electromagnetic field configuration is a 2-form $F$ that obeys the Maxwell equations \eqref{eq:Maxwell} in the metric background \eqref{eq:Kerr}.
In particular, $F$ is necessarily closed ($\ed F=0$) and must therefore be (locally) exact, so that $F=\ed A$ for some gauge potential 1-form $A=A_\mu\ed x^\mu$.

Such an electromagnetic field is \textit{stationary} if $\mathcal{L}_KF=0$ and \textit{axisymmetric} if $\mathcal{L}_RF=0$.
A field that is stationary \textit{and} axisymmetric always admits a 1-form potential $A$ whose components $A_\mu(r,\theta)$ are independent of $(t,\phi)$.\footnote{$A=A_\mu(r,\theta)\ed x^\mu$ is the general solution to $\mathcal{L}_KA=\mathcal{L}_RA=0$, which implies $\mathcal{L}_KF=\mathcal{L}_RF=0$ by Cartan's formula $\br{\ed,\mathcal{L}_\xi}=0$.
Conversely, $\mathcal{L}_\xi F=0$ implies $\ed\mathcal{L}_\xi A=0$, so $\mathcal{L}_\xi A=\ed\lambda_\xi$ for a scalar $\lambda_\xi$.
Since $[K,R]=0$, $\mathcal{L}_K\mathcal{L}_RA-\mathcal{L}_R\mathcal{L}_KA=\mathcal{L}_{[K,R]}A=0$, so $\ed\pa{\mathcal{L}_K\lambda_R-\mathcal{L}_R\lambda_K}=0$.
This is the integrability condition for the linear system $\mathcal{L}_K\ed\tilde{\lambda}=\ed\lambda_K$, $\mathcal{L}_R\ed\tilde{\lambda}=\ed\lambda_R$, which therefore admits a simultaneous solution $\tilde{\lambda}$.
Thus, the gauge-transformed potential $\tilde{A}=A-\ed\tilde{\lambda}$ is such that $F=\ed\tilde{A}$ and $\mathcal{L}_K\tilde{A}=\mathcal{L}_R\tilde{A}=0$.}

The electric and magnetic field components \eqref{eq:LabFrameFields} in the Boyer-Lindquist lab frame take the explicit form
\begin{subequations}
\label{eq:BoyerLindquistFrameFields}
\begin{align}
    E^r&=\br{2aMr+\Pi\frac{A_{t,r}}{A_{\phi,r}}}\!\frac{A_{\phi,r}}{\Sigma^2},
    &&B^r=\frac{A_{\phi,\theta}}{\sqrt{-g}},\\
	E^\theta&=\br{2aMr+\Pi\frac{A_{t,\theta}}{A_{\phi,\theta}}}\!\frac{A_{\phi,\theta}}{\Delta\Sigma^2},
    &&B^\theta=-\frac{A_{\phi,r}}{\sqrt{-g}},\\
	E^\phi&=0,\quad
	B^\phi=\frac{A_{\theta,r}-A_{r,\theta}}{\sqrt{-g}},
    &&A_{\mu,\nu}\equiv\pd_\nu A_\mu.
\end{align}
\end{subequations}
For future reference, we also record the inverse relations%
\begin{subequations}
\label{eq:PotentialsAndFields}
\begin{align}
    A_{t,r}&=\frac{\Sigma}{\Pi}\pa{\Sigma\,E^r+2aMr\sin{\theta}\,B^\theta},\\
    A_{t,\theta}&=\frac{\Sigma}{\Pi}\pa{\Delta\Sigma\,E^\theta-2aMr\sin{\theta}\,B^r},\\
    A_{\phi,r}&=-\Sigma\sin{\theta}\,B^\theta,\quad
    A_{\phi,\theta}=\Sigma\sin{\theta}\,B^r.
\end{align}
\end{subequations}

The field $F=\ed A$ is not automatically degenerate, since \eqref{eq:DegenerateVolume} does not generically vanish:
\begin{align}
    \label{eq:DegeneracyCondition}
    \frac{1}{4}F_{\mu\nu}(\star F)^{\mu\nu}=\frac{A_{t,r}A_{\phi,\theta}-A_{t,\theta}A_{\phi,r}}{\Sigma\sin{\theta}}.
\end{align}

In the Kerr-Schild coordinates $\bar{x}^\mu$, a stationary and axisymmetric gauge potential $A=A_\mu(r,\theta)\ed x^\mu$ becomes $A=\bar{A}_\mu(r,\theta)\ed\bar{x}^\mu$ with components
\begin{subequations}
\label{eq:PotentialTransformation}
\begin{gather}
	\bar{A}_r=A_r-\frac{2Mr}{\Delta}A_t-\frac{a}{\Delta}A_\phi,\quad
	\bar{A}_\theta=A_\theta,\\
	\bar{A}_{\bar{t}}=A_t,\quad
	\bar{A}_{\bar{\phi}}=A_\phi.
\end{gather}
\end{subequations}
In terms of the associated electromagnetic field strength $F=\ed A=\bar{F}_{\mu\nu}\ed\bar{x}^\mu\ed\bar{x}^\nu$ with Kerr-Schild components $\bar{F}_{\mu\nu}=\bar{\nabla}_\mu\bar{A}_\nu-\bar{\nabla}_\nu\bar{A}_\mu$, the local electric and magnetic fields \eqref{eq:LocalFields} in the Kerr-Schild lab frame of $\bar{\zeta}=-\ed\bar{t}$ are
\begin{align}
    \label{eq:KerrSchildFields}
	\bar{E}^i=\bar{F}^{0i},\quad
	\bar{B}^i=(\star\bar{F})^{i0}.
\end{align}
Using \eqref{eq:PotentialTransformation}, these are related to the fields \eqref{eq:BoyerLindquistFrameFields} in the Boyer-Lindquist lab frame by
\begin{subequations}
\label{eq:BLtoKS}
\begin{align}
	\bar{E}^r&=E^r,\quad
	\bar{B}^r=B^r,\quad
	\bar{B}^\theta=B^\theta,\\
	\bar{E}^{\bar{\phi}}&=E^\phi+\frac{a\pa{2Mr+\Sigma}}{\Pi}E^r-\frac{2Mr}{\sin{\theta}}\frac{\Sigma}{\Pi}B^\theta,\\
	\bar{E}^\theta&=E^\theta+\frac{2Mr\sin{\theta}}{\Sigma}B^\phi,\\
	\bar{B}^{\bar{\phi}}&=B^\phi+\frac{a+2Mr\Omega}{\Delta}B^r,
\end{align}
\end{subequations}
where we introduced a convenient quantity
\begin{align}
    \label{eq:FirstDefinition}
    \Omega\equiv-\frac{A_{t,\theta}}{A_{\phi,\theta}}.
\end{align}
The inverse transformation is\footnote{This agrees with (50)--(52) of \cite{McKinney2004} given that their $\omega$ (defined in (26) therein) is our $\Omega$ in \eqref{eq:FirstDefinition}.}
\begin{subequations}
\label{eq:KStoBL}
\begin{align}
	E^r&=\bar{E}^r,\quad
	B^r=\bar{B}^r,\quad
	B^\theta=\bar{B}^\theta,\\
	E^\theta&=\bar{E}^\theta+\frac{2Mr\sin{\theta}}{\Sigma}\br{\frac{a-2Mr\Omega}{\Delta}\bar{B}^r-\bar{B}^{\bar{\phi}}}\!,\\
	E^\phi&=\bar{E}^{\bar{\phi}}-\frac{a\pa{2Mr+\Sigma}}{\Pi}\bar{E}^r+\frac{2Mr}{\sin{\theta}}\frac{\Sigma}{\Pi}\bar{B}^\theta,\\
	B^\phi&=\bar{B}^{\bar{\phi}}-\frac{a-2Mr\Omega}{\Delta}\bar{B}^r.
\end{align}
\end{subequations}

\subsection{Electromagnetic fluxes and (conserved) currents}
\label{app:Efluxes}

Given a Killing vector field $\xi$, its associated \textit{Noether current} in the field configuration $F$ is defined in terms of the associated electromagnetic stress-tensor \eqref{eq:StressTensor} as
\begin{align}
    \mathcal{J}_\xi^\mu\equiv-T_{\rm EM}^{\mu\nu}\xi_\nu.
\end{align}
By construction, such a current obeys\footnote{Since $T_{\rm EM}^{\mu\nu}$ is symmetric, the term $T_{\rm EM}^{\mu\nu}\nabla_\mu\xi_\nu=2T_{\rm EM}^{\mu\nu}\nabla_{(\mu}\xi_{\nu)}$ missing in the first equation vanishes by the Killing equation.\footref{fn:Killing}}
\begin{align}
    \label{eq:Conservation}
    \nabla_\mu\mathcal{J}_\xi^\mu=-\pa{\nabla_\mu T_{\rm EM}^{\mu\nu}}\xi_\nu
    =-f_\nu\xi^\nu,
\end{align}
where $f$ is the Lorentz force density \eqref{eq:LorentzForce}.
Therefore, the Noether current $\mathcal{J}_\xi$ associated to the isometry along $\xi$ is conserved if and only if $f\cdot\xi=0$.
This is always the case in FFE, where the Lorentz force $f_\nu=J^\mu F_{\mu\nu}$ vanishes identically by the force-free condition \eqref{eq:ForceFree}.
It can also happen whenever dissipation vanishes along the symmetry direction (that is, when $F_{\mu\nu}J^\mu\xi^\nu=0$).

The Kerr geometry \eqref{eq:Kerr} admits two Killing vector fields \eqref{eq:Killing}, with associated Noether currents\footnote{The signs here are conventional and match the ADM formulas.}
\begin{align}
    \label{eq:Currents}
    \mathcal{J}_\mathcal{E}^\mu\equiv-T^\mu_\nu K^\nu
    =-T^\mu_t,\quad
    \mathcal{J}_\mathcal{L}^\mu\equiv T^\mu_\nu R^\nu
    =T^\mu_\phi,
\end{align}
corresponding to the flows of electromagnetic energy and angular momentum, respectively.
Here and henceforth, we suppress the subscript `EM' on $T$ and need no longer track the ordering of its indices, as they are symmetric.

The electromagnetic energy and angular momentum of a field $F$ on Kerr are conserved if and only if the currents \eqref{eq:Currents} are conserved, which by \eqref{eq:Conservation} happens if and only if $f_t=f_\phi=0$.
Otherwise, the field exerts a force (or a force is exerted on it) and it loses (or gains) energy (if $f_t\neq0$) and/or angular momentum (if $f_\phi\neq0$).

Regardless of whether the currents $\mathcal{J}_{\mathcal{E},\mathcal{L}}$ are conserved, they always measure the flow of electromagnetic energy and angular momentum.
For fields that are stationary and axisymmetric, these flows only exhibit nontrivial behavior in the poloidal directions $r$ and $\theta$.
To see this, let $\mathcal{P}$ denote a curve in the poloidal plane $(r,\theta)$, and let $\mathcal{P}\times S^1$ denote the 2-surface that is obtained by revolving $\mathcal{P}$ around the spin axis (that is, in the direction $R=\pd_\phi$).

\noindent Finally, let $\mathcal{S}=\mathcal{P}\times S^1\times\Delta t$ be the extension of this surface by a time interval $\Delta t$ along $K=\pd_t$.
The energy and angular momentum fluxes through $\mathcal{P}\times S^1$ are then
\begin{align}
    \dot{\mathcal{E}}\equiv\frac{d}{dt}\int_\mathcal{S}\star\mathcal{J}_\mathcal{E},\quad
    \dot{\mathcal{L}}\equiv\frac{d}{dt}\int_\mathcal{S}\star\mathcal{J}_\mathcal{L},
\end{align}
where the integral over the 3-surface $\mathcal{S}$ is performed over the 3-currents that are Hodge-dual to \eqref{eq:Currents},\footnote{By definition,\footref{fn:HodgeDual} the Hodge dual of a $p$-form in $\eta$ in $n$ dimensions has components $(\star\eta)_{\mu_1\cdots\mu_{n-p}}=\frac{1}{p!}\epsilon_{\mu_1\cdots\mu_n}\eta^{\mu_{n-p+1}\cdots\mu_n}$.}
\begin{align}
    \label{eq:HodgeCurrent}
    (\star\mathcal{J}_{\mathcal{E},\mathcal{L}})_{\mu\nu\rho}=\epsilon_{\mu\nu\rho\kappa}\mathcal{J}_{\mathcal{E},\mathcal{L}}^\kappa.
\end{align}
By the same manipulations as in \eqref{eq:VolumeForm}, on $\mathcal{S}$ we have\footnote{$(\star\mathcal{J}_i)=(\star\mathcal{J}_i)_{\mu\nu\rho}\ed x^\mu\ed x^\nu\ed x^\rho=\frac{1}{3!}(\star\mathcal{J}_i)_{\mu\nu\rho}\ed x^\mu\wedge\ed x^\nu\wedge\ed x^\rho$.
When integrating over $\ed t$ and $\ed\phi$, $(\star\mathcal{J}_i)=(\star\mathcal{J}_i)_{t\phi\rho}\ed t\wedge\ed\phi\wedge\ed x^\rho$ which is $(\star\mathcal{J}_i)_{t\phi r}\ed t\wedge\ed\phi\wedge\ed r+(\star\mathcal{J}_i)_{t\phi\theta}\ed t\wedge\ed\phi\wedge\ed\theta$.
By \eqref{eq:HodgeCurrent}, this is $\epsilon_{t\phi r\theta}\mathcal{J}_i^\theta\ed t\wedge\ed\phi\wedge\ed r+\epsilon_{t\phi\theta r}\mathcal{J}_i^r\ed t\wedge\ed\phi\wedge\ed\theta$, where by \eqref{eq:VolumeForm}, $\epsilon_{t\phi r\theta}=-\epsilon_{t\phi\theta r}=\sqrt{-g}$.
Positive orientation has $\ed\phi$ to the right.}
\begin{align}
    (\star\mathcal{J}_{\mathcal{E},\mathcal{L}})\big|_\mathcal{S}=\sqrt{-g}\ed t\wedge\pa{\mathcal{J}_{\mathcal{E},\mathcal{L}}^r\ed\theta-\mathcal{J}_{\mathcal{E},\mathcal{L}}^\theta\ed r}\wedge\ed\phi.
\end{align}
As such, $\frac{d}{dt}\int_\mathcal{S}\star\mathcal{J}_{\mathcal{E},\mathcal{L}}=\int_{\mathcal{P}\times S^1}\pa{\mathcal{J}_{\mathcal{E},\mathcal{L}}^r\ed\theta-\mathcal{J}_{\mathcal{E},\mathcal{L}}^\theta\ed r}\wedge\ed\phi$.
The $\phi$ integral is trivial by axisymmetry, so the \textit{outward} fluxes of energy and angular momentum through $\mathcal{P}$ are
\begin{subequations}
\label{eq:PoloidalFluxes}
\begin{align}
    \dot{\mathcal{E}}&=2\pi\int_\mathcal{P}\sqrt{-g}\pa{\mathcal{J}_\mathcal{E}^r\ed\theta-\mathcal{J}_\mathcal{E}^\theta\ed r},\\
    \dot{\mathcal{L}}&=2\pi\int_\mathcal{P}\sqrt{-g}\pa{\mathcal{J}_\mathcal{L}^r\ed\theta-\mathcal{J}_\mathcal{L}^\theta\ed r}.
\end{align}
\end{subequations}
An explicit computation---using \eqref{eq:PotentialsAndFields} to eliminate the gauge potential in favor of the electric and magnetic field components---results in the flux densities
\begin{subequations}
\label{eq:CurrentComponents}
\begin{align}
    \mathcal{J}_\mathcal{E}^r&=-T_t^r
    =\frac{\Delta}{\Sigma^2}\pa{A_{\theta,r}-A_{r,\theta}}A_{t,\theta}\\
    &=\frac{\Delta\sin{\theta}}{\Pi}\pa{\Delta\Sigma\,E^\theta-2aMr\sin{\theta}\,B^r}B^\phi,\\
    \mathcal{J}_\mathcal{E}^\theta&=-T_t^\theta
    =-\frac{\Delta}{\Sigma^2}\pa{A_{\theta,r}-A_{r,\theta}}A_{t,r}\\
    &=-\frac{\Delta\sin{\theta}}{\Pi}\pa{\Sigma\,E^r+2aMr\sin{\theta}\,B^\theta}B^\phi,\\
    \mathcal{J}_\mathcal{L}^r&=T_\phi^r
    =-\frac{\Delta}{\Sigma^2}\pa{A_{\theta,r}-A_{r,\theta}}A_{\phi,\theta}\\
    &=-\Delta\sin^2{\theta}\,B^rB^\phi,\\
    \mathcal{J}_\mathcal{L}^\theta&=T_\phi^\theta
    =\frac{\Delta}{\Sigma^2}\pa{A_{\theta,r}-A_{r,\theta}}A_{\phi,r}\\
    &=-\Delta\sin^2{\theta}\,B^\theta B^\phi.
\end{align}
\end{subequations}
We emphasize that we have not yet assumed that $F$ is degenerate (we will do so only in \autoref{app:KerrDegenerateFields} below).

\subsection{Horizon fluxes}

We will be particularly interested in the radial fluxes through spheres of constant radius $r$, which correspond to circular poloidal curves $\mathcal{P}$.
In such cases, it follows from \eqref{eq:PoloidalFluxes} that the (outward) radial fluxes are\footnote{Our $\dot{\mathcal{E}}_r$ agrees with (32) of \cite{McKinney2004}.}
\begin{align}
    \label{eq:RadialFluxes}
    \frac{\dot{\mathcal{E}}_r}{2\pi}=\int_0^\pi\sqrt{-g}\mathcal{J}_\mathcal{E}^r\ed\theta,\quad
    \frac{\dot{\mathcal{L}}_r}{2\pi}=\int_0^\pi\sqrt{-g}\mathcal{J}_\mathcal{L}^r\ed\theta,
\end{align}
with the radial flux densities $\mathcal{J}_\mathcal{E}^r$, $\mathcal{J}_\mathcal{E}^\theta$ given in \eqref{eq:CurrentComponents}.
Naively, $\mathcal{J}_\mathcal{E}^r$ and $\mathcal{J}_\mathcal{E}^\theta$ both appear to vanish identically at the event horizon \eqref{eq:HorizonRadius}, where $\Delta=0$.
However, since the Boyer-Lindquist coordinates are singular across the horizon, the electric and magnetic field components \eqref{eq:BoyerLindquistFrameFields} can diverge at $r=r_+$ even if $F$ is in fact regular there (so that the horizon fluxes are actually finite).

This issue is remedied by working with the Kerr-Schild lab-frame electric and magnetic fields $\bar{E}^i$ and $\bar{B}^i$ defined in \eqref{eq:KerrSchildFields}, which are always regular across the horizon and are related to the singular Boyer-Lindquist fields \eqref{eq:BoyerLindquistFrameFields} by \eqref{eq:BLtoKS}--\eqref{eq:KStoBL}.
Using \eqref{eq:KStoBL}, the horizon radial flux densities of energy and angular momentum take the manifestly regular Kerr-Schild form\footnote{This agrees with (34) of \cite{McKinney2004}.}
\begin{subequations}
\label{eq:HorizonFluxes}
\begin{align}
    \mathcal{J}_\mathcal{E}^r\big|_{r=r_+}&=\Omega_+\mathcal{J}_\mathcal{L}^r\big|_{r=r_+},\\
    \mathcal{J}_\mathcal{L}^r\big|_{r=r_+}
    &=2Mr_+\pa{\Omega_{\rm H}-\Omega_+}\!\pa{\bar{B}^r\sin{\theta}}^2.
\end{align}
\end{subequations}
where $\Omega_{\rm H}$ is the angular velocity \eqref{eq:HorizonVelocity} of the horizon and $\Omega_+$ is the quantity \eqref{eq:FirstDefinition} evaluated at $r=r_+$.

\section{Stationary, axisymmetric, and degenerate fields on Kerr}
\label{app:KerrDegenerateFields} 

\subsection{General form of the field strength}

A Kerr electromagnetic field $F$ is necessarily closed.
If $F$ is also degenerate, then as discussed above \eqref{eq:EulerPotentials}, there exist (infinitely many) pairs of Euler potentials $(\lambda_1,\lambda_2)$ such that $F=\ed\lambda_1\wedge\ed\lambda_2$ (recall \autoref{app:ClosedDegenerateForms}).
Moreover, if $F$ is also stationary and axisymmetric, then as shown in (61) of \cite{Gralla2014}, one may always choose a pair of Euler potentials of the form\footnote{The product rule $\mathcal{L}_\xi\pa{A\wedge B}=\pa{\mathcal{L}_\xi A}\wedge B+A\wedge\pa{\mathcal{L}_\xi B}$ and Cartan's formula imply that $\mathcal{L}_\xi F=\ed\mathcal{L}_\xi\lambda_1\wedge\ed\lambda_2+\ed\lambda_1\wedge\ed\mathcal{L}_\xi\lambda_2$.
Since $\mathcal{L}_R\lambda_1=0$ and $\ed\mathcal{L}_R\lambda_2=\ed\mathcal{L}_R\phi=\ed(1)=0$, $\mathcal{L}_RF=0$ so $F$ is automatically axisymmetric.
Likewise, $\mathcal{L}_K\lambda_1=0$ while $\ed\mathcal{L}_K\lambda_2=-\ed\Omega(\psi)$, so $\mathcal{L}_KF=-\ed\psi\wedge\ed\Omega(\psi)=0$ and hence, $F$ is also stationary (note that this conclusion crucially depends on $\Omega$ being a function of $\psi$).
Closure and degeneracy are manifest.}
\begin{align}
    \label{eq:KerrPotentials}
    \lambda_1=\psi(r,\theta),\quad
    \lambda_2=\psi_2(r,\theta)+\phi-\Omega(\psi)t.
\end{align}

Technically, the derivation of these potentials assumes that $\pd_\phi\cdot F\neq0$, excluding all-toroidal magnetic fields.\footnote{By \eqref{eq:BoyerLindquistFrameFields}, $\pd_\phi^\mu F_{\mu\nu}\ed x^\nu=-\pd_rA_\phi\ed r-\pd_\theta A_\phi\ed\theta=-\ed A_\phi(r,\theta)$ vanishes if and only if $B^r=B^\theta=0$, leaving only $B^\phi$ nonzero.}

If $\pd_\phi\cdot F=0$, the Euler potentials can be taken to be of the form (D15) in \cite{Gralla2014}, namely $\lambda_1=\chi(r,\theta)$ and $\lambda_2=\chi_2(r,\theta)+t$.
Since these can be obtained as a limit of \eqref{eq:KerrPotentials},\footnote{Take $\psi\to0$ and $\Omega(\psi)\to\infty$ with $\psi\Omega(\psi)\to-\chi$ kept finite and $\ed\psi\wedge\ed\psi_2\to\ed\chi\wedge\ed\chi_2$, as in (103) of \cite{Gralla2014}.}
it follows that the potentials \eqref{eq:KerrPotentials} are in fact still general.

Next, we define a function $I(r,\theta)$ as in (64) of \cite{Gralla2014}:
\begin{align}
    \frac{I}{2\pi}\ed t\wedge\ed\phi=\star\pa{\ed\psi\wedge\ed\psi_2}.
\end{align}
Explicitly, this function takes the form
\begin{align}
    \label{eq:CurrentDefinition}
    I=\frac{2\pi\Delta\sin{\theta}}{\Sigma}\pa{\pd_r\psi\pd_\theta\psi_2-\pd_\theta\psi\pd_r\psi_2}.
\end{align}
Eliminating $\psi_2$ in favor of $I$, we may then express $F$ as
\begin{align}
    \label{eq:GeneralForm}
    F=\frac{\Sigma I}{2\pi\Delta\sin{\theta}}\ed r\wedge\ed\theta+\ed\psi\wedge\br{\ed\phi-\Omega(\psi)\ed t}.
\end{align}
We conclude that an electromagnetic field $F$ on Kerr that is stationary, axisymmetric, and degenerate must be of the general form \eqref{eq:GeneralForm}.
In particular, such a field can be fully specified by three poloidal functions $\psi(r,\theta)$, $I(r,\theta)$, and $\Omega(\psi)$.
These functions all admit intuitive physical interpretations, which we will describe in the next sections following \S7 of \cite{Gralla2014}.

To do so, we first note that $F=\ed\psi\wedge\ed\lambda_2=\ed\pa{\psi\ed\lambda_2}$, so we may take the associated gauge potential to be
\begin{align}
    \label{eq:GaugePotential}
    A=\psi\ed\lambda_2.
\end{align}
Then a direct comparison of \eqref{eq:GeneralForm} against $F=\ed A$ with components $A=A_\mu(r,\theta)\ed x^\mu$ quickly shows that
\begin{subequations}
\label{eq:Identifications}
\begin{align}
    \label{eq:MagneticFluxFunction}
    \psi(r,\theta)&=A_\phi(r,\theta),\\
    \label{eq:Isorotation}
    \Omega(\psi)&=-\frac{A_{t,\theta}}{A_{\phi,\theta}}
    =-\frac{A_{t,r}}{A_{\phi,r}},\\
    I(r,\theta)&=\frac{2\pi\Delta\sin{\theta}}{\Sigma}\pa{A_{\theta,r}-A_{r,\theta}}.
\end{align}
\end{subequations}
The MHD version of this decomposition, as well as the energy fluxes and boundary conditions discussed in the next sections, were analyzed in detail by \cite{Phinney1983}.

\subsection{Poloidal field lines and magnetic flux function \texorpdfstring{$\psi$}{}}

First, recall that the 2-surfaces of constant $\lambda_1$ and $\lambda_2$ define field sheets, and that these sheets are timelike provided that $F=\ed\lambda_1\wedge\ed\lambda_2$ is magnetically dominated (\autoref{app:ClosedDegenerateForms}).
In that case, these field sheets may be viewed as the world sheets of \textit{magnetic field lines}, which are defined with respect to a given time coordinate $t$ as the intersection of a field sheet with a hypersurface of constant $t$ (a spatial slice).

For fields that are also stationary and axisymmetric, we can further define \textit{poloidal field lines} as the level sets of $\lambda_1=\psi(r,\theta)$.
These are projections in the poloidal plane $(r,\theta)$ of the magnetic field lines, whose bending in the azimuthal direction is controlled by $\psi_2(r,\theta)$.
Thus, $\psi$ may be viewed as a coordinate on poloidal field lines.

At this stage, we note that the spin axis of the black hole must always be a poloidal field line---that is, $\psi(r,0)$ and $\psi(r,\pi)$ must be constant---in order to ensure that the gauge potential \eqref{eq:GaugePotential} stays regular on the spin axis, where $\ed\lambda_2$ diverges (because $\ed\phi$ is singular at the poles).
It is clear from \eqref{eq:MagneticFluxFunction} that by a suitable gauge transformation, we can arrange for $\psi$ to vanish on axis,
\begin{align}
    \label{eq:MagneticFluxCondition}
    \psi(r,0)=\psi(r,\pi)
    =0,
\end{align}
a choice that we will always assume from now on.

Another physical interpretation of $\psi$ is as a \textit{magnetic flux function}.
Consider a loop of revolution obtained by rotating a poloidal point $(r,\theta)$ in the azimuthal direction $\phi$ at a fixed time $t$.
The magnetic flux through this loop is the integral of $F=\ed A$ over any 2-surface $\mathcal{S}$ enclosed by this loop $\pd\mathcal{S}$.
By \eqref{eq:GaugePotential} and Stokes' theorem, this is
\begin{align}
    \int_\mathcal{S}F=\int_\mathcal{S}\ed\pa{\psi\ed\lambda_2}
    =\psi\int_{\pd\mathcal{S}}\ed\lambda_2
    =2\pi\psi.
\end{align}
(Note that $\ed\lambda_2=\ed\phi$ on $\pd\mathcal{S}$.)
Hence, $2\pi\psi(r,\theta)$ measures the magnetic flux through the loop of revolution at $(r,\theta)$.
We note that mathematically, this argument relies on the regularity of $A=\psi\ed\lambda_2$ everywhere on $S$, including at the poles, which requires the imposition of \eqref{eq:MagneticFluxCondition}.
Physically, this condition ensures that the magnetic flux vanishes as the loop $\pd S$ shrinks to zero, whereupon it only encloses the spin axis.

\subsection{Field-line angular velocity \texorpdfstring{$\Omega(\psi)$}{}}

Since the poloidal function $\Omega(\psi)$ is a function of $\psi(r,\theta)$ only, it may be viewed as a function on poloidal field lines.
By \eqref{eq:KerrPotentials}, it may be interpreted as the angular velocity of field lines as they rotate about the spin axis.

The fact that every field line of fixed $\psi$ rotates at the same rate is encapsulated in \eqref{eq:Isorotation}, which is known as \textit{Ferraro's law of isorotation}.
The first equality in \eqref{eq:Isorotation} is consistent with our earlier definition \eqref{eq:FirstDefinition}, which is now extended by the second equality, which is required to ensure that \eqref{eq:DegeneracyCondition} vanishes and hence that $F$ is degenerate.

The surfaces where the norm of $\ed\phi-\Omega(\psi)\ed t$ vanishes are known as \textit{light surfaces}.
They correspond to critical surfaces where the field lines rotate at the speed of light.

\subsection{Field-line (conserved) current \texorpdfstring{$I(\psi)$}{}}

The quantity $I(r,\theta)$ is known as the \textit{poloidal current} for two reasons.
First, it completely controls the poloidal components of the Maxwell current $J^\mu=\nabla_\nu F^{\mu\nu}$:
\begin{align}
    J^r=-\frac{\pd_\theta I}{2\pi\Sigma\sin{\theta}},\quad
    J^\theta=\frac{\pd_rI}{2\pi\Sigma\sin{\theta}}.
\end{align}
Second, if the Noether currents \eqref{eq:Currents} associated with flows of electromagnetic energy and angular momentum are conserved, then $f_t=f_\phi=0$.
A direct computation reveals that this is equivalent to $\ed\psi\wedge\ed I=0$, that is,
\begin{align}
    \label{eq:CurrentConservation}
    I(r,\theta)=I(\psi(r,\theta)).
\end{align}
In other words, $I(\psi)$ is the conserved current flowing along the poloidal field line labeled by $\psi$.

\subsection{Electromagnetic fluxes}

For the field \eqref{eq:GeneralForm}, the flux densities \eqref{eq:CurrentComponents} are just%
\begin{subequations}
\begin{align}
    \mathcal{J}_\mathcal{E}^r&=-\frac{\Omega(\psi)I\pd_\theta\psi}{2\pi\Sigma\sin{\theta}},
    &&\mathcal{J}_\mathcal{E}^\theta=\frac{\Omega(\psi)I\pd_r\psi}{2\pi\Sigma\sin{\theta}},\\
    \mathcal{J}_\mathcal{L}^r&=-\frac{I\pd_\theta\psi}{2\pi\Sigma\sin{\theta}},
    &&\mathcal{J}_\mathcal{L}^\theta=\frac{I\pd_r\psi}{2\pi\Sigma\sin{\theta}},
\end{align}
\end{subequations}
as can be obtained either via direct computation or, as a useful consistency check, by plugging in \eqref{eq:FieldsToPoloidal} below.

As a result (recall from \eqref{eq:MetricDeterminant} that $\sqrt{-g}=\Sigma\sin{\theta}$), the energy and angular momentum fluxes \eqref{eq:PoloidalFluxes} become\footnote{This agrees with (80)--(81) of \cite{Gralla2014}.}
\begin{align}
    \label{eq:SimpleFluxes}
    \dot{\mathcal{E}}=-\int_\mathcal{P}\Omega(\psi)I\ed\psi,\quad
    \dot{\mathcal{L}}=-\int_\mathcal{P}I\ed\psi,
\end{align}
These fluxes vanish when integrated along a contour of fixed $\psi$, which makes it clear that the electromagnetic energy and angular momentum both flow along poloidal field lines.
Moreover, when the Noether currents \eqref{eq:Currents} are conserved, so that \eqref{eq:CurrentConservation} holds, $\Omega$ and $I$ are both functions of $\psi$ only and so these fluxes become ordinary 1-dimensional integrals, which makes it clear that energy and angular momentum are conserved along field lines.

\subsection{Znajek condition for regularity at the horizon}
\label{app:znajek}

Under the coordinate transformation to Kerr-Schild coordinates \eqref{eq:CoordinateTransformation}, the general stationary, axisymmetric, and degenerate electromagnetic field \eqref{eq:GeneralForm} becomes
\begin{subequations}
\begin{align}
    F&=\ed\psi\wedge\!\br{\ed\bar{\phi}-\Omega(\psi)\ed\bar{t}}\!+\frac{\Sigma Z}{2\pi\Delta\sin{\theta}}\ed r\wedge\ed\theta,\\
    Z&=\br{I-2\pi\pa{\Omega(\psi)-\frac{a}{2Mr}}\pd_\theta\psi\frac{2Mr\sin{\theta}}{\Sigma}}\!.
\end{align}
\end{subequations}
Since these coordinates are regular, the field $F$ is regular at the (future) horizon (where $\Delta\to0$) if and only if $Z$ vanishes at $r=r_+$.
Thus, the Kerr field \eqref{eq:GeneralForm} is regular on the (future) horizon if and only if it obeys the \textit{Znajek condition} \citep{Znajek1977} on the horizon,\footnote{This assumes that the black hole is subextremal: $\ab{a}<M$.
At extremality ($\ab{a}=M$), regularity imposes a second Znajek condition at the horizon: (120) of \cite{Gralla2014}.}
\begin{align}
    \label{eq:Znajek}
    I=2\pi\pa{\Omega-\Omega_{\rm H}}\pd_\theta\psi\frac{\sqrt{\Pi}\sin{\theta}}{\Sigma}.
\end{align}
We emphasize that every quantity in this expression is to be evaluated at $r=r_+$, where $\sqrt{\Pi}=r_+^2+a^2=2Mr_+$.

Using the Znajek condition to evaluate the fluxes \eqref{eq:SimpleFluxes} on the horizon results in
(126)--(127) of \cite{Gralla2014}.
Explicitly, we obtain the expressions
\begin{subequations}
\begin{align}
    \dot{\mathcal{E}}&=2\pi\int_0^\pi\Omega_+\pa{\Omega_{\rm H}-\Omega_+}\!\pa{\pd_\theta\psi}^2\frac{\sin{\theta}\sqrt{\Pi}}{\Sigma}\ed\theta,\\
    \dot{\mathcal{L}}&=2\pi\int_0^\pi\pa{\Omega_{\rm H}-\Omega_+}\!\pa{\pd_\theta\psi}^2\frac{\sin{\theta}\sqrt{\Pi}}{\Sigma}\ed\theta,
\end{align}
\end{subequations}
which are consistent with \eqref{eq:RadialFluxes}--\eqref{eq:HorizonFluxes}.
There, we had to use Kerr-Schild coordinates to obtain regular horizon flux densities.
Here, we directly derived regular expressions by working with the variables \eqref{eq:Identifications}, but to do so we had to apply the Znajek condition \eqref{eq:Znajek}, whose derivation also involved Kerr-Schild coordinates.

\subsection{Field-line angular velocity from Kerr symmetry}

Besides the Killing vectors \eqref{eq:Killing}, the Kerr metric \eqref{eq:Kerr} also possesses an antisymmetric rank-2 tensor $Y$ that obeys the Killing-Yano equation $\nabla_{(\lambda}Y_{\mu)\nu}=0$:
\begin{align}
	Y&=a\cos{\theta}\ed r\wedge\pa{\ed t-a\sin^2{\theta}\ed\phi}\\
    &\phantom{=}-r\sin{\theta}\ed\theta\wedge\br{a\ed t-\pa{r^2+a^2}\ed\phi}.
\end{align}
The existence of this Killing 2-form underlies many of the special properties of the Kerr geometry, including: the separability of its wave equation, the integrability of its geodesics, and the existence of the Penrose-Walker constant $\kappa$, which reduces to simple algebra the problem of parallel transport of polarization along null rays.

Here, we show how to use this higher-rank symmetry of Kerr to project out the field-line angular velocity $\Omega(\psi)$ from the full 2-form \eqref{eq:GeneralForm}.
To do so, we first define
\begin{align}
    \tilde{K}_\mu\equiv Y_{\mu\nu}K^\nu
    =Y_{\mu t},\quad
    \tilde{R}_\mu\equiv Y_{\mu\nu}R^\nu
    =Y_{\mu\phi}.
\end{align}
Raising indices, we obtain the explicit vector fields
\begin{subequations}
\begin{align}
    \tilde{K}&=\frac{a}{\Sigma}\br{\Delta\cos{\theta}\pd_r-r\sin{\theta}\pd_\theta},\\
    \tilde{R}&=\frac{\sin{\theta}}{\Sigma}\br{-\frac{a^2}{2}\Delta\sin{2\theta}\pd_r+r\pa{r^2+a^2}\pd_\theta}\!.
\end{align}
\end{subequations}
Then a direct calculation reveals that
\begin{align}
    \label{eq:SymmetryProjection}
	\Omega(\psi)=-\frac{F_{\mu\nu}\tilde{K}^\mu K^\nu}{F_{\mu\nu}\tilde{K}^\mu R^\nu}
	=-\frac{F_{\mu\nu}\tilde{R}^\mu K^\nu}{F_{\mu\nu}\tilde{R}^\mu R^\nu},
\end{align}
which shows that $\Omega(\psi)$ can be extracted from $F$ using only the Killing symmetries of Kerr.
This also confirms that $\Omega(\psi)$ is an invariant scalar.

It is also possible to extract $\Omega(\psi)$ from $\star F$ using only the Killing symmetries.
Define four projections
\begin{align}
	\hat{K}_\mu&=(\star F)_{\mu\nu}K^\nu,
	&&\hat{R}_\mu=(\star F)_{\mu\nu}R^\nu,\\
	\mathring{K}_\mu&=(\star F)_{\mu\nu}\tilde{K}^\nu,
	&&\mathring{R}_\mu=(\star F)_{\mu\nu}\tilde{R}^\nu.
\end{align}
Then a direct computation shows that
\begin{align}
    \Omega(\psi)=-\frac{\hat{K}\cdot\mathring{R}}{\hat{R}\cdot\mathring{R}}
    =\frac{(\star F)_{\mu\nu}(\star F)^{\mu\tau}\pd_t^\nu J_{\tau\rho}\pd_\phi^\rho}{(\star F)_{\mu\nu}(\star F)^{\mu\tau}\pd_\phi^\nu J_{\tau\rho}\pd_\phi^\rho},
\end{align}
and likewise, that
\begin{align}
    \Omega(\psi)=-\frac{\hat{K}\cdot\mathring{K}}{\hat{R}\cdot\mathring{K}}
    =\frac{(\star F)_{\mu\nu}(\star F)^{\mu\tau}\pd_t^\nu J_{\tau\rho}\pd_t^\rho}{(\star F)_{\mu\nu}(\star F)^{\mu\tau}\pd_\phi^\nu J_{\tau\rho}\pd_t^\rho}.
\end{align}
To the best of our knowledge, these expressions for $\Omega(\psi)$ as scalar symmetry projections are new in the literature.

For completeness, we also note that the current $I$ can be similarly projected out, for instance as
\begin{align}
    I=(\star F)_{\mu\nu}K^\mu R^\nu.
\end{align}

\subsection{Boyer-Lindquist electric and magnetic fields}

Using \eqref{eq:MetricDeterminant} and \eqref{eq:Identifications}, we can re-express the electric and magnetic fields in the Boyer-Lindquist lab frame \eqref{eq:BoyerLindquistFrameFields} in terms of $(\psi,\Omega,I)$ as
\begin{subequations}
\label{eq:FieldsToPoloidal}
\begin{align}
    E^r&=\frac{\Pi\br{\omega-\Omega(\psi)}}{\Sigma^2}\pd_r\psi,
    &&B^r=\frac{\pd_\theta\psi}{\Sigma\sin{\theta}},\\
	E^\theta&=\frac{\Pi\br{\omega-\Omega(\psi)}}{\Delta\Sigma^2}\pd_\theta\psi,
    &&B^\theta=-\frac{\pd_r\psi}{\Sigma\sin{\theta}},\\
	E^\phi&=0,
	&&B^\phi=\frac{I}{2\pi\Delta\sin^2{\theta}},
\end{align}
\end{subequations}
with $\omega$ the angular velocity \eqref{eq:ZAMO} of the Boyer-Lindquist ZAMO.
We can also write the $E^i$ in terms of the $B^i$ as
\begin{subequations}
\begin{align}
    E^r&=\frac{\Pi\sin{\theta}}{\Sigma}\br{\Omega(\psi)-\omega}B^\theta,\\
    E^\theta&=-\frac{\Pi\sin{\theta}}{\Delta\Sigma}\br{\Omega(\psi)-\omega}B^r,\\
    E^\phi&=0.
\end{align}
\end{subequations}

If the field \eqref{eq:GeneralForm} is magnetically dominated, then we showed that its kernel contains a family $u_{(\gamma)}$ of timelike 4-velocities \eqref{eq:TimelikeVector}, which are conveniently parameterized by their Lorentz factor $\gamma=-\eta\cdot u_{(\gamma)}$ relative to the normal observer.
The one with minimal Lorentz boost \eqref{eq:MinimalLorentzBoost} is perpendicular to the lab-frame magnetic field,
\begin{align}
    \label{eq:DriftVelocity}
    u_\perp^\mu=\sqrt{\frac{B^2}{B^2-E^2}}\pa{\eta^\mu-\frac{\epsilon^{\mu\nu\alpha\beta}\eta_\nu E_\alpha B_\beta}{B^2}}\!,
\end{align}
while the other ones all have a nonvanishing field-parallel component.
Here, we chose to express \eqref{eq:TimelikeVector} in terms of the lab-frame fields \eqref{eq:LabFrameFields} rather than the normal fields \eqref{eq:NormalFields}.\footref{fn:NormalLabFrame}
Next, recall the decomposition \eqref{eq:VelocityDecomposition}:
\begin{align}
    u_\perp=\gamma_\perp\pa{\eta+\tilde{v}_\perp},\quad
    \gamma_\perp=\sqrt{\frac{B^2}{B^2-E^2}}.
\end{align}
We see that $u_\perp=u_\perp^\mu\pd_\mu$ has the explicit components\footref{fn:PerpendicularVelocity}
\begin{subequations}
\begin{align}
    u_\perp^t&=\gamma_\perp\eta^t
    =\frac{\gamma_\perp}{\alpha}
    =\sqrt{\frac{B^2}{B^2-E^2}\frac{\Pi}{\Delta\Sigma}},\\
    \frac{u_\perp^i}{\gamma_\perp}&=\tilde{v}_\perp^i
    =\frac{\alpha}{\sqrt{-g}B^2}\tilde{\epsilon}^{ijk}E_jB_k,
\end{align}
\end{subequations}
where the field-perpendicular velocity \eqref{eq:FieldPerpendicularVelocity} is explicitly%
\begin{subequations}
\begin{align}
    \tilde{v}_\perp^r&=-\frac{\Pi\sin^2{\theta}}{\alpha\Sigma}\br{\Omega(\psi)-\omega}\frac{B^rB^\phi}{B^2},\\
    \tilde{v}_\perp^\theta&=-\frac{\Pi\sin^2{\theta}}{\alpha\Sigma}\br{\Omega(\psi)-\omega}\frac{B^\theta B^\phi}{B^2},\\
    \tilde{v}_\perp^\phi&=\frac{\Omega(\psi)-\omega}{\alpha}\br{1-\frac{\Pi\sin^2{\theta}}{\Sigma}\frac{\pa{B^\phi}^2}{B^2}}\!,\\
    B^2&=\frac{\Sigma}{\Delta}\pa{B^r}^2+\Sigma\pa{B^\theta}^2+\frac{\Pi\sin^2{\theta}}{\Sigma}\pa{B^\phi}^2.
\end{align}
\end{subequations}
These components differ from those defined in \eqref{eq:AvoidConfusion},
\begin{subequations}
\label{eq:PerpendicularFrameComponents}
\begin{align}
    \frac{u_\perp^r}{u_\perp^t}&=\alpha \tilde{v}_\perp^r
    =-\frac{\Pi\sin^2{\theta}}{\Sigma}\br{\Omega(\psi)-\omega}\frac{B^rB^\phi}{B^2},\\
    \frac{u_\perp^\theta}{u_\perp^t}&=\alpha \tilde{v}_\perp^\theta
    =-\frac{\Pi\sin^2{\theta}}{\Sigma}\br{\Omega(\psi)-\omega}\frac{B^\theta B^\phi}{B^2},\\
    \frac{u_\perp^\phi}{u_\perp^t}&=-\beta^\phi+\alpha \tilde{v}_\perp^\phi
    =\omega+\alpha \tilde{v}_\perp^\phi\\
    &=\Omega(\psi)-\frac{\Pi\sin^2{\theta}}{\Sigma}\br{\Omega(\psi)-\omega}\frac{\pa{B^\phi}^2}{B^2},
\end{align}
\end{subequations}
with the latter being the angular velocity $\Omega_\perp$ of the flow.

The more general flows $u_{(\gamma)}=\gamma\pa{u_\perp/\gamma_\perp+\tilde{v}_\parallel}$, which have nonzero field-parallel components $\tilde{v}_\parallel\neq0$, are then
\begin{align}
    u_{(\gamma)}^\mu=\gamma\pa{\frac{u_\perp^\mu}{\gamma_\perp}\pm\sqrt{\frac{1}{\gamma_\perp^2}-\frac{1}{\gamma^2}}\frac{B^\mu}{\sqrt{B^2}}}\!,\quad
    \gamma\geq\gamma_\perp.
\end{align}
Each such flow defines an associated magnetic field $b_{(\gamma)}$ in the frame of $u_{(\gamma)}$, which is explicitly given by \eqref{eq:DriftFrameMagneticField},
\begin{align}
    b_{(\gamma)}^\mu=\frac{B^\mu+\pa{u_{(\gamma)}\cdot B}u_{(\gamma)}^\mu}{u_{(\gamma)}^t},\quad
    u_{(\gamma)}^t=\frac{\gamma}{\alpha}.
\end{align}
In particular, for the purely field-perpendicular flow,
\begin{align}
    b_\perp^\mu=\frac{B^\mu}{u_\perp^t}.
\end{align}
Meanwhile, the corresponding electric field vanishes in each of these frames, so that \eqref{eq:GRMHD} applies; in particular,
\begin{align}
    \label{eq:DecompositionGRMHD}
    (\star F)^{\mu\nu}=b_{(\gamma)}^\mu u_{(\gamma)}^\nu-u_{(\gamma)}^\mu b_{(\gamma)}^\nu
    =b_\perp^\mu u_\perp^\nu-u_\perp^\mu b_\perp^\nu.
\end{align}
Likewise, the stress-tensor $T_{\rm EM}^{\mu\nu}$ takes the form \eqref{eq:DriftStressTensor}, or equivalently \eqref{eq:PerpendicularDriftStressTensor}, from which the flux densities \eqref{eq:Currents} are obtained by plugging in the components \eqref{eq:FieldsToPoloidal}.

\subsection{Field-line angular velocity from accretion flow}

The dual of the stationary, axisymmetric, degenerate field \eqref{eq:GeneralForm} obeys $(\star F)^{it}=B^i$ (by definition), and
\begin{align}
    \label{eq:DualFieldComparison}
    (\star F)^{r\theta}=0,\quad
    \frac{(\star F)^{r\phi}}{B^r}=\frac{(\star F)^{\theta\phi}}{B^\theta}
    =\Omega(\psi).
\end{align}
Given a timelike frame $u=u_{(\gamma)}$ in which the electric field vanishes, \eqref{eq:DecompositionGRMHD} gives
\begin{align}
    \label{eq:HodgeDualDecomposition}
    (\star F)^{\mu\nu}=B^\mu\frac{u^\nu}{u^t}-B^\nu\frac{u^\mu}{u^t}.
\end{align}
If $u=u_\perp$, the components $u^\mu/u^t$ take the form \eqref{eq:PerpendicularFrameComponents}.
Comparing this with \eqref{eq:DualFieldComparison} immediately implies that
\begin{subequations}
\label{eq:FieldlineVelocity}
\begin{align}
    \frac{u^\theta}{u^r}&=\frac{B^\theta}{B^r},\\
    \Omega(\psi)&=\frac{u^\phi}{u^t}-\frac{u^r}{u^t}\frac{B^\phi}{B^r},\\
    \Omega(\psi)&=\frac{u^\phi}{u^t}-\frac{u^\theta}{u^t}\frac{B^\phi}{B^\theta}.
\end{align}
\end{subequations}
The first relation guarantees the compatibility of the last two, and it can be used to prove their equivalence to the symmetry projections in \eqref{eq:SymmetryProjection}.

\section{Stationary and axisymmetric force-free electrodynamics in Kerr}
\label{app:KerrSolutions}

As we showed in \autoref{app:FFE}, a force-free field is always degenerate.
Hence, stationary, axisymmetric, force-free fields on Kerr must necessarily be of the form \eqref{eq:GeneralForm}.

The force-free condition \eqref{eq:ForceFree} requires the Lorentz force density $f_\nu$ to vanish.
This imposes two constraints $f_t=f_\phi=0$ that are satisfied if and only if the currents \eqref{eq:Currents} are conserved (or equivalently, if \eqref{eq:CurrentConservation} holds).
Thus, a force-free field on Kerr is entirely specified by three poloidal functions: the magnetic flux $\psi(r,\theta)$, and the field-line angular velocity $\Omega(\psi)$ and current $I(\psi)$.
The two other constraints $f_r=f_\theta=0$ are equivalent to each other and to the (Grad-Shafranov) \textit{stream equation}, which we write as in (3.4) of \cite{Camilloni2022}:
\begin{align}
    \label{eq:Stream}
    &\xi_\mu\pd_r\pa{\xi^\mu\Delta\sin{\theta}\pd_r\psi}+\xi_\mu\pd_\theta\pa{\xi^\mu\sin{\theta}\pd_\theta\psi}\\
    &+\frac{\Sigma}{\Delta\sin{\theta}}\frac{I(\psi)I'(\psi)}{\pa{2\pi}^2}=0,\quad
    \xi=\ed\phi-\Omega(\psi)\ed t.\notag
\end{align}

\subsection{Exact analytic force-free fields in Kerr}

In the Kerr spacetime, only two exact stationary and axisymmetric force-free fields are known analytically.
The first, due to \cite{Menon2007}, allows for the magnetic flux to be any invertible function $f(\theta)$,
\begin{align}
    \psi_{\rm MD}(r,\theta)=f(\theta),
\end{align}
with inverse $\theta(\psi)=f^{(-1)}(\psi)$.
As such, the field lines are purely radial, and so it is possible to satisfy \eqref{eq:MagneticFluxCondition}.
The angular velocity of field lines is always the same regardless of the specific choice of $f$,
\begin{align}
    \Omega_{\rm MD}(\psi)=\frac{1}{a\sin^2{\theta(\psi)}},
\end{align}
while the field-line current is (up to a sign and constant)
\begin{align}
    I_{\rm MD}(\psi)=\pm2\pi\sqrt{I_0^2+\br{\frac{f'(\theta(\psi))}{a\sin{\theta(\psi)}}}^2}.
\end{align}
In order to avoid an unphysical line current along the spin axis, we must set $I_0=0$ and $f'(\theta)=0$ at the poles.
In that case, the field becomes null,
\begin{align}
    F^2=\frac{2I_0^2}{\Delta(r)\sin^2{\theta}}
    =0,
\end{align}
and the current is ingoing and given by
\begin{align}
    J^\mu=\frac{\pd_\theta\pa{\frac{\pd_\theta f(\theta)}{\sin{\theta}}}}{a\Sigma\sin{\theta}}\pa{\frac{r^2+a^2}{\Delta}\pd_t-\pd_r+\frac{a}{\Delta}\pd_\phi}\!.
\end{align}
The flux densities \eqref{eq:Currents} are $\mathcal{J}_\mathcal{E}^\theta=\mathcal{J}_\mathcal{L}^\theta=0$ and
\begin{align}
    \mathcal{J}_\mathcal{E}^r=-\frac{\br{f'(\theta)}^2}{a^2\Sigma\sin^4{\theta}},\quad
    \mathcal{J}_\mathcal{L}^r=\frac{\br{f'(\theta)}^2}{a\Sigma\sin^2{\theta}}.
\end{align}
Besides not being magnetically dominated, this solution also fails to describe energy extraction from the black hole, since $\mathcal{J}_\mathcal{E}^r<0$ for any choice of $f$.
There exists a non-stationary, nonaxisymmetric generalization of this solution with the same properties \citep{Brennan2013}.

The second exact solution is due to \cite{Menon2015}.
It is ``dual'' to the first, in the sense that it has the opposite (perpendicular) field geometry, with  field lines that are circular rather than radial.
The magnetic flux of this dual solution can be any invertible function $g(r)$,
\begin{align}
    \psi_{\rm M*}(r,\theta)=g(r),
\end{align}
with inverse $r(\psi)=g^{(-1)}(\psi)$.
The angular velocity of field lines is always the same regardless of $g$,
\begin{align}
    \Omega_{\rm M*}(\psi)=\frac{a}{r^2(\psi)+a^2},
\end{align}
while the field-line current is (up to a sign and constant)
\begin{align}
    I_{\rm M*}(\psi)=\pm2\pi\sqrt{I_0^2-\br{\frac{\Delta(r(\psi))f'(r(\psi))}{r^2(\psi)+a^2}}^2}.
\end{align}
This solution is not physically admissible as it cannot satisfy \eqref{eq:MagneticFluxCondition}, so we will not investigate it any further.

Since neither of these exact solutions is realistic, we must turn to perturbative force-free solutions to study electromagnetic energy extraction from Kerr black holes.

\subsection{Perturbing the force-free equations in small spin}

We follow \cite{BlandfordZnajek1977} and solve the force-free equations on Kerr perturbatively in slow spin.
That is, we expand the stream equation \eqref{eq:Stream} in small $\ab{a}\ll M$, as reviewed in \S4 of \cite{Gralla2016}.

At leading order, $a=0$ and so we are in Schwarzschild.
In that case, we ought to have no rotation and therefore no current.
Hence, to order $\mathcal{O}\!\pa{a^0}$, we expect $\Omega=I=0$,
and as such, $\psi(r,\theta)$ ought to correspond to a vacuum Maxwell field in a Schwarzschild background.

This motivates the perturbative expansion
\begin{subequations}
    \label{eq:PerturbativeAnsatz}
    \begin{align}
        \psi&=\psi_0(r,\theta)+\mathcal{O}\!\pa{a^2}\!,\\
        \Omega&=\frac{a}{M^2}\Omega_1(\psi_0)+\mathcal{O}\!\pa{a^3}\!,\\
        I&=\frac{a}{M^2}I_1(\psi_0)+\mathcal{O}\!\pa{a^3}\!,
    \end{align}
\end{subequations}
valid to subleading order $\mathcal{O}\!\pa{a}$ in spin.
Here, $\psi_0$, $\Omega_1$, and $I_1$ are all independent of the spin $a$, while the scaling of the error terms is fixed by the Znajek condition \eqref{eq:Znajek}.
The leading-order magnetic flux function $\psi_0$ must solve \eqref{eq:Stream} with $a=\Omega=I=0$, that is, the linear equation
\begin{align}
    \label{eq:MaxwellSchwarzschild}
    r^2\pd_r\br{\pa{1-\frac{2M}{r}}\pd_r\psi}+\sin{\theta}\pa{\frac{\pd_\theta\psi}{\sin{\theta}}}=0.
\end{align}
Its solutions are vacuum Maxwell fields in Schwarzschild.
They are reviewed in Appendix~B of \cite{Gralla2016}.

At subleading order $\mathcal{O}(a)$, the stream equation \eqref{eq:Stream} is identically satisfied by the perturbative ansatz \eqref{eq:PerturbativeAnsatz}: the only requirement imposed by FFE is that $\Omega_1$ and $I_1$ be functions of $\psi_0$ alone.
The nontrivial content of the full Kerr stream equation \eqref{eq:Stream} only kicks in at the higher order $\mathcal{O}\!\pa{a^2}$ in perturbation theory, which we do not review here as the details become quite complicated; see, e.g., \cite{Armas2020} for an in-depth treatment.

Thus, to solve the force-free equations to linear order in $a$, one needs a vacuum Schwarzschild flux $\psi_0$ together with two additional relations to determine $\Omega_1$ and $I_1$.\footnote{If one only wishes to solve the force-free equations to linear order in $a$, then in principle one can freely choose to impose any second relation between $\psi_0$, $\Omega_1$, and $I_1$.
However, almost all the resulting fields will be spurious linearized solutions that do not correspond to any family of exact solutions, and hence cannot be extended to solutions at higher order in the spin perturbation.}

One such relation is universal and follows from the Znajek condition \eqref{eq:Znajek}, which must always hold at the horizon $r_+=2M+\mathcal{O}(a)$: to leading-order in $a$, we have
\begin{align}
    \label{eq:PerturbativeZnajek}
    I_1=2\pi\pa{\Omega_1-\frac{1}{4}}\sin{\theta}\pd_\theta\psi_0\quad
    \text{at }r=2M.
\end{align}
As for a second relation, \cite{BlandfordZnajek1977} had the key insight to demand that the ansatz \eqref{eq:PerturbativeAnsatz} match onto an exact force-free solution in Minkowski spacetime at radii $r\gg M$, where the Kerr geometry becomes flat.
In other words, at large radii, the ansatz \eqref{eq:PerturbativeAnsatz} should solve the stream equation \eqref{eq:Stream} with $a=M=0$, while $\Omega=\Omega_1$ and $I=I_1$: more precisely, in cylindrical coordinates $(\rho,z)=r(\sin{\theta},\cos{\theta})$, \eqref{eq:PerturbativeAnsatz} should solve
\begin{align}
    \label{eq:FlatFFE}
    &\pa{1-\rho^2\Omega^2}\pa{\pd_\rho^2\psi+\pd_z^2\psi}-\frac{1+\rho^2\Omega^2}{\rho}\pd_\rho\psi\\
    &-\rho^2\br{\pa{\pd_\rho\psi}^2+\pa{\pd_z\psi}^2}\Omega\Omega'+\frac{II'}{(2\pi)^2}=0.\notag
\end{align}

\subsection{Flat force-free fields}

The force-free equation in flat spacetime \eqref{eq:FlatFFE} is highly nontrivial to solve---even numerically---since it is both nonlinear and involves three unknowns \citep[see, e.g.,][and the many references therein]{Mahlmann2018}.

Thankfully, there exist some exactly known analytic solutions, reviewed for instance in \cite{Compere2016}.

The most important one is due to \cite{Michel1973} and corresponds to a monopolar (radial) magnetic field:
\begin{subequations}
\label{eq:FlatMonopole}
\begin{align}
    \label{eq:Monopoles}
    \psi(r,\theta)&=1\mp_\psi\cos{\theta},\\
    I(\psi)&=\pm_I2\pi\psi\pa{2-\psi}\Omega(\psi),
\end{align}
\end{subequations}
with $\Omega(\psi)$ and the two signs $\pm_\psi$ and $\pm_I$ arbitrary.

Another important solution is due to \cite{Blandford1976} and corresponds to a parabolic magnetic field:
\begin{subequations}
\label{eq:FlatParabolic}
\begin{align}
    u(r,\theta)&=r\pa{1\mp_\psi\cos{\theta}},\\
    \label{eq:ParabolicFlux}
    \psi(u)&=\int\frac{\ed u}{\sqrt{1+\br{u\Omega(u)}^2}},\\
    \label{eq:ParabolicCurrent}
    I(\psi)&=\pm_I\frac{4\pi u\Omega(u)}{\sqrt{1+\br{u\Omega(u)}^2}},
\end{align}
\end{subequations}
with $\Omega(u(\psi))$ and the two signs $\pm_\psi$ and $\pm_I$ arbitrary.

In each case, to satisfy \eqref{eq:MagneticFluxCondition} we need a different sign
\begin{align}
    \mp_\psi=-\sign\pa{\cos{\theta}}
\end{align}
in each hemisphere.
This results in the ``split monopole''
\begin{align}
    \psi_{\rm split}=1-\ab{\cos{\theta}},
\end{align}
which is a force-free solution everywhere except in the equatorial plane, where the two opposite-sign solutions \eqref{eq:Monopoles} are stitched, producing a disk-like discontinuity (a ``current sheet''); see \cite{Gralla2014} for further details.
Likewise, only the ``split'' parabolic field
\begin{align}
    u_{\rm parabolic}=r\pa{1-\ab{\cos{\theta}}}
\end{align}
is physical.
Unlike the monopole solution \eqref{eq:FlatMonopole}, which has a realistic field-line geometry even when it is not split, the solution \eqref{eq:FlatParabolic} only has a parabolic field-line geometry when it is split (and is otherwise unphysical).

Lastly, we note that \cite{Gralla2016} found one more exact solution with hyperbolic field-line geometry.
This solution is valid everywhere without splitting and provides a model for a configuration with an equatorial disk terminating at an inner edge $r=b$: in terms of $(\rho,z)=r(\sin{\theta},\cos{\theta})$, it is given by
\begin{subequations}
\label{eq:FlatHyperbolic}
\begin{align}
    u(r,\theta)&=\frac{\sqrt{z^2+\pa{\rho-b}^2}-\sqrt{z^2+\pa{\rho+b}^2}}{2b},\\
    \psi(u)&=\int\frac{u\ed u}{\sqrt{1-u^2}\sqrt{1-\br{bu^2\Omega(u)}^2}},\\
    I(\psi)&=\pm_I\frac{2\pi u^2\Omega(u)}{\sqrt{1-\br{bu^2\Omega(u)}^2}}.
\end{align}
\end{subequations}
Again, $\Omega(u(\psi))$ and the sign $\pm_I$ are arbitrary, as is the parameter $b\ge0$.
This family smoothly interpolates between the radial monopole field \eqref{eq:FlatMonopole} when $b=0$ and the vertical uniform magnetic field when $b\to\infty$,
\begin{subequations}
\label{eq:FlatVertical}
\begin{align}
    \psi(r,\theta)&=r^2\sin^2{\theta},\\
    I(\psi)&=\pm_I4\pi\psi\Omega(\psi),
\end{align}
\end{subequations}
with $\Omega_1(\psi_0)$ and the sign $\pm_I$ still arbitrary.

Henceforth, we consider only the northern hemisphere $0\le\theta\le\pi/2$ where $\mp_\psi=-$, and we choose the sign
\begin{align}
    \pm_I=-,
\end{align}
appropriate to outgoing flux in the northern hemisphere.

\subsection{Vacuum Maxwell fields in Schwarzschild}

Given a force-free field in flat spacetime---a solution to \eqref{eq:FlatFFE}---we can construct a perturbative force-free solution \eqref{eq:PerturbativeAnsatz} in Kerr with similar field-line geometry.

To do so, we must find a vacuum Maxwell field in Schwarzschild---a solution to \eqref{eq:MaxwellSchwarzschild}---whose magnetic flux function shares the same asymptotic behavior as $r\to\infty$.
For the monopole field, this task is trivial, as
\begin{align}
    \label{eq:SchwarzschildMonopole}
    \psi_0(r,\theta)=1-\cos{\theta}
\end{align}
is also an exact solution of \eqref{eq:MaxwellSchwarzschild}.
The same is true of
\begin{align}
    \psi_0(r,\theta)=r^2\sin^2{\theta},
\end{align}
as the vertical field is also an exact solution of \eqref{eq:MaxwellSchwarzschild}.
On the other hand, the parabolic field $\psi_0=r(1-\cos{\theta})$ no longer solves \eqref{eq:MaxwellSchwarzschild}.
It would be natural to search for a solution of the form $\psi_0=f(r)(1-\cos{\theta})$ with $f(r)\stackrel{r\to\infty}{\approx}r$.
Such a solution exists, but it is singular on the horizon.\footnote{The solution $f(r)=c_1\br{r+2M\log\pa{r-2M}}+c_2$ diverges at $r=2M$ for any $c_1,c_2$.}
Instead, the appropriate solution is\footnote{This agrees with (7.1) in \cite{BlandfordZnajek1977}, up to a shift by $4M\pa{1-\log{2}}$ to ensure that $\psi(r,0)=0$.}
\begin{align}
    \label{eq:SchwarzschildParabolicField}
    \psi_0(r,\theta)&=r\pa{1-\cos{\theta}}-4M\pa{1-\log{2}}\\
    &\phantom{=}+2M\pa{1+\cos{\theta}}\br{1-\log\pa{1+\cos{\theta}}}\!.\notag
\end{align}
This reproduces the flat-space parabolic field as $r\to\infty$, is regular on the Schwarzschild horizon $r_+=2M$, and moreover respects the boundary condition \eqref{eq:MagneticFluxCondition}.

This matching procedure may seem rather haphazard, but it can be approached more systematically.
Such an approach is indispensable for the hyperbolic field, whose extension to Schwarzschild is much more complicated.

The key idea is that all the flat-space force-free fields have magnetic flux functions $\psi_0$ that are also solutions to the vacuum Maxwell equation in flat spacetime (since one can choose $\Omega=I=0$), which is \eqref{eq:MaxwellSchwarzschild} with $M=0$.

Since \eqref{eq:MaxwellSchwarzschild} is separable, one
can canonically identify its mode solutions when $M=0$ (the flat-space modes) with its mode solutions when $M\neq0$ (in Schwarzschild).

These sets of modes and their canonical identification are described in Appendix~B of \cite{Gralla2016}.
As an application, the vacuum magnetic flux function for the flat-space hyperbolic field, corresponding to \eqref{eq:FlatHyperbolic} with $\Omega=I=0$, is explicitly given in (B.18) therein, and its canonical extension to a vacuum Maxwell field in Schwarzschild is explicitly given in (B.22)--(B.25).

\subsection{Perturbative force-free monopole field on Kerr}
\label{app:Monopole}

We now have all the ingredients needed to solve for the perturbative force-free solution \eqref{eq:PerturbativeAnsatz} with a radial field-line geometry: the Blandford-Znajek monopole.

The leading-order magnetic flux function $\psi_0(r,\theta)$ is given in \eqref{eq:SchwarzschildMonopole}, so only the field-line angular velocity $\Omega_1(\psi_0)$ and current $I(\psi_0)$ remain to be determined.
The perturbative Znajek condition \eqref{eq:PerturbativeZnajek} yields one relation between $\psi_0$, $\Omega_1$ and $I_1$,
\begin{align}
    I_1(\psi_0)=2\pi\psi_0\pa{2-\psi_0}\pa{\Omega_1(\psi_0)-\frac{1}{4}}\!,
\end{align}
which is technically imposed at the horizon but must in fact hold everywhere, as both $\Omega_1$ and $I_1$ are only functions of $\psi_0$.
Matching onto the flat-space solution \eqref{eq:FlatMonopole} yields a second relation between $\psi_0$, $\Omega_1$ and $I_1$,
\begin{align}
    \label{eq:MonopoleCondition}
    I_1(\psi_0)=-2\pi\psi_0\pa{2-\psi_0}\Omega_1(\psi_0),
\end{align}
which is technically imposed at $r\to\infty$ but must again hold everywhere, as both $\Omega_1$ and $I_1$ are only functions of $\psi_0$.
Equating these relations lets us solve for
\begin{align}
    \Omega_1(\psi_0)=\frac{1}{8},
\end{align}
in terms of which $I_1(\psi_0)$ is then given by \eqref{eq:MonopoleCondition}.

Therefore, to leading order in perturbation theory, the force-free monopole solution is approximately
\begin{subequations}
\label{eq:PerturbativeMonopole}
    \begin{align}
    \psi(r,\theta)&=1-\cos{\theta},\\
    \Omega(\psi)&=\frac{1}{2}\Omega_{\rm H},\\
    I(\psi)&=-2\pi\psi\pa{2-\psi}\Omega(\psi).
\end{align}
\end{subequations}
Although we will not reproduce the details, it is possible to continue this perturbative expansion to higher orders \citep{Armas2020}.
For instance, at the next order,
\begin{subequations}
\label{eq:SubleadingMonopole}
\begin{align}
    \psi&=\pa{1-\cos\theta}+\frac{a^2}{M^2}\hat{R}\!\pa{\frac{r}{M}}\sin^2{\theta}\cos{\theta},\\
    \Omega&=\frac{a}{8M^2}+\frac{a^3}{M^4}\pa{\frac{1}{32}-\frac{4\hat{R}(2)-1}{64}\sin^2{\theta}}\!,\\
    I&=-2\pi\psi\pa{2-\psi}\Omega,
\end{align}
\end{subequations}
where $\hat{R}(x)$ is a complicated function given in (4.49) of \cite{Armas2020}, and such that $72\hat{R}(2)=6\pi^2-49$.

\subsection{Perturbative force-free parabolic field on Kerr}

Next, we solve for the perturbative force-free solution \eqref{eq:PerturbativeAnsatz} on Kerr with a parabolic field-line geometry.

The leading-order magnetic flux function $\psi_0(r,\theta)$ is given in \eqref{eq:SchwarzschildParabolicField}, so only the field-line angular velocity $\Omega_1(\psi_0)$ and current $I(\psi_0)$ remain to be determined.
The perturbative Znajek condition \eqref{eq:PerturbativeZnajek} yields the first needed relation between $\psi_0$, $\Omega_1$ and $I_1$ (which, again, is only imposed at the horizon, but must hold everywhere as both $\Omega_1$ and $I_1$ are only functions of $\psi_0$):
\begin{subequations}
\begin{align}
    \label{eq:ParabolicCondition1}
    I_1(\psi_0)&=\frac{4\pi\psi_0}{Y(\psi_0)-1}\pa{\Omega_1(\psi_0)-\frac{1}{4}}\!,\\
    Y(\psi_0)&=1+\frac{\psi_0}{M}\frac{\br{1+\log{X(\psi_0)}}^{-1}}{X(\psi_0)\br{2-X(\psi_0)}},\\
    X(\psi_0)&=\exp\br{W_k\!\pa{-\frac{\psi_0}{2M}+\log{4}}}\!.
\end{align}
\end{subequations}
Here, $W_k(z)$ is the $k^\text{th}$ branch of Lambert's W-function, defined such that $w=W_k(z)$ solves $z=we^w$.\footnote{It is also known as the product logarithm and is implemented in \textsc{Mathematica} as \texttt{ProductLog}$[k,z]$.}
For real $x$ and $y$, the equation $x=ye^y$ only admits solutions if $x\ge-e^{-1}$, in which case the two branches $W_0$ and $W_{-1}$ suffice to represent them: if $x\ge0$, then there is a unique solution $y=W_0(x)$, whereas if $-e^{-1}\le x<0$, then there is a second solution $y=W_{-1}(x)$.

To obtain a second relation between $\psi_0$, $\Omega_1$ and $I_1$, we match onto the flat-space solution \eqref{eq:FlatParabolic} with arbitrary field-line velocity $\Omega(\psi)$ and field-line current \eqref{eq:ParabolicCurrent}:
\begin{align}
    I(\psi)=-\frac{4\pi u\Omega(u)}{\sqrt{1+\br{u\Omega(u)}^2}}
    \stackrel{\Omega\ll1}{\approx}-4\pi u\Omega(u),
\end{align}
where in the last step, we expanded to leading order in $\Omega\sim\mathcal{O}(a)$.
The resulting truncation is consistent with the perturbative ansatz \eqref{eq:PerturbativeAnsatz}, which assumes that $\Omega$ and $I$ are both linear in the small spin $0<\ab{a}\ll M$.

Since to leading order in $\Omega$, the magnetic flux \eqref{eq:ParabolicFlux} is $\psi_0(u)\approx u$, we thus obtain a second relation
\begin{align}
    \label{eq:ParabolicCondition2}
    I_1(\psi_0)\approx-4\pi\psi_0\Omega_1(\psi_0).
\end{align}
This condition is only imposed at $r\to\infty$, but must hold everywhere as both $\Omega_1$ and $I_1$ are only functions of $\psi_0$.
Equating this expression to \eqref{eq:ParabolicCondition1} lets us solve for\footnote{This is consistent with (14) of \cite{Gralla2015}, where this expression first appeared (and $I$ is defined with the opposite sign).}
\begin{align}
    \label{eq:ParabolicVelocity}
    \Omega_1(\psi_0)=\frac{1}{4Y(\psi_0)}.
\end{align}

Therefore, to leading order in perturbation theory, the force-free parabolic solution is approximately
\begin{subequations}
\label{eq:PerturbativeParabolicField}
    \begin{align}
    \psi(r,\theta)&=r\pa{1-\cos{\theta}}-4M\pa{1-\log{2}}\\
    &\phantom{=}+2M\pa{1+\cos{\theta}}\br{1-\log\pa{1+\cos{\theta}}},\notag\\
    \Omega(\psi)&=\frac{\Omega_{\rm H}}{Y(\psi)},\\
    I(\psi)&=-4\pi\psi\Omega(\psi).
\end{align}
\end{subequations}

We note that our derivation differs from the treatment by \cite{BlandfordZnajek1977}: while their (7.4) agrees with our \eqref{eq:ParabolicCondition2}, their (7.2) is only the restriction to the horizon of our more general \eqref{eq:ParabolicCondition1}.
For this reason, we are able to find the field-line angular velocity \eqref{eq:ParabolicVelocity} everywhere, while they only obtain its restriction to the horizon, $\Omega_+(\theta)=\Omega(\psi(r_+,\theta))$, which they give in (7.5):
\begin{align}
    \label{eq:ParabolicHorizonFlux}
    \Omega_+(\theta)=\frac{\frac{\sin^2{\theta}}{16}\br{1+\log\pa{1+\cos{\theta}}}}{\log{2}+\frac{\sin^2{\theta}}{4}-\log\pa{1+\cos{\theta}}\cos^4{\frac{\theta}{2}}}.
\end{align}
As a consistency check, one can plug the horizon flux
\begin{align}
    \frac{\psi(r_+,\theta)}{2M}=\log{4}-\pa{1+\cos{\theta}}\log\pa{1+\cos{\theta}},
\end{align}
into \eqref{eq:ParabolicVelocity}, and verify whether this reproduces \eqref{eq:ParabolicHorizonFlux}.
This is the case provided that one uses $W_0$ for $\theta\in[0,\theta_e]$ and $W_{-1}$ for $\theta\in[\theta_e,\pi/2]$, where $\theta_e=\arccos\pa{e^{-1}-1}$ is the unique angle such that $\psi(r_+,\theta_e)=-e^{-1}$.

\subsection{Perturbative force-free hyperbolic field on Kerr}

For the perturbative force-free solution \eqref{eq:PerturbativeAnsatz} on Kerr with a hyperbolic field-line geometry, the above steps are repeated in detail in \S4 of \cite{Gralla2016}.

\section{Parametric model for accretion flow velocities}
\label{app:AART}

Here, we briefly summarize the model for accretion flow introduced in \cite{CardenasAvendano2022}---full derivations are provided in Appendix~B therein.
We consider an equatorial 4-velocity of the form
\begin{align}    \tilde{u}=\tilde{u}^t\pd_t+\tilde{u}^r\pd_r+\tilde{u}^\phi\pd_\phi.
\end{align}
Such a flow has angular and radial-infall velocities
\begin{align}
     \Omega=\frac{u^\phi}{u^t},\quad
     \iota=-\frac{u^r}{u^t}\geq0.
\end{align}
We work in the Kerr metric \eqref{eq:Kerr} with Boyer-Lindquist coordinates.
\cite{Bardeen1972} first showed that the innermost stable circular-equatorial orbit (ISCO) lies at a radius (here, the sign $s=+1$ corresponds to prograde orbits and $s=-1$ to retrograde orbits)
\begin{subequations}
\label{eq:ISCO}
\begin{align}
 \frac{r_{\rm ms}}{M}&=3+Z_2-s\sqrt{\pa{3-Z_1}\!\pa{3+Z_1+2Z_2}},\\
 Z_1&=1+\sqrt[3]{1-a_*^2}\br{\sqrt[3]{1+a_*}+\sqrt[3]{1-a_*}},\\
 Z_2&=\sqrt{3a_*^2+Z_1^2},\quad a_*=\frac{a}{M}.
\end{align}
\end{subequations}
Following \cite{CardenasAvendano2022}, we define\footnote{Here, we also include retrograde orbits with $s=-1$.}
\begin{align}
     \hat{\ell}=
     \begin{cases}
         \xi\frac{sa}{\ab{a}}\frac{\sqrt{M}\pa{r^2+a^2-2s\ab{a}\sqrt{Mr}}}{\sqrt{r}\pa{r-2M}+s\ab{a}\sqrt{M}}&\text{if }r\geq r_{\rm ms},\\
         \hat{\ell}_{\rm ms}\equiv\left.\hat{\ell}\right|_{r=r_{\rm ms}}&\text{if }r<r_{\rm ms},
     \end{cases}
\end{align}
as well as
\begin{align}
    \hat{\mathcal{E}}_{\rm ms}=\sqrt{\frac{\Delta(r_{\rm ms})}{\frac{\Pi(r_{\rm ms})}{r_{\rm ms}^2}-\frac{4aM\hat{\ell}_{\rm ms}}{r_{\rm ms}}-\pa{1-\frac{2M}{r_{\rm ms}}}\hat{\ell}_{\rm ms}^2}},
\end{align}
where $\xi\in(0,1]$ is a sub-Keplerianity parameter.
When $\xi=1$, $\hat{\ell}$ is the specific angular momentum of Keplerian circular orbits down to the ISCO, while $\hat{\mathcal{E}}_{\rm ms}$ and $\hat{\ell}_{\rm ms}$ are the specific energy and angular momentum of the ISCO.

As in \cite{Pu2016} and \cite{Vincent2022}, we linearly superpose a prograde circular flow $\hat{u}$ and a radial inflow $\bar{u}$ to obtain the general flow
\begin{subequations}
\begin{align}
     \tilde{u}^r&=\hat{u}^r+\pa{1-\beta_r}\pa{\bar{u}^r-\hat{u}^r},\\
     \tilde{\Omega}&=\hat{\Omega}+\pa{1-\beta_\phi}\pa{\bar{\Omega}-\hat{\Omega}},
\end{align}
\end{subequations}
where $\beta_r$ and $\beta_\phi$ are parameters in $[0,1]$, and
\begin{subequations}
\begin{align}
     \bar{u}^r&=-\frac{\sqrt{2Mr\pa{r^2+a^2}}}{r^2},\quad
     \bar{\Omega}=\omega
     =\frac{2aMr}{\Pi},\\
     \hat{u}^r&=
     \begin{cases}
         0&\text{if }r\geq r_{\rm ms},\\
         -\pa{1-\frac{2Mr}{r}+\frac{a^2}{r^2}}\hat{\nu}\hat{\mathcal{E}}_{\rm ms}&\text{if }r<r_{\rm ms},
     \end{cases}\\
     \hat{\Omega}&=\frac{\pa{1-\frac{2M}{r}}\hat{\ell}-\frac{2Ma}{r}}{\frac{\Pi}{r^2}-\frac{2aM\hat{\ell}}{r}},
\end{align}
\end{subequations}
with
\begin{align}
     \hat{\nu}=\frac{r}{\Delta}\sqrt{\frac{\Pi}{r^2}-\frac{4aM\hat{\ell}_{\rm ms}}{r}-\pa{1-\frac{2M}{r}}\hat{\ell}_{\rm ms}^2-\frac{\Delta}{\hat{\mathcal{E}}_{\rm ms}}}.\notag
\end{align}

Given $\tilde{u}^r$ and $\tilde{\Omega}=\tilde{u}^\phi/\tilde{u}^t$, the normalization condition $\tilde{u}\cdot\tilde{u}=-1$ fixes the remaining component $\tilde{u}^t$ of $\tilde{u}$ to be
\begin{align}
     \label{eq:ur}
     \tilde{u}^t=\sqrt{\frac{1+\frac{r^2}{\Delta(r)}\pa{\tilde{u}^r}^2}{1-\pa{r^2+a^2}\tilde{\Omega}^2-\frac{2M}{r}\pa{1-a\tilde{\Omega}}^2}}.
\end{align}
Since $\tilde{u}^\phi=\tilde{\Omega}\tilde{u}^t$, this completes the definition of $\tilde{u}$.

For most values of the parameters $(\xi,\beta_r,\beta_\phi)$, the flow $\tilde{u}$ is not geodesic.
However, if $\beta_r=\beta_\phi=0$, then $\tilde{u}$ reduces to the radial inflow $\bar{u}$, which is $\xi$-independent and geodesic, corresponding to an equatorial inflow with zero angular momentum.\footnote{This explains why $\tilde{\Omega}$ equals the ZAMO angular velocity \eqref{eq:ZAMO}.}
Likewise, if $\beta_r=\beta_\phi=1$, then $\tilde{u}$ reduces to the circular flow $\tilde{u}(\xi)$, which is sub-Keplerian and hence non-geodesic for $\xi<1$, but reduces to geodesic circular Keplerian motion when $\xi=1$.

\section{Semi-Analytic model description}
\label{app:ModelDescription}

\subsection{Magnetic field, fluid velocity, and emissivity}

This appendix describes in detail the construction of the semi-analytic models used to generate the simulated images of M87* in \autoref{sec:RingModels}.
As discussed therein, in all models, we take the lab-frame magnetic field $B^i=\star F^{i0}$ to be that of the Blandford-Znajek monopole solution \eqref{eq:SubleadingMonopole}, and then we specify some fluid 4-velocity $u^\mu$ to complete the definition of $\star F^{\mu\nu}$.
Since by degeneracy, the electric field is assumed to vanish in the fluid frame,
\begin{align}
\label{eq:MaxwellTensor}
    \star F^{\mu\nu}=b^\mu u^\nu-b^\nu u^\mu,
\end{align}
where in this formula, the ``fluid-frame'' magnetic field is
\begin{align}
    b^t=u_iB^i,\quad
    b^i=\frac{B^i+b^tu^i}{u^t}.
\end{align}

As mentioned in \autoref{sec:RingModels}, we use two different models for $u^\mu$.
First, we adopt the drift-frame velocity \eqref{eq:DriftVelocity} that is determined by the full Blandford-Znajek solution for both $B^i=\star F^{i0}$ and $E^i=F^{0i}$.
Given a fixed $\star F^{\mu\nu}$, \eqref{eq:DriftVelocity} provides the unique frame $u_\perp^\mu$ in which the local electric field vanishes ($e^\mu=0$) and $u\cdot B=0$.
As we will show in \citetalias{Paper2}, replacing $u_\perp^\mu$ by a general member of the family \eqref{eq:TimelikeVector}---which differs from $u_\perp^\mu$ by a boost along the magnetic field---would not affect the direction of linear polarization for synchrotron radiation.

In order to survey different models of fluid inflow, we also explore the family of parameterized accretion flow models for $u^\mu$ introduced by \citet{CardenasAvendano2022} and reproduced in \autoref{app:AART}.
That is, we fix the lab-frame magnetic field components $B^i=\star F^{i0}$ to their values in the Blandford-Znajek monopole solution \eqref{eq:SubleadingMonopole}, and then we freely prescribe some choice of $u^\mu$ drawn from \autoref{app:AART}.
Generically, such a choice is inconsistent with the allowed family of flows \eqref{eq:TimelikeVector} and therefore modifies the spatial components $\star F^{ij}$ (and hence the electric field $E^i=F^{0i}$) according to \eqref{eq:MaxwellTensor}.

Although commonly used in semi-analytic modeling of near-horizon flows, this approach suffers from a major drawback: it typically results in a field $F_{\mu\nu}$ that does not satisfy Maxwell's equations \eqref{eq:Maxwell}.
In particular, it is the Bianchi identity $\ed F=0$, or equivalently, the ``induction equation'' $\nabla_\mu(\star F)^{\mu\nu}=0$, that generically fails to hold.

More precisely, while the induction equation's time component (the ``no-monopoles'' constraint $\nabla_i B^i=0$) is always satisfied, on the other hand, its spatial part
\begin{align}
    \pd_t\pa{\sqrt{-g}B^i}=-\pd_j\br{\sqrt{-g}\pa{B^j\frac{u^i}{u^t}-B^i\frac{u^j}{u^t}}}
\end{align}
will typically not be.
Thus, if one wishes to regard the resulting tensor $F_{\mu\nu}$ as a valid electromagnetic field, it can no longer be viewed as a stationary configuration, but rather as the initial state of a time-evolving field.

Finally, having specified $B^i$ and $u^\mu$ everywhere in the equatorial plane of the black hole, we need only define the emissivity $J(r_{\rm eq})$ for emission received at 230\,GHz.
We consider a thin ring centered at $r_{\rm ring}=2r_{+}$ and use the same emissivity as in (12) of \citet{Gralla2020}: 
\begin{align}
    J(r_{\rm eq})=\frac{e^{-\frac{1}{2}\br{\mathrm{arcsinh}\pa{\frac{r_{\rm eq}-r_{\rm ring}}{\sigma}}^2}}}{\sqrt{\pa{r_{\rm eq}-r_{\rm ring}}^2+\sigma^2}},
\end{align}
with a ring width of $\sigma=0.3$.

\subsection{Ray tracing and parallel transport}

For a given spin $a$, viewing angle $\theta_{\rm o}$, magnetic field $B^i$, and 4-velocity $u^\mu$, we generate polarized synchrotron images of an equatorial emission ring using the public code \texttt{kgeo} \citep{kgeo}.\footnote{https://github.com/achael/kgeo}
We parameterize the image plane with coordinates $(\alpha,\beta)$ defined by \citet{Bardeen1972} such that the $\beta$ axis is aligned with the black hole spin projected onto the plane perpendicular to the line of sight.
\texttt{kgeo} analytically computes null geodesics by tracing backwards from each point on the observer screen using the formalism of \citet{GrallaLupsasca2020}.
To determine the EVPA of the received emission, \texttt{kgeo} also solves for the parallel transport along each ray of the polarization vector $f^\mu$, which is initially taken to point in a direction locally perpendicular to both the magnetic field and photon momentum at the source. 

In particular, for axisymmetric, equatorial emission models, \texttt{kgeo} solves for the radius $r_{\rm eq}(\alpha,\beta;n)$ where the null geodesic that reaches the observer at coordinate $(\alpha,\beta)$ makes its $(n+1)^\text{th}$ pass through the equatorial plane $\theta=\pi/2$.
This radius, together with the geodesic's conserved specific angular momentum $\lambda=-\alpha\sin{\theta_{\rm o}}$ and Carter constant $\eta=(\alpha^2-a^2)\cos^2{\theta_{\rm o}}+\beta^2$, determines everywhere along the trajectory the local momentum $k^\mu$ of a photon with energy $E$ loaded onto the ray:
\begin{subequations}
\begin{align}
    k_t&=-E,\quad
    k_\phi=E\lambda,\\
    \frac{k_r}{E}\!&=\pm\frac{\sqrt{\pa{r^2+a^2-a\lambda}^2-\Delta\br{\eta+(\lambda-a)^2}}}{\Delta},\\
    \frac{k_\theta}{E}\!&=\pm\sqrt{\eta+a^2\cos^2{\theta}-\frac{\lambda^2}{\tan^2{\theta}}}.
\end{align}
\end{subequations}

Having determined the first equatorial emission radius $r_{\rm eq}(\alpha,\beta;0)$ for a given pixel $(\alpha,\beta)$, the observed total intensity in the direct ($n=0$) image at this position is given by \citep{Narayan2021}
\begin{align}
\label{eq:Iobs}
    \mathcal{I}(\alpha,\beta)=(1+z)^{-(2+\alpha_\nu)}\pa{\sin\theta_B}^{1+\alpha_\nu}J(r_{\rm eq}),
\end{align}
where $\alpha_\nu$ is the emitter frame spectral index (for M87*, we assume $\alpha_\nu=1$), $z$ is the total (gravitational and Doppler) redshift, and $\theta_B$ is the pitch angle between the wavevector $k^\mu$ and magnetic field $b^\mu$ in the fluid frame.
The redshift and pitch angle can be explicitly computed from the wavevector $k^\mu$, the fluid 4-velocity $u^\mu$, and the fluid-frame magnetic field $b^\mu$ (see (73) of \cite{Dexter2016}):
\begin{align}
    1+z&=\frac{u\cdot k}{k_t},\\
    \cos\theta_B&=\frac{b\cdot k}{\sqrt{\pa{b\cdot b}\pa{u\cdot k}^2}}.
\end{align}

In our simplified model, we assume a constant degree of polarization.
That is, we take the observed linearly polarized intensity $\mathcal{P}$ in each image-plane pixel to be a constant fraction $m$ of the total observed intensity $\mathcal{I}$:
\begin{align}
    \label{eq:ObservedPolarization}
    \mathcal{P}=m\mathcal{I}e^{2i\chi}.
\end{align}
We compute the observed linear polarization EVPA $\chi$ in each pixel following \citet{Dexter2016} and \citet{Himwich2020}. In particular, we use (36)--(39) of \citet{Dexter2016} to define tetrad matrices $e_a^\mu$ that take the wavevector $k^\mu$ and magnetic field vector $b^\mu$ into local vectors defined in the frame comoving with the fluid 4-velocity $u^\mu$:
\begin{align}
    k_a'=e_a^\mu k_\mu,\quad
    b_a'=e_a^\mu b_\mu.
\end{align}
We treat the spatial fluid-frame vectors $\vec{k'}$ and $\vec{b'}$ as Cartesian 3-vectors.
The local polarization direction of the synchrotron emission in the fluid frame is then
\begin{align}
    \vec{f'}=\frac{\vec{k'}\times\vec{b'}}{\big|\vec{k'}\big|}.
\end{align}
The same tetrad matrix yields $f^\mu$ in global coordinates:
\begin{align}
    f^\mu=e_a^\mu f'^a.
\end{align}
To parallel transport the polarization $f^\mu$ from the source to the observer screen, we compute the complex Penrose-Walker constant $\kappa$ \citep[see, e.g.,][]{Himwich2020}:
\begin{align}
    \kappa&=\kappa_1+i\kappa_2
    =(r-ia\cos\theta)(A-iB),\\
    A&=(k^tf^r-k^rf^t)+a\sin^2\theta(k^rf^\phi-k^\phi f^r),\notag\\
    B&=\left[(r^2+a^2)(k^\phi f^\theta-k^\theta f^\phi)-a(k^tf^\theta - k^\theta f^t)\right]\sin\theta.\notag
\end{align}
Finally, $\kappa$ determines the observed EVPA $\chi$: in terms of $\mu=-(\alpha+a\sin{\theta_{\rm o}})$,
\begin{align}
    \label{eq:EVPA}
    e^{2i\chi}=\frac{(\beta+i\mu)(\kappa_1-i\kappa_2)}{(\beta-i\mu)(\kappa_1+i\kappa_2)}.
\end{align}

In summary, to generate a polarized image from our analytic model, we specify the spin $a$, magnetic field $B^i$ (from the BZ split monopole), velocity field $u^\mu$, and emission radius $r_{\rm eq}=2r_+$.
We choose a viewing angle $\theta_{\rm o}=163\deg$ \citep{Mertens2016} for prograde flows ($a>0$), and $\theta_{\rm o}=17\deg$ for retrograde flows ($a<0$).
We consider only the direct $n=0$ image, and neglect contributions from the ``photon ring'' images consisting of photons that pass through the equatorial plane more than once \citep{JohnsonLupsasca2020}.
The total-intensity image is determined by \eqref{eq:Iobs}, while the image in linear polarization is determined by \eqref{eq:ObservedPolarization}.
The fractional linear polarization $m$ is a constant whose choice does not affect the values of $\angle\beta_2$.
The EVPA is determined by parallel transporting the polarization vector from source to observer, resulting in \eqref{eq:EVPA}.  

In \citetalias{Paper2}, we will introduce an equivalent, but far simpler, procedure for computing the observed EVPA $\chi$ from point-like synchrotron emitters in the Kerr spacetime.
That method provides identical results to the well-established procedure that is summarized here and which is implemented in \texttt{kgeo}.

\section{GRMHD Library Details}
\label{app:GRMHDLibrary}

In \autoref{sec:GRMHD}, we generate 230\,GHz images from the nine GRMHD simulations taken from \citet{Narayan2022} using the public code \texttt{ipole} \citep{Noble2007,Moscibrodzka2018}.\footnote{https://github.com/moscibrodzka/ipole}
These simulations were run to a maximal time $t_{\rm max}=10^5\,t_{\rm g}$; we use the range $50,000\,t_{\rm g}$ to $100,000\,t_{\rm g}$ to generate model images of M87* from each simulation.

To set the electron temperature, we follow the $R_{\rm high}$ prescription from \cite{Moscibrodzka2016}, where $R_{\rm high}$ denotes the ion-to-electron temperature ratio in weakly magnetized regions of the accretion flow.
We cover values of $R_{\rm high}\in[1,10,20,40,80,160]$, and we compute 230\,GHz images keeping the black hole mass, distance, viewing angle, and total flux density fixed to values appropriate for M87*.
As for the temperature ratio in strongly magnetized regions $R_{\rm low}$, we fix it to unity.
In M87*, electrons may be significantly cooled by radiation even in highly magnetized regions, so $R_{\rm low}>1$ may be a better model of the real electron temperature in these regions (\citetalias{PaperVIII}; \citealt{Chael2019}).

GRMHD simulations can develop numerical artifacts whenever the plasma temperature or internal energy are evolved in regions where the magnetic field becomes very strong relative to the plasma density, as measured by the magnetization $\sigma=b^2/\rho$.
As a result, when generating images, we cut off all emission in regions where $\sigma>1$.

Because GRMHD simulations are scale-free, we may scale the mass density in the ray tracing step to achieve an average total flux density of 0.5\,Jy in our 230\,GHz simulated images of M87*.

The simulation images were generated keeping track of full polarization, using radiative transfer that includes both Faraday rotation and conversion internal to the emission region.
Our image library is computed using the same parameters as in \citet{Ricarte2022} and \citet{Qiu2023}, where more details on the radiative transfer procedure are presented.
Finally, we blur each snapshot image with a $20\,\mu$as full-width half-max Gaussian kernel and compute $\beta_2$ using \eqref{eq:Beta2}.

\section{Time-averaging GRMHD simulations}
\label{app:Averaging}

Electromagnetic fields are always degenerate in GRMHD.
Therefore, by linearity, the average over time and azimuth of a GRMHD-simulated field $F$ should fall in this class and take the general form \eqref{eq:GeneralForm}.
However, some care is required in computing this average,
\begin{align}
    \label{eq:FieldAverage}
    \av{F^{\mu\nu}}=\frac{1}{2\pi T}\int_0^T\ed t\int_0^{2\pi}\ed\phi\,F^{\mu\nu}.
\end{align}
The reason, which we already mentioned above \eqref{eq:LabFrameFields}, is that numerical GRMHD codes often use the lab-frame fields $B^i$ and the accretion flow $u^\mu$ as primitive variables, and the electromagnetic field is then retrieved from the formula \eqref{eq:HodgeDualDecomposition} for its Hodge dual.
While it is tempting to define a time-and-azimuth-averaged flow
\begin{align}
    \label{eq:FlowAverage}
    \av{u^\mu}=\frac{1}{2\pi T}\int_0^T\ed t\int_0^{2\pi}\ed\phi\,u^\mu,
\end{align}
and likewise a time-and-azimuth-averaged magnetic field
\begin{align}
    \label{eq:MagneticAverage}
    \av{B^\mu}=\frac{1}{2\pi T}\int_0^T\ed t\int_0^{2\pi}\ed\phi\,B^\mu,
\end{align}
this can be misleading because, in contrast to \eqref{eq:HodgeDualDecomposition},
\begin{align}
    \av{(\star F)^{\mu\nu}}\neq\av{B^\mu}\frac{\av{u^\nu}}{\av{u^t}}-\av{B^\nu}\frac{\av{u^\mu}}{\av{u^t}}.
\end{align}

Typically, one averages $\tilde{v}$ rather than $u$, as this keeps the latter timelike and normalized (while averaging the components of $u$ independently, as in \eqref{eq:FlowAverage}, does not).
Still, the same problem occurs at the level of $\star F$ (but this type of averaging is appropriate for other purposes, such as computing the mean angular velocity of the fluid).

In summary, to properly average a GRMHD-simulated electromagnetic field, one must first form the snapshot Hodge dual $(\star F)^{\mu\nu}$ from each of the snapshot magnetic fields $B^i$ and snapshot flows $u^\mu$ via \eqref{eq:HodgeDualDecomposition}, and only then perform the average \eqref{eq:FieldAverage}.

If instead, these operations are performed in reverse order, by first averaging the snapshot magnetic fields and flows as in \eqref{eq:FlowAverage}--\eqref{eq:MagneticAverage}, and afterwards forming the product \eqref{eq:HodgeDualDecomposition}, then the result will not yield the field average \eqref{eq:FieldAverage}---since multiplication and averaging do not commute---and will typically fail to even solve the Maxwell equations \eqref{eq:Maxwell}.
Likewise, the last two equations in \eqref{eq:FieldlineVelocity} may disagree.

%% file: paper1.bbl
\begin{thebibliography}{}
\expandafter\ifx\csname natexlab\endcsname\relax\def\natexlab#1{#1}\fi
\providecommand{\url}[1]{\href{#1}{#1}}
\providecommand{\dodoi}[1]{doi:~\href{http://doi.org/#1}{\nolinkurl{#1}}}
\providecommand{\doeprint}[1]{\href{http://ascl.net/#1}{\nolinkurl{http://ascl.net/#1}}}
\providecommand{\doarXiv}[1]{\href{https://arxiv.org/abs/#1}{\nolinkurl{https://arxiv.org/abs/#1}}}

\bibitem[{{Armas} {et~al.}(2020){Armas}, {Cai}, {Comp{\`e}re}, {Garfinkle}, \&
  {Gralla}}]{Armas2020}
{Armas}, J., {Cai}, Y., {Comp{\`e}re}, G., {Garfinkle}, D., \& {Gralla}, S.~E.
  2020, \jcap, 2020, 009, \dodoi{10.1088/1475-7516/2020/04/009}

\bibitem[{{Bardeen} {et~al.}(1972){Bardeen}, {Press}, \&
  {Teukolsky}}]{Bardeen1972}
{Bardeen}, J.~M., {Press}, W.~H., \& {Teukolsky}, S.~A. 1972, \apj, 178, 347,
  \dodoi{10.1086/151796}

\bibitem[{{Baumgarte} \& {Shapiro}(2003)}]{Baumgarte2003}
{Baumgarte}, T.~W., \& {Shapiro}, S.~L. 2003, \apj, 585, 921,
  \dodoi{10.1086/346103}

\bibitem[{{Begelman} {et~al.}(1984){Begelman}, {Blandford}, \&
  {Rees}}]{Begelman1984}
{Begelman}, M.~C., {Blandford}, R.~D., \& {Rees}, M.~J. 1984, Reviews of Modern
  Physics, 56, 255, \dodoi{10.1103/RevModPhys.56.255}

\bibitem[{{Bisnovatyi-Kogan} \& {Ruzmaikin}(1974)}]{Bisnovatyi1974}
{Bisnovatyi-Kogan}, G.~S., \& {Ruzmaikin}, A.~A. 1974, \apss, 28, 45,
  \dodoi{10.1007/BF00642237}

\bibitem[{{Blandford} \& {Globus}(2022)}]{Blandford2022}
{Blandford}, R., \& {Globus}, N. 2022, \mnras, 514, 5141,
  \dodoi{10.1093/mnras/stac1682}

\bibitem[{{Blandford}(1976)}]{Blandford1976}
{Blandford}, R.~D. 1976, \mnras, 176, 465, \dodoi{10.1093/mnras/176.3.465}

\bibitem[{{Blandford} \& {Payne}(1982)}]{BlandfordPayne1982}
{Blandford}, R.~D., \& {Payne}, D.~G. 1982, \mnras, 199, 883,
  \dodoi{10.1093/mnras/199.4.883}

\bibitem[{{Blandford} \& {Znajek}(1977)}]{BlandfordZnajek1977}
{Blandford}, R.~D., \& {Znajek}, R.~L. 1977, \mnras, 179, 433,
  \dodoi{10.1093/mnras/179.3.433}

\bibitem[{{Brennan} {et~al.}(2013){Brennan}, {Gralla}, \&
  {Jacobson}}]{Brennan2013}
{Brennan}, T.~D., {Gralla}, S.~E., \& {Jacobson}, T. 2013, Classical and
  Quantum Gravity, 30, 195012, \dodoi{10.1088/0264-9381/30/19/195012}

\bibitem[{{Camilloni} {et~al.}(2022){Camilloni}, {Dias}, {Grignani}, {Harmark},
  {Oliveri}, {Orselli}, {Placidi}, \& {Santos}}]{Camilloni2022}
{Camilloni}, F., {Dias}, O. J.~C., {Grignani}, G., {et~al.} 2022, \jcap, 2022,
  032, \dodoi{10.1088/1475-7516/2022/07/032}

\bibitem[{{C{\'a}rdenas-Avenda{\~n}o}
  {et~al.}(2023){C{\'a}rdenas-Avenda{\~n}o}, {Lupsasca}, \&
  {Zhu}}]{CardenasAvendano2022}
{C{\'a}rdenas-Avenda{\~n}o}, A., {Lupsasca}, A., \& {Zhu}, H. 2023, \prd, 107,
  043030, \dodoi{10.1103/PhysRevD.107.043030}

\bibitem[{{Carter}(1979)}]{Carter1979}
{Carter}, B. 1979, in Active Galactic Nuclei, ed. C.~{Hazard} \& S.~{Mitton}
  (Cambridge University Press), 273--300

\bibitem[{{Chael}(2022)}]{eht-imaging}
{Chael}, A. 2022, {eht-imaging}, v1.2.4,  Zenodo,
  \dodoi{10.5281/zenodo.7226661}

\bibitem[{{Chael}(2023)}]{kgeo}
---. 2023, {kgeo}, v0.1,  Zenodo, \dodoi{10.5281/zenodo.8092969}

\bibitem[{{Chael} {et~al.}(2021){Chael}, {Johnson}, \& {Lupsasca}}]{Chael2021}
{Chael}, A., {Johnson}, M.~D., \& {Lupsasca}, A. 2021, \apj, 918, 6,
  \dodoi{10.3847/1538-4357/ac09ee}

\bibitem[{{Chael} {et~al.}(2019){Chael}, {Narayan}, \& {Johnson}}]{Chael2019}
{Chael}, A., {Narayan}, R., \& {Johnson}, M.~D. 2019, \mnras, 486, 2873,
  \dodoi{10.1093/mnras/stz988}

\bibitem[{{Chael} {et~al.}(2016){Chael}, {Johnson}, {Narayan}, {Doeleman},
  {Wardle}, \& {Bouman}}]{Chael2016}
{Chael}, A.~A., {Johnson}, M.~D., {Narayan}, R., {et~al.} 2016, \apj, 829, 11,
  \dodoi{10.3847/0004-637X/829/1/11}

\bibitem[{{Comp{\`e}re} {et~al.}(2016){Comp{\`e}re}, {Gralla}, \&
  {Lupsasca}}]{Compere2016}
{Comp{\`e}re}, G., {Gralla}, S.~E., \& {Lupsasca}, A. 2016, \prd, 94, 124012,
  \dodoi{10.1103/PhysRevD.94.124012}

\bibitem[{{Cruz-Osorio} {et~al.}(2022){Cruz-Osorio}, {Fromm}, {Mizuno},
  {Nathanail}, {Younsi}, {Porth}, {Davelaar}, {Falcke}, {Kramer}, \&
  {Rezzolla}}]{Cruz2022}
{Cruz-Osorio}, A., {Fromm}, C.~M., {Mizuno}, Y., {et~al.} 2022, Nature
  Astronomy, 6, 103, \dodoi{10.1038/s41550-021-01506-w}

\bibitem[{{Dexter}(2016)}]{Dexter2016}
{Dexter}, J. 2016, \mnras, 462, 115, \dodoi{10.1093/mnras/stw1526}

\bibitem[{{Doeleman} {et~al.}(2019){Doeleman}, {Blackburn}, {Dexter}, {Gomez},
  {Johnson}, {Palumbo}, {Weintroub}, {Farah}, {Fish}, {Loinard}, {Lonsdale},
  {Narayanan}, {Patel}, {Pesce}, {Raymond}, {Tilanus}, {Wielgus}, {Akiyama},
  {Bower}, {Broderick}, {Deane}, {Fromm}, {Gammie}, {Gold}, {Janssen},
  {Kawashima}, {Krichbaum}, {Marrone}, {Matthews}, {Mizuno}, {Rezzolla},
  {Roelofs}, {Ros}, {Savolainen}, {Yuan}, {Zhao}, {Blackburn}, {Doeleman},
  {Dexter}, {Gomez}, {Johnson}, {Palumbo}, {Weintroub}, {Farah}, {Fish},
  {Loinard}, {Lonsdale}, {Narayanan}, {Patel}, {Pesce}, {Raymond}, {Tilanus},
  {Wielgus}, {Akiyama}, {Bower}, {Broderick}, {Deane}, {Fromm}, {Gammie},
  {Gold}, {Janssen}, {Kawashima}, {Krichbaum}, {Marrone}, {Matthews}, {Mizuno},
  {Rezzolla}, {Roelofs}, {Ros}, {Savolainen}, {Yuan}, \& {Zhao}}]{Doeleman2019}
{Doeleman}, S., {Blackburn}, L., {Dexter}, J., {et~al.} 2019, in Bulletin of
  the American Astronomical Society, Vol.~51, 256.
\newblock \doarXiv{1909.01411}

\bibitem[{{EHT MWL Science Working Group} {et~al.}(2021)}]{EHTMWL}
{EHT MWL Science Working Group}, {et~al.} 2021, \apjl, 911, L11,
  \dodoi{10.3847/2041-8213/abef71}

\bibitem[{{Emami} {et~al.}(2023){Emami}, {Ricarte}, {Wong}, {Palumbo}, {Chang},
  {Doeleman}, {Broderick}, {Narayan}, {Wielgus}, {Blackburn}, {Prather},
  {Chael}, {Anantua}, {Chatterjee}, {Marti-Vidal}, {G{\'o}mez}, {Akiyama},
  {Liska}, {Hernquist}, {Tremblay}, {Vogelsberger}, {Alcock}, {Smith},
  {Steiner}, {Tiede}, \& {Roelofs}}]{Emami2022}
{Emami}, R., {Ricarte}, A., {Wong}, G.~N., {et~al.} 2023, \apj, 950, 38,
  \dodoi{10.3847/1538-4357/acc8cd}

\bibitem[{{Gammie} {et~al.}(2003){Gammie}, {McKinney}, \&
  {T{\'o}th}}]{Gammie2003}
{Gammie}, C.~F., {McKinney}, J.~C., \& {T{\'o}th}, G. 2003, \apj, 589, 444,
  \dodoi{10.1086/374594}

\bibitem[{{Gralla} \& {Jacobson}(2014)}]{Gralla2014}
{Gralla}, S.~E., \& {Jacobson}, T. 2014, \mnras, 445, 2500,
  \dodoi{10.1093/mnras/stu1690}

\bibitem[{{Gralla} \& {Lupsasca}(2020)}]{GrallaLupsasca2020}
{Gralla}, S.~E., \& {Lupsasca}, A. 2020, \prd, 101, 044032,
  \dodoi{10.1103/PhysRevD.101.044032}

\bibitem[{{Gralla} {et~al.}(2020){Gralla}, {Lupsasca}, \&
  {Marrone}}]{Gralla2020}
{Gralla}, S.~E., {Lupsasca}, A., \& {Marrone}, D.~P. 2020, \prd, 102, 124004,
  \dodoi{10.1103/PhysRevD.102.124004}

\bibitem[{{Gralla} {et~al.}(2015){Gralla}, {Lupsasca}, \&
  {Rodriguez}}]{Gralla2015}
{Gralla}, S.~E., {Lupsasca}, A., \& {Rodriguez}, M.~J. 2015, \prd, 92, 044053,
  \dodoi{10.1103/PhysRevD.92.044053}

\bibitem[{{Gralla} {et~al.}(2016){Gralla}, {Lupsasca}, \&
  {Rodriguez}}]{Gralla2016}
---. 2016, \prd, 93, 044038, \dodoi{10.1103/PhysRevD.93.044038}

\bibitem[{{Harris} {et~al.}(2020){Harris}, {Millman}, {van der Walt},
  {Gommers}, {Virtanen}, {Cournapeau}, {Wieser}, {Taylor}, {Berg}, {Smith},
  {Kern}, {Picus}, {Hoyer}, {van Kerkwijk}, {Brett}, {Haldane}, {del R{\'\i}o},
  {Wiebe}, {Peterson}, {G{\'e}rard-Marchant}, {Sheppard}, {Reddy}, {Weckesser},
  {Abbasi}, {Gohlke}, \& {Oliphant}}]{NumPy2020}
{Harris}, C.~R., {Millman}, K.~J., {van der Walt}, S.~J., {et~al.} 2020, \nat,
  585, 357, \dodoi{10.1038/s41586-020-2649-2}

\bibitem[{{Himwich} {et~al.}(2020){Himwich}, {Johnson}, {Lupsasca}, \&
  {Strominger}}]{Himwich2020}
{Himwich}, E., {Johnson}, M.~D., {Lupsasca}, A., \& {Strominger}, A. 2020,
  \prd, 101, 084020, \dodoi{10.1103/PhysRevD.101.084020}

\bibitem[{{Hunter}(2007)}]{Matplotlib}
{Hunter}, J.~D. 2007, Computing in Science and Engineering, 9, 90,
  \dodoi{10.1109/MCSE.2007.55}

\bibitem[{{Jia} {et~al.}(2022){Jia}, {White}, {Quataert}, \&
  {Ressler}}]{Jia2022}
{Jia}, H., {White}, C.~J., {Quataert}, E., \& {Ressler}, S.~M. 2022, \mnras,
  515, 1392, \dodoi{10.1093/mnras/stac1517}

\bibitem[{{Jim{\'e}nez-Rosales} {et~al.}(2021){Jim{\'e}nez-Rosales}, {Dexter},
  {Ressler}, {Tchekhovskoy}, {Baub{\"o}ck}, {Dallilar}, {de Zeeuw}, {Drescher},
  {Eisenhauer}, {von Fellenberg}, {Gao}, {Genzel}, {Gillessen}, {Habibi},
  {Ott}, {Stadler}, {Straub}, \& {Widmann}}]{Jimenez2021}
{Jim{\'e}nez-Rosales}, A., {Dexter}, J., {Ressler}, S.~M., {et~al.} 2021,
  \mnras, 503, 4563, \dodoi{10.1093/mnras/stab784}

\bibitem[{{Johnson} {et~al.}(2020){Johnson}, {Lupsasca}, {Strominger}, {Wong},
  {Hadar}, {Kapec}, {Narayan}, {Chael}, {Gammie}, {Galison}, {Palumbo},
  {Doeleman}, {Blackburn}, {Wielgus}, {Pesce}, {Farah}, \&
  {Moran}}]{JohnsonLupsasca2020}
{Johnson}, M.~D., {Lupsasca}, A., {Strominger}, A., {et~al.} 2020, Science
  Advances, 6, eaaz1310, \dodoi{10.1126/sciadv.aaz1310}

\bibitem[{{Komissarov}(2002)}]{Komissarov2002}
{Komissarov}, S.~S. 2002, \mnras, 336, 759,
  \dodoi{10.1046/j.1365-8711.2002.05313.x}

\bibitem[{{Liepold} {et~al.}(2023){Liepold}, {Ma}, \& {Walsh}}]{Liepold2023}
{Liepold}, E.~R., {Ma}, C.-P., \& {Walsh}, J.~L. 2023, \apjl, 945, L35,
  \dodoi{10.3847/2041-8213/acbbcf}

\bibitem[{{Lu} {et~al.}(2023){Lu}, {Asada}, {Krichbaum}, {Park}, {Tazaki},
  {Pu}, {Nakamura}, {Lobanov}, {Hada}, {Akiyama}, {Kim}, {Marti-Vidal},
  {G{\'o}mez}, {Kawashima}, {Yuan}, {Ros}, {Alef}, {Britzen}, {Bremer},
  {Broderick}, {Doi}, {Giovannini}, {Giroletti}, {Ho}, {Honma}, {Hughes},
  {Inoue}, {Jiang}, {Kino}, {Koyama}, {Lindqvist}, {Liu}, {Marscher},
  {Matsushita}, {Nagai}, {Rottmann}, {Savolainen}, {Schuster}, {Shen}, {de
  Vicente}, {Walker}, {Yang}, {Zensus}, {Algaba}, {Allardi}, {Bach},
  {Berthold}, {Bintley}, {Byun}, {Casadio}, {Chang}, {Chang}, {Chang}, {Chen},
  {Chen}, {Chilson}, {Chuter}, {Conway}, {Crew}, {Dempsey}, {Dornbusch},
  {Faber}, {Friberg}, {Garc{\'\i}a}, {Garrido}, {Han}, {Han}, {Hasegawa},
  {Herrero-Illana}, {Huang}, {Huang}, {Impellizzeri}, {Jiang}, {Jinchi},
  {Jung}, {Kallunki}, {Kirves}, {Kimura}, {Koay}, {Koch}, {Kramer}, {Kraus},
  {Kubo}, {Kuo}, {Li}, {Lin}, {Liu}, {Liu}, {Lo}, {Lu}, {MacDonald},
  {Martin-Cocher}, {Messias}, {Meyer-Zhao}, {Minter}, {Nair}, {Nishioka},
  {Norton}, {Nystrom}, {Ogawa}, {Oshiro}, {Patel}, {Pen}, {Pidopryhora},
  {Pradel}, {Raffin}, {Rao}, {Ruiz}, {Sanchez}, {Shaw}, {Snow}, {Sridharan},
  {Srinivasan}, {Tercero}, {Torne}, {Traianou}, {Wagner}, {Walther}, {Wei},
  {Yang}, \& {Yu}}]{Lu2023}
{Lu}, R.-S., {Asada}, K., {Krichbaum}, T.~P., {et~al.} 2023, \nat, 616, 686,
  \dodoi{10.1038/s41586-023-05843-w}

\bibitem[{{Lupsasca} {et~al.}(in prep){Lupsasca}, {Chael}, {Wong}, \&
  {Quataert}}]{Paper2}
{Lupsasca}, A., {Chael}, A., {Wong}, G., \& {Quataert}, E. in prep

\bibitem[{{MacDonald} \& {Thorne}(1982)}]{MacDonald1982}
{MacDonald}, D., \& {Thorne}, K.~S. 1982, \mnras, 198, 345,
  \dodoi{10.1093/mnras/198.2.345}

\bibitem[{{Mahlmann} {et~al.}(2018){Mahlmann}, {Cerd{\'a}-Dur{\'a}n}, \&
  {Aloy}}]{Mahlmann2018}
{Mahlmann}, J.~F., {Cerd{\'a}-Dur{\'a}n}, P., \& {Aloy}, M.~A. 2018, \mnras,
  477, 3927, \dodoi{10.1093/mnras/sty858}

\bibitem[{{McKinney}(2006)}]{McKinney2006}
{McKinney}, J.~C. 2006, \mnras, 367, 1797,
  \dodoi{10.1111/j.1365-2966.2006.10087.x}

\bibitem[{{McKinney} \& {Gammie}(2004)}]{McKinney2004}
{McKinney}, J.~C., \& {Gammie}, C.~F. 2004, \apj, 611, 977,
  \dodoi{10.1086/422244}

\bibitem[{{McKinney} \& {Narayan}(2007)}]{McKinney2007}
{McKinney}, J.~C., \& {Narayan}, R. 2007, \mnras, 375, 531,
  \dodoi{10.1111/j.1365-2966.2006.11220.x}

\bibitem[{{McKinney} {et~al.}(2012){McKinney}, {Tchekhovskoy}, \&
  {Blandford}}]{McKinney2012}
{McKinney}, J.~C., {Tchekhovskoy}, A., \& {Blandford}, R.~D. 2012, \mnras, 423,
  3083, \dodoi{10.1111/j.1365-2966.2012.21074.x}

\bibitem[{{Menon}(2015)}]{Menon2015}
{Menon}, G. 2015, \prd, 92, 024054, \dodoi{10.1103/PhysRevD.92.024054}

\bibitem[{{Menon} \& {Dermer}(2007)}]{Menon2007}
{Menon}, G., \& {Dermer}, C.~D. 2007, General Relativity and Gravitation, 39,
  785, \dodoi{10.1007/s10714-007-0418-2}

\bibitem[{{Mertens} {et~al.}(2016){Mertens}, {Lobanov}, {Walker}, \&
  {Hardee}}]{Mertens2016}
{Mertens}, F., {Lobanov}, A.~P., {Walker}, R.~C., \& {Hardee}, P.~E. 2016,
  \aap, 595, A54, \dodoi{10.1051/0004-6361/201628829}

\bibitem[{{Michel}(1973)}]{Michel1973}
{Michel}, F.~C. 1973, \apjl, 180, L133, \dodoi{10.1086/181169}

\bibitem[{{Mizuno} {et~al.}(2021){Mizuno}, {Fromm}, {Younsi}, {Porth},
  {Olivares}, \& {Rezzolla}}]{Mizuno2021}
{Mizuno}, Y., {Fromm}, C.~M., {Younsi}, Z., {et~al.} 2021, \mnras, 506, 741,
  \dodoi{10.1093/mnras/stab1753}

\bibitem[{{Mo{\'s}cibrodzka} {et~al.}(2016){Mo{\'s}cibrodzka}, {Falcke}, \&
  {Shiokawa}}]{Moscibrodzka2016}
{Mo{\'s}cibrodzka}, M., {Falcke}, H., \& {Shiokawa}, H. 2016, \aap, 586, A38,
  \dodoi{10.1051/0004-6361/201526630}

\bibitem[{{Mo{\'s}cibrodzka} \& {Gammie}(2018)}]{Moscibrodzka2018}
{Mo{\'s}cibrodzka}, M., \& {Gammie}, C.~F. 2018, \mnras, 475, 43,
  \dodoi{10.1093/mnras/stx3162}

\bibitem[{{Narayan} {et~al.}(2022){Narayan}, {Chael}, {Chatterjee}, {Ricarte},
  \& {Curd}}]{Narayan2022}
{Narayan}, R., {Chael}, A., {Chatterjee}, K., {Ricarte}, A., \& {Curd}, B.
  2022, \mnras, 511, 3795, \dodoi{10.1093/mnras/stac285}

\bibitem[{{Narayan} {et~al.}(2003){Narayan}, {Igumenshchev}, \&
  {Abramowicz}}]{Narayan2003}
{Narayan}, R., {Igumenshchev}, I.~V., \& {Abramowicz}, M.~A. 2003, \pasj, 55,
  L69, \dodoi{10.1093/pasj/55.6.L69}

\bibitem[{{Narayan} {et~al.}(2021){Narayan}, {Palumbo}, {Johnson}, {Gelles},
  {Himwich}, {Chang}, {Ricarte}, {Dexter}, {Gammie}, {Chael}, \& {The Event
  Horizon Telescope Collaboration}}]{Narayan2021}
{Narayan}, R., {Palumbo}, D. C.~M., {Johnson}, M.~D., {et~al.} 2021, \apj, 912,
  35, \dodoi{10.3847/1538-4357/abf117}

\bibitem[{{Noble} {et~al.}(2006){Noble}, {Gammie}, {McKinney}, \& {Del
  Zanna}}]{Noble2006}
{Noble}, S.~C., {Gammie}, C.~F., {McKinney}, J.~C., \& {Del Zanna}, L. 2006,
  \apj, 641, 626, \dodoi{10.1086/500349}

\bibitem[{{Noble} {et~al.}(2007){Noble}, {Leung}, {Gammie}, \&
  {Book}}]{Noble2007}
{Noble}, S.~C., {Leung}, P.~K., {Gammie}, C.~F., \& {Book}, L.~G. 2007,
  Classical and Quantum Gravity, 24, S259, \dodoi{10.1088/0264-9381/24/12/S17}

\bibitem[{{Palumbo} \& {Wong}(2022)}]{Palumbo2022}
{Palumbo}, D. C.~M., \& {Wong}, G.~N. 2022, \apj, 929, 49,
  \dodoi{10.3847/1538-4357/ac59b4}

\bibitem[{{Palumbo} {et~al.}(2023){Palumbo}, {Wong}, {Chael}, \&
  {Johnson}}]{Palumbo2023}
{Palumbo}, D. C.~M., {Wong}, G.~N., {Chael}, A., \& {Johnson}, M.~D. 2023,
  \apjl, 952, L31, \dodoi{10.3847/2041-8213/ace630}

\bibitem[{{Palumbo} {et~al.}(2020){Palumbo}, {Wong}, \&
  {Prather}}]{Palumbo2020}
{Palumbo}, D. C.~M., {Wong}, G.~N., \& {Prather}, B.~S. 2020, \apj, 894, 156,
  \dodoi{10.3847/1538-4357/ab86ac}

\bibitem[{{Parfrey} {et~al.}(2019){Parfrey}, {Philippov}, \&
  {Cerutti}}]{Parfrey2019}
{Parfrey}, K., {Philippov}, A., \& {Cerutti}, B. 2019, \prl, 122, 035101,
  \dodoi{10.1103/PhysRevLett.122.035101}

\bibitem[{{Penna} {et~al.}(2013){Penna}, {Narayan}, \&
  {S{\k{a}}dowski}}]{Penna2013}
{Penna}, R.~F., {Narayan}, R., \& {S{\k{a}}dowski}, A. 2013, \mnras, 436, 3741,
  \dodoi{10.1093/mnras/stt1860}

\bibitem[{{Penrose}(1969)}]{Penrose1969}
{Penrose}, R. 1969, Nuovo Cimento Rivista Serie, 1, 252

\bibitem[{{Penrose} \& {Rindler}(1984)}]{Penrose1984}
{Penrose}, R., \& {Rindler}, W. 1984, {Spinors and space-time. Vol. 1:
  Two-spinor calculus and relativistic fields.} (Cambridge University Press)

\bibitem[{{Phinney}(1983)}]{Phinney1983}
{Phinney}, E.~S. 1983, PhD thesis, -

\bibitem[{{Pu} {et~al.}(2016){Pu}, {Akiyama}, \& {Asada}}]{Pu2016}
{Pu}, H.-Y., {Akiyama}, K., \& {Asada}, K. 2016, \apj, 831, 4,
  \dodoi{10.3847/0004-637X/831/1/4}

\bibitem[{{Qiu} {et~al.}(2023){Qiu}, {Ricarte}, {Narayan}, {Wong}, {Chael}, \&
  {Palumbo}}]{Qiu2023}
{Qiu}, R., {Ricarte}, A., {Narayan}, R., {et~al.} 2023, \mnras, 520, 4867,
  \dodoi{10.1093/mnras/stad466}

\bibitem[{{Raymond} {et~al.}(2021){Raymond}, {Palumbo}, {Paine}, {Blackburn},
  {C{\'o}rdova Rosado}, {Doeleman}, {Farah}, {Johnson}, {Roelofs}, {Tilanus},
  \& {Weintroub}}]{Raymond2021}
{Raymond}, A.~W., {Palumbo}, D., {Paine}, S.~N., {et~al.} 2021, \apjs, 253, 5,
  \dodoi{10.3847/1538-3881/abc3c3}

\bibitem[{{Ricarte} {et~al.}(2022){Ricarte}, {Palumbo}, {Narayan}, {Roelofs},
  \& {Emami}}]{Ricarte2022}
{Ricarte}, A., {Palumbo}, D. C.~M., {Narayan}, R., {Roelofs}, F., \& {Emami},
  R. 2022, \apjl, 941, L12, \dodoi{10.3847/2041-8213/aca087}

\bibitem[{{Ricarte} {et~al.}(2020){Ricarte}, {Prather}, {Wong}, {Narayan},
  {Gammie}, \& {Johnson}}]{Ricarte2020}
{Ricarte}, A., {Prather}, B.~S., {Wong}, G.~N., {et~al.} 2020, \mnras, 498,
  5468, \dodoi{10.1093/mnras/staa2692}

\bibitem[{{Schnittman}(2015)}]{Schnittman2015}
{Schnittman}, J.~D. 2015, \apj, 806, 264, \dodoi{10.1088/0004-637X/806/2/264}

\bibitem[{{S{\k{a}}dowski} {et~al.}(2014){S{\k{a}}dowski}, {Narayan},
  {McKinney}, \& {Tchekhovskoy}}]{Sadowski2014}
{S{\k{a}}dowski}, A., {Narayan}, R., {McKinney}, J.~C., \& {Tchekhovskoy}, A.
  2014, \mnras, 439, 503, \dodoi{10.1093/mnras/stt2479}

\bibitem[{{S{\k{a}}dowski} {et~al.}(2013){S{\k{a}}dowski}, {Narayan},
  {Tchekhovskoy}, \& {Zhu}}]{Sadowski2013}
{S{\k{a}}dowski}, A., {Narayan}, R., {Tchekhovskoy}, A., \& {Zhu}, Y. 2013,
  \mnras, 429, 3533, \dodoi{10.1093/mnras/sts632}

\bibitem[{{Tchekhovskoy} {et~al.}(2012){Tchekhovskoy}, {McKinney}, \&
  {Narayan}}]{Tchekhovskoy2012}
{Tchekhovskoy}, A., {McKinney}, J.~C., \& {Narayan}, R. 2012, in Journal of
  Physics Conference Series, Vol. 372, Journal of Physics Conference Series,
  012040, \dodoi{10.1088/1742-6596/372/1/012040}

\bibitem[{{Tchekhovskoy} {et~al.}(2010){Tchekhovskoy}, {Narayan}, \&
  {McKinney}}]{Tchekhovskoy2010}
{Tchekhovskoy}, A., {Narayan}, R., \& {McKinney}, J.~C. 2010, \apj, 711, 50,
  \dodoi{10.1088/0004-637X/711/1/50}

\bibitem[{{Tchekhovskoy} {et~al.}(2011){Tchekhovskoy}, {Narayan}, \&
  {McKinney}}]{Tchekhovskoy2011}
---. 2011, \mnras, 418, L79, \dodoi{10.1111/j.1745-3933.2011.01147.x}

\bibitem[{{The Event Horizon Telescope Collaboration}
  {et~al.}(2019{\natexlab{a}})}]{PaperI}
{The Event Horizon Telescope Collaboration}, {et~al.} 2019{\natexlab{a}},
  \apjl, 875, L1, \dodoi{10.3847/2041-8213/ab0ec7}

\bibitem[{{The Event Horizon Telescope Collaboration}
  {et~al.}(2019{\natexlab{b}})}]{PaperV}
---. 2019{\natexlab{b}}, \apjl, 875, L5, \dodoi{10.3847/2041-8213/ab0f43}

\bibitem[{{The Event Horizon Telescope Collaboration}
  {et~al.}(2019{\natexlab{c}})}]{PaperVI}
---. 2019{\natexlab{c}}, \apjl, 875, L6, \dodoi{10.3847/2041-8213/ab1141}

\bibitem[{{The Event Horizon Telescope Collaboration}
  {et~al.}(2021{\natexlab{a}})}]{PaperVII}
---. 2021{\natexlab{a}}, \apjl, 910, L12, \dodoi{10.3847/2041-8213/abe71d}

\bibitem[{{The Event Horizon Telescope Collaboration}
  {et~al.}(2021{\natexlab{b}})}]{PaperVIII}
---. 2021{\natexlab{b}}, \apjl, 910, L13, \dodoi{10.3847/2041-8213/abe4de}

\bibitem[{{The Event Horizon Telescope Collaboration}
  {et~al.}(2022)}]{PaperI_SgrA}
---. 2022, \apjl, 930, L12, \dodoi{10.3847/2041-8213/ac6674}

\bibitem[{{Thorne} {et~al.}(1986){Thorne}, {Price}, \&
  {MacDonald}}]{Thorne1986}
{Thorne}, K.~S., {Price}, R.~H., \& {MacDonald}, D.~A. 1986, {Black holes: The
  membrane paradigm} (New Haven: Yale University Press)

\bibitem[{{Vincent} {et~al.}(2022){Vincent}, {Gralla}, {Lupsasca}, \&
  {Wielgus}}]{Vincent2022}
{Vincent}, F.~H., {Gralla}, S.~E., {Lupsasca}, A., \& {Wielgus}, M. 2022, \aap,
  667, A170, \dodoi{10.1051/0004-6361/202244339}

\bibitem[{{Virtanen} {et~al.}(2020){Virtanen}, {Gommers}, {Oliphant},
  {Haberland}, {Reddy}, {Cournapeau}, {Burovski}, {Peterson}, {Weckesser},
  {Bright}, {van der Walt}, {Brett}, {Wilson}, {Millman}, {Mayorov}, {Nelson},
  {Jones}, {Kern}, {Larson}, {Carey}, {Polat}, {Feng}, {Moore}, {VanderPlas},
  {Laxalde}, {Perktold}, {Cimrman}, {Henriksen}, {Quintero}, {Harris},
  {Archibald}, {Ribeiro}, {Pedregosa}, {van Mulbregt}, \& {SciPy 1. 0
  Contributors}}]{SciPy2020}
{Virtanen}, P., {Gommers}, R., {Oliphant}, T.~E., {et~al.} 2020, Nature
  Methods, 17, 261, \dodoi{10.1038/s41592-019-0686-2}

\bibitem[{{Williams}(1995)}]{Williams1995}
{Williams}, R.~K. 1995, \prd, 51, 5387, \dodoi{10.1103/PhysRevD.51.5387}

\bibitem[{{Wong} {et~al.}(in prep){Wong}, {Lupsasca}, {Chael}, \&
  {Quataert}}]{Paper3}
{Wong}, G., {Lupsasca}, A., {Chael}, A., \& {Quataert}, E. in prep

\bibitem[{{Wong} {et~al.}(2021){Wong}, {Du}, {Prather}, \& {Gammie}}]{Wong2021}
{Wong}, G.~N., {Du}, Y., {Prather}, B.~S., \& {Gammie}, C.~F. 2021, \apj, 914,
  55, \dodoi{10.3847/1538-4357/abf8b8}

\bibitem[{{Znajek}(1977)}]{Znajek1977}
{Znajek}, R.~L. 1977, \mnras, 179, 457, \dodoi{10.1093/mnras/179.3.457}

\end{thebibliography}
